\documentclass[aps,prd,reprint,11pt,onecolumn,amsmath,amssymb,nofootinbib,groupedaddress,superscriptaddress,floatfix,tightenlines,eqsecnum]{revtex4-2}

\usepackage{macros}
\newcommand{\medi}{MEDI}

\graphicspath{{./figs/}}

\begin{document}

\title{High-redshift physics from the acoustic scale}

\author{Zachary J. Weiner}\email[]{zweiner@perimeterinstitute.ca}
\affiliation{Perimeter Institute for Theoretical Physics, Waterloo, Ontario N2L 2Y5, Canada}

\date{\today}
\begin{abstract}
We present a simplified and general description of the high-redshift information in acoustic scale
measurements from the cosmic microwave background and large-scale structure.
The transverse distance interval between photon--baryon decoupling and a late epoch in the matter
era provides an analytically tractable summary statistic thereof and a general diagnostic of the
current tension between the Dark Energy Spectroscopic Instrument and the CMB.
We show that this ``matter-era distance excess'' is unlikely to be explained by modified dynamics at
low redshift.
We then analytically derive the matter-era distance interval's sensitivity to new physics at high
redshift, including nonstandard recombination, nonminimal dark matter dynamics, and spatial
curvature; in particular, we explain how this observable represents a direct geometric measurement
of (and underlies the current incompatibility with) neutrino masses.
Finally, we demonstrate that phenomenological models of dynamical dark energy mediate the matter-era
distance excess in a manner reliant on their unphysical, extrapolated behavior at high redshift.
Invoking alternative explanations of the excess removes the CMB's contribution to the evidence for
these models; the residual preference of around $1.7\sigma$ mostly derives from DESI's two
lowest-redshift measurements of the Alcock--Paczynski distortion, without which it drops to
$0.5 \sigma$.
\end{abstract}

\maketitle
\makeatletter
\let\oldl@section\l@section
\renewcommand{\l@section}[2]{\vspace{-0.25\baselineskip}\oldl@section{#1}{#2}}
\def\l@subsubsection#1#2{}
\makeatother
\vspace{-\baselineskip}
\tableofcontents

\section{Introduction}
\label{sec:introduction}

Cosmological distances derived from the acoustic scale are the leading source of contemporary
information on the expansion history from high to low redshift.
The angular extent of the sound horizon on the sky as inferred from the photon distribution
represents the second-best measured number in cosmology~\cite{Planck:2018vyg, AtacamaCosmologyTelescope:2025blo,
SPT-3G:2025bzu} (next to the cosmic microwave background temperature~\cite{Fixsen:1996nj,
Mather:1993ij}), while analogous measurements in the distribution of galaxies and other tracers of
large-scale structure have achieved subpercent precision~\cite{DESI:2025zgx}.
Both are reasonably independent of the underlying cosmology and are modeled from first principles.
Taking seriously any microphysical implications of these measurements demands a rigorous
understanding of not just their potential systematics but also all cosmological physics relevant to
their predictions---including possible alternative explanations of the data---and what complementary
observational information such inference relies on.

In this work, we present a comprehensive and analytic description of the high-redshift information
in the acoustic scale, i.e., distances from baryon acoustic oscillations (BAO) and the cosmic
microwave background (CMB).
While we primarily focus on physics outside of the redshift coverage of BAO measurements (which is
presently $z \lesssim 2-3$), our treatment does not ignore possible dynamics beyond the $\Lambda$
cold dark matter (\LCDM{}) model in this redshift range.
Rather, we devise an effective summary of the information in acoustic scale data on
intermediate- and high-redshift physics in the form of a single measurement that is nearly
independent of low-redshift dynamics and whose predictions can be understood analytically.
This description substantially sharpens the typical, vague appeal to degeneracy breaking in
explaining how BAO data contribute to the inference of cosmological parameters (both in their
reduced uncertainties and potentially shifted central values), which often can be incomplete or
simply incorrect.

Our approach builds on insight from Ref.~\cite{Loverde:2024nfi} on how the acoustic scale
contributes to measurements of neutrino masses in an otherwise \LCDM{} cosmology, motivated by the
incompatibility of the Dark Energy Spectroscopic Instrument's BAO data (combined with CMB data) with
nonzero neutrino masses~\cite{DESI:2024mwx, DESI:2025zgx}.
We extend this analysis beyond \LCDM{} and neutrino masses, explaining how the neutrino mass tension
serves as a proxy for the general tension between DESI and the CMB.
We quantify the relevant information as the distance photons travel during matter domination
(relative to the sound horizon), which reduces to a direct measurement of the matter density in the
simplest cases.
This observable was in fact proposed long ago as a tool to ``search for dynamical shenanigans'' at
high redshift in Ref.~\cite{Eisenstein:2004an}, and its value in constraining curvature was
recognized in Ref.~\cite{Knox:2005hx}; in this work, we demonstrate its paramount role in
constraining the cosmological model in general.
In \cref{sec:matter-era-distance-interval}, we define the matter-era distance interval (\medi{}),
derive its full parameter dependence in \LCDM{}, and obtain its measurement from the acoustic scale.
\Cref{sec:matter-density} explains how the \medi{} represents the key contribution of the CMB's
geometric information to the constraining power of acoustic scale data.
We then show in \cref{sec:medi-tension} that the tension between the acoustic scale and the CMB is
most generally summarized by the \medi{} as a matter-era distance excess.

In \cref{sec:high-redshift-solutions}, we demonstrate the matter-era distance interval's efficacy in
summarizing the sensitivity of CMB and BAO data to various modifications to the cosmological model
at intermediate and high redshift, including not just massive neutrinos~\cite{Lesgourgues:2012uu}
(and hyperlight scalar fields, which are phenomenologically similar~\cite{Amendola:2005ad,
Marsh:2011bf, Hlozek:2014lca, Hlozek:2016lzm, Baryakhtar:2024rky}) but also dark matter that
decays~\cite{Turner:1984ff, BOSS:2014hhw, Poulin:2016nat, Bringmann:2018jpr, McCarthy:2022gok,
Lynch:2025ine, Montandon:2026vuc} or whose mass evolves (as realized in scalar-mediated dark force
models~\cite{Archidiacono:2022iuu, Bottaro:2023wkd, Bottaro:2024pcb, Costa:2025kwt}).
Beyond modifications to the postdecoupling expansion history, the \medi{}
summarizes most of the information on spatial curvature and any other scenario featuring a geometric
degeneracy in the CMB---namely, early recombination via a time-varying electron
mass~\cite{Sekiguchi:2020teg, Baryakhtar:2024rky, Loverde:2024nfi} and the (closely
related~\cite{Baryakhtar:2024rky}) hypothesis that the present-day CMB temperature is
mismeasured~\cite{Ivanov:2020mfr}.
We also explain why high-density recombination models (like dark radiation~\cite{Wyman:2013lza,
Dvorkin:2014lea, Bernal:2016gxb} or early dark energy~\cite{Poulin:2018cxd, Poulin:2018dzj,
Smith:2019ihp, Hill:2020osr, Poulin:2023lkg}) have a parametrically less relevant effect on acoustic
scale predictions and therefore are not candidates to mediate the BAO--CMB tension.
In addition, we identify the physical mechanism by which a larger-than-measured optical depth to
reionization~\cite{Loverde:2024nfi, Sailer:2025lxj, Jhaveri:2025neg} alleviates the matter-era
distance tension.
We summarize the status of these various proposals by presenting their impact on the neutrino mass
tension in \cref{sec:mnu-vs-highz}.

Though designed to be maximally independent of dark energy, the matter-era distance interval
is presently affected by its density at the subpercent level.
In \cref{sec:dde-alternatives}, we show that the CMB's contribution to putative preferences for
dynamical dark energy~\cite{DESI:2025zgx} derives fully from mediating the matter-era distance
excess.
Moreover, our analytic description demonstrates that its resolution relies specifically on
extrapolating linearized dark-energy equations of state to high redshift, where the behavior is
nominally the most unphysical (i.e., as interpreted for fluidlike dark energy alone).
We show that any alternative resolution to the matter-era distance tension removes the evidence for
dark energy dynamics added by the CMB; these high-redshift alternatives also have greater potential
to resolve the geometric side of the neutrino mass tension.
Finally, we show that the residual, ``direct'' evidence for dark energy dynamics from DESI derives
mostly from the Alcock--Paczynski distortion measured in its two lowest-redshift bins.
We summarize our analysis in \cref{sec:conclusions} and conclude by discussing the outlook for
resolving the matter-era distance and neutrino mass tensions and for disentangling genuine evidence
of dark energy dynamics from high-redshift alternatives.

\section{The matter-era distance interval}
\label{sec:matter-era-distance-interval}

Were late-time observations consistent with an Einstein--de Sitter Universe, i.e., zero dark energy
and curvature, the CMB, even treated as unlensed, would uniquely measure the matter abundance at
early \textit{and} late times---that in CDM and baryons at recombination (via the shape of the
anisotropy spectra) and in all nonrelativistic matter present at late times (via the distance to
last scattering).
Disagreement between these measurements either would suggest the baryon and CDM abundances were
miscalibrated or would simply measure the change in matter abundance after recombination---in
particular, an increase from massive neutrinos, but also changes due to a range of nonminimal dark
matter phenomena.
Our premise is to combine measurements to realize this idealized scenario as robustly as possible.

We start in \cref{sec:theory} by precisely defining the observable---heuristically, the distance
photons travel in the matter era---and explaining its dependence on the energy content of the
standard cosmological model; we comment on what generalizations thereof this matter-era distance
interval (\medi{}) is sensitive to, deferring dedicated analyses to
\cref{sec:high-redshift-solutions}.
We then derive measurements of the \medi{} from acoustic scale observations, discussing their
robustness to the low-redshift expansion history and to nonminimal physics around photon--baryon
decoupling.
Finally, \cref{sec:medi-tension} demonstrates that the tension between CMB data and BAO data from
DESI is generally described as a matter-era distance excess.
In particular, we show how this tension manifests as a matter density deficit in standard cosmology
that penalizes nonzero neutrino masses.

\subsection{Theoretical description}\label{sec:theory}

The photons observed by large-scale structure surveys traversed a large fraction of their distance
during dark energy domination, despite it beginning only a fraction of an $e$-fold ago.
CMB photons travel a uniquely large distance during matter domination, but still retain a sizeable
sensitivity to dark energy.
To isolate the (majority of the) impact of matter on distances, we separate the comoving distance
$\chi(a)$ into contributions from during and after the matter era as
\begin{align}
    \chi(a)
    = \int_{a}^{1} \frac{\ud \ln \tilde{a}}{\tilde{a} H(\tilde{a})}
    = \int_{a}^{a_m} \frac{\ud \ln \tilde{a}}{\tilde{a} H(\tilde{a})}
        + \int_{a_m}^{1} \frac{\ud \ln \tilde{a}}{\tilde{a} H(\tilde{a})}
    &\equiv \chi(a, a_m) + \chi(a_m)
    \label{eqn:distance-interval-def}
    ,
\end{align}
with $\chi(a_1, a_2)$ denoting the comoving distance interval between $a_1$ and $a_2$,
$\chi(a) = \chi(a, 1)$, and $a_m$ some yet-unspecified scale factor late in the matter era.
The matter-era distance interval $\chi(a, a_m)$ may be approximated (for illustrative if
not quantitative purposes) by a Universe without dark energy, while the distance to matter
domination $\chi(a_m)$, so to speak, requires the full energy content from $a_m$ to today.
We may generalize $\chi(a_1, a_2)$ in terms of the
transverse distance $D_M \equiv \chi \sinc \left[ \chi / R_k \right]$,
with $R_k^2 = - 1 / \omega_k H_{100}^2$ the curvature radius and $\omega_k$ the (signed)
present density in curvature,\footnote{
    We parametrize present-day densities with
    $\omega_X \equiv \bar{\rho}_{X, 0} / 3 H_{100}^2 \Mpl^2$ where
    $H_0 = h \cdot 100~\mathrm{Mpc}^{-1} \mathrm{km} / \mathrm{s} \equiv h H_{100}$.
}
via
\begin{align}
    D_M(a_1, a_2) \equiv D_M(a_1) - D_M(a_2);
    \label{eqn:tranvserse-medi-def}
\end{align}
except for in \cref{sec:theory-curvature}, we fix a flat Universe and refer to $\chi$ in such
calculations, but we denote observables by $D_M$.

\subsubsection{Matter and radiation}

In a standard matter--radiation Universe with matter density $\omega_m$ that equals the radiation
density at an equality scale factor $a_\mathrm{eq} = \omega_r / \omega_m$,
\begin{align}
    \chi_{mr}(a_\mathrm{d}, a_m)
    &= \frac{
            \sqrt{a_m + a_\mathrm{eq}}
            - \sqrt{a_\mathrm{d} + a_\mathrm{eq}}
        }{
            H_{100} \sqrt{\omega_m} / 2
        }
    .
    \label{eqn:medi-mr}
\end{align}
Here we take photon--baryon decoupling ($a_\mathrm{d}$) as the high-redshift anchor, anticipating
arguments made in \cref{sec:inference}.
At fixed radiation density (or just $a_\mathrm{eq}$ as constrained by the CMB with
$\omega_r$ free), \cref{eqn:medi-mr} is parametrized by $\omega_m$ alone.
To the extent that low-redshift distances directly constrain the expansion history between the
present and the end of the matter era [and therefore $\chi(a_m)$], a geometric measurement of the
matter density reduces to inverting \cref{eqn:medi-mr} for $\omega_m$, analogous to that which the
distance to last scattering would provide in an Einstein--de Sitter Universe.

If no additional matter components become relevant after decoupling, the size of the Universe in the
matter era is determined by the density $\bar{\rho}_{cb}$ in baryons and CDM.
The shape of the temperature and polarization spectra measures
$\bar{\rho}_{cb}(a) / \bar{\rho}_r(a) = a / a_\mathrm{eq}$ around last
scattering~\cite{Eisenstein:2004an, Hu:2004kn, Weinberg:2008zzc, Ilic:2020onu, Baryakhtar:2024rky,
Costa:2025kwt}, setting a baseline expectation for the acoustic scale with $\omega_m$ determined by
$a_\star^3 \bar{\rho}_{cb}(a_\star) = \omega_{cb}$ with $a_\star$ the scale factor of peak
visibility.
In more generality, $\chi(a_\mathrm{d}, a_m)$ is sensitive to physics that modifies
the extrapolation of the CMB-calibrated density in dark matter and baryons to the present, i.e.,
that the total matter density $\bar{\rho}_m(a \gg a_\star)$ might differ from
$\bar{\rho}_{cb}(a_\star) / (a / a_\star)^3$.
Photons travel furthest when the Universe is largest, making the transverse distances out to any
redshift most sensitive to the density in matter at late times.
We consider specific models featuring nonminimal dynamics of matter in \cref{sec:postdecoupling} and
treat the most important application, massive neutrinos, in \cref{sec:mnu}; for our present,
illustrative purposes, it suffices to treat the ``late-time'' matter density (that measured by
distances) as an independent parameter without specifying its physical makeup.

To build intuition for these more general scenarios, we employ a simplified treatment in which
the matter density instantaneously changes at some scale factor $a_\times$ after decoupling.
Simply treating $\omega_m = a^3 \bar{\rho}_m(a) / 3 H_{100}^2 \Mpl^2$ (at $a > a_\times$) as a free
parameter not necessarily equal to $\omega_{cb}$, we apply \cref{eqn:medi-mr} to write
$\chi(a_\mathrm{d}, a_m) = \chi(a_\mathrm{d}, a_\times) + \chi(a_\times, a_m)$ and
\begin{align}
   \chi(a_\mathrm{d}, a_m)
    &\simeq \frac{
                \sqrt{a_m}
                - \sqrt{a_\mathrm{d} + a_\mathrm{eq}}
            }{H_{100} \sqrt{\omega_m} / 2}
        \left[
            1
            + \left( \sqrt{\frac{\omega_m}{\omega_{cb}}} - 1 \right)
            \frac{
                \sqrt{a_\times}
                - \sqrt{a_\mathrm{d} + a_\mathrm{eq}}
            }{
                \sqrt{a_m}
                - \sqrt{a_\mathrm{d} + a_\mathrm{eq}}
            }
        \right]
    .
    \label{eqn:medi-mr-piecewise}
\end{align}
For simplicity, we take $a_m > a_\times \gg a_\mathrm{eq}$ (which also hides any small sensitivity
to model-dependent changes in the relative amount of matter and radiation).
The dependence of \cref{eqn:medi-mr-piecewise} on $a_\times$ is suppressed by the smallness of the
change in matter density and any hierarchy in $a_\times$ and $a_m$; however, the larger
$\sqrt{a_\times / a_m}$, the more sensitive a measurement of $\omega_m / \omega_{cb}$ is to the
transition (including its noninstantaneous dynamics in any particular scenario).

The variability of $a_\mathrm{eq}$ and $a_\mathrm{d}$ is a highly subdominant source of uncertainty
in \cref{eqn:medi-mr}.
(Per above, $\omega_m$ is not set by $\omega_r / a_\mathrm{eq}$ in general.)
That is, since $a_\mathrm{eq} < a_\mathrm{d} \ll a_m$, the relative sensitivities of
\cref{eqn:medi-mr} are
$\partial \ln \chi_{mr}(a_\mathrm{d}, a_m) / \partial \ln a_\mathrm{eq}
\simeq - a_\mathrm{eq} / 2 \sqrt{a_m (a_\mathrm{d} + a_\mathrm{eq})}
\approx - 0.008$
and $\partial \ln \chi_{mr}(a_\mathrm{d}, a_m) / \partial \ln a_\mathrm{d}
\simeq - a_\mathrm{d} / 2 \sqrt{a_m (a_\mathrm{d} + a_\mathrm{eq})}
\approx - 0.03$.
The subpercent and (well) subpermille constraints on $a_\mathrm{eq}$ and $a_\mathrm{d}$,
respectively, from current CMB data in \LCDM{} thus propagate uncertainty to the \medi{} that is
subdominant even to the uncertainty in the CMB's acoustic scale information (by respective factors
of $\sim 5$ and $\sim 50$) and do not impede the extraction of matter-era information.

\subsubsection{Massive neutrinos}\label{sec:mnu}

An increase in the matter abundance after decoupling is an irreducible prediction of standard
cosmology and particle physics~\cite{Lesgourgues:2012uu},\footnote{
    The postdecoupling dynamics of hyperlight scalar fields closely resemble those of massive
    neutrinos; we leave a discussion thereof to \cref{app:scalars}.
} as neutrinos with mass $m_{\nu_i}$ (that are light enough to satisfy even the most conservative
bounds from the CMB~\cite{Planck:2018vyg, Hou:2012xq}) become nonrelativistic after decoupling at
\begin{align}
    z_{\nu_i} + 1
    = 1 / a_{\nu_i}
    = \frac{m_{\nu_i}}{3.15 T_\nu(a_0)}
    = 94.3 \,
        \frac{m_{\nu_i}}{50~\mathrm{meV}}
    \label{eqn:massive-neutrino-z-NR}
\end{align}
and have a present-day mass density
\begin{align}
    \omega_{\nu_i}
    = \frac{3 \zeta(3)}{2 \pi^2} \frac{m_{\nu_i} T_\nu(a_0)^3}{3 H_{100}^2 \Mpl^2}
    &= \frac{m_{\nu_i}}{93.1~\mathrm{eV}}.
    \label{eqn:massive-neutrino-density}
\end{align}
Neutrino oscillation experiments require their summed mass $\summnu$ be at least
$0.0588~\mathrm{eV}$ or $0.099~\mathrm{eV}$ in the normal and inverted hierarchies,
respectively~\cite{deSalas:2020pgw, Esteban:2020cvm}, for a respective total mass density
$\omega_\nu = \summnu / 93.1 = 6.23 \times 10^{-4}$ or $1.06 \times 10^{-3}$.
For both hierarchies, at least one mass eigenstate is heavier than $50~\mathrm{meV}$ and therefore
becomes nonrelativistic around redshift $94.3$.
In practice, we follow convention in taking a degenerate mass hierarchy for simplicity and in order
to consider mass sums as small as zero, for which reason we do not distinguish between eigenstates
below.

Since $a_\nu / a_m$ is not especially small, the impact of neutrino masses on the \medi{}
decreases superlinearly with smaller neutrino fraction
$f_\nu \equiv \omega_\nu / (\omega_{cb} + \omega_\nu)
\approx \summnu / 13.2~\mathrm{eV} \cdot (\omega_{cb} / 0.142)^{-1}$.
A piecewise approximation in the style of \cref{eqn:medi-mr-piecewise} suggests the sensitivity
coefficient is roughly
$\partial \ln \chi(a_\mathrm{d}, a_m) / \partial f_\nu = - (1 - \sqrt{a_\nu / a_m}) / 2$, which
does provide a good estimate.
The neutrinos' nonrelativistic transition is rather gradual, however; in \cref{app:mnu} we derive an
improved analytic result that accounts for this transition, following a perturbative approach
described in \cref{sec:postdecoupling}:
\begin{align}
\begin{split}
    \frac{\Delta \chi(a_\mathrm{d}, a_m)}{1 / H_{100} \sqrt{\omega_{cb}}}
    &\simeq
        - f_\nu \sqrt{a_m}
        \left[
            1
            + \frac{\sqrt{a_\nu / a_m}}{\sqrt{a_\times / a_\nu}}
            \left(
                \frac{7 \pi^6 \left( a_\times / a_\nu \right)^2}{19440 \zeta(3)^2}
                - \frac{a_\times}{a_\nu}
                - 1
            \right)
            + a_\nu / a_m
        \right]
        .
    \label{eqn:medi-mnu-apx}
\end{split}
\end{align}
Here $a_\times$ is a matching scale factor chosen near $a_\nu$ to minimize errors; \cref{app:mnu}
describes the rather good accuracy of \cref{eqn:medi-mnu-apx} for $a_m = 1 / (1 + 2.330)$ with
$a_\times = 1.32 a_\nu$, in which case the numerical coefficient multiplying $\sqrt{a_\nu / a_m}$ is
about $-1.66$.
\Cref{eqn:medi-mnu-apx} [but not \cref{eqn:medi-mnu}] takes $a_\mathrm{d}$ and $a_\mathrm{eq}$ to be
negligible compared to $a_m$ and $a_\nu$ for simplicity's sake.

\subsubsection{Cosmological constant}\label{sec:theory-cc}

The express purpose of the matter-era distance interval is to maximally sequester the impact of dark
energy on distances.
Nevertheless, precision inference is sensitive to the nominally small effects of dark energy,
especially since its presence only increases the total density (and decreases distances).
Fortunately, the smallness of the effect on the \medi{} (in contrast to any individual distance)
permits analytic approximations.
The comoving distance in a Universe with matter and cosmological constant whose densities are
equal at scale factor
$\amL = \sqrt[3]{\omega_m / \omega_\Lambda} = \sqrt[3]{\Omega_m / (1 - \Omega_m)}$
is~\cite{Baes:2017rfj}
\begin{align}
    \chi(a)
    &= \frac{2}{\sqrt{\omega_m} H_{100} / c}
        \left[ F_M(1; \amL) - F_M(a; \amL) \right].
    \label{eqn:comoving-distance-matter-Lambda}
\end{align}
Here $F_M(a; \amL)$ encodes the correction from $\Lambda$ to the result for a pure matter Universe
[$F_M(a; \amL) \to \sqrt{a}$ for $\amL \to \infty$]
and is given in terms of the hypergeometric function ${}_2{F}_1$ by
\begin{align}
    F_M(a; \amL)
    &\equiv \sqrt{a} \cdot {}_2{F}_1(1/6, 1/2; 7/6, - [a / \amL]^3)
    \label{eqn:matter-lambda-distance-function}
    \\
    &= \sqrt{a} \left[ 1 - (a / \amL)^3 / 14 + \mathcal{O}([a / \amL]^6) \right]
    ,
    \label{eqn:hyp-series}
\end{align}
the latter expression indicating the leading-order result in small $a / \amL$.
Comparing \cref{eqn:comoving-distance-matter-Lambda,eqn:matter-lambda-distance-function} for a
matter-$\Lambda$ Universe and \cref{eqn:medi-mr} for a matter--radiation Universe
motivates a factorized ansatz for $\chi(a)$ that replaces $\sqrt{a}$ in
\cref{eqn:matter-lambda-distance-function} with $\sqrt{a + a_\mathrm{eq}}$, which is in fact
accurate at the $0.03\%$ level or better.
The error from applying the factorized approximation to the \medi{} in the form
\begin{align}
    \chi(a_\mathrm{d}, a_m)
    &\simeq \frac{
            \left[ 1 - (a_m / \amL)^3 / 14 \right] \sqrt{a_m + a_\mathrm{eq}}
            - \sqrt{a_\mathrm{d} + a_\mathrm{eq}}
        }{
            H_{100} \sqrt{\omega_m}/ 2
        }
    \label{eqn:medi-with-cc}
\end{align}
[neglecting $(a_\mathrm{d} / \amL)^3 \sim 10^{-9}$] is below $10^{-4}$ at all
redshifts $z_m = 1 / a_m - 1 \gtrsim 2$ (i.e., those of interest).
Even simply multiplying the matter--radiation result $\chi_{mr}(a_\mathrm{d}, a_m)$
[\cref{eqn:medi-mr}] by the approximate correction in \cref{eqn:hyp-series} achieves comparable
accuracy.
We generalize this result beyond a cosmological constant in \cref{sec:dde}.

The coefficient $1 / 14$ in \cref{eqn:medi-with-cc} affirms the suppressed sensitivity of the
integrated distance to dark energy's instantaneous contribution to the energy density.
For instance, the fiducial $\sim 5.7\%$ contribution at $z_m = 2.330$ (for DESI DR2's Lyman-$\alpha$
sample) corresponds to a $0.4\%$ effect on $\chi(a_\mathrm{d}, a_m)$.
Taking the matching redshift $z_m$ to be the highest redshift measured by BAO distances thus
serendipitously compartmentalizes the support of the distance to decoupling into the interval
that is directly measured from tracers of large-scale structure [$\chi(a)$ for $a < a_m$] and that
which depends almost exclusively on matter [$\chi(a_\mathrm{d}, a_m)$].
Dark energy's nominal half-percent effect is of course not irrelevant for deriving subpercent
measurements of the matter density (or other effects on the \medi{}), but it does indicate a large
hierarchy in information.\footnote{
    The logarithmic sensitivities we quote translate the relative precision of the
    measurement to the relative precision of the resulting parameter constraint.
}
That is, while
$\partial \ln \chi(a_\mathrm{d}, a_m) / \partial \ln \omega_m = -1/2$,
the sensitivity to dark energy drops with its density fraction and is further suppressed by the
smallness of its contribution to the full \medi{}:
$\partial \ln \chi(a_\mathrm{d}, a_m) / \partial \ln \Omega_\mathrm{DE}(a_m)
\simeq \Omega_\mathrm{DE}(a_m) / 14$, smaller than the former by more
than a factor of $100$.
Lower-redshift distances $\chi(a)$, however parametrized, necessarily have order-unity sensitivity
to dark energy.

\subsection{Measurement from the acoustic scale}
\label{sec:inference}

We proceed by devising model-independent measurements of the matter-era distance interval.
Redshift surveys extract distances from the acoustic scale via the transverse and longitudinal scales
\begin{subequations}\label{eqn:theta-bao}
\begin{align}
    \theta_\perp(a)
    &\equiv r_\mathrm{d} / D_M(a)
    \label{eqn:theta-bao-perp}
    \\
    \theta_\parallel(a)
    &\equiv r_\mathrm{d} / c H(a)^{-1}
    \equiv r_\mathrm{d} / D_H(a)
    \label{eqn:theta-bao-parallel}
    ,
\end{align}
or, when sample statistics preclude separate measurements thereof, a volume average of the two,
\begin{align}
    \theta_\mathrm{BAO}(a)
    &\equiv \sqrt[3]{
                \theta_\perp(a)^2
                \theta_\parallel(a)
                \left( 1 / a - 1 \right)
        }
    \label{eqn:theta-bao-DV}.
\end{align}
\end{subequations}
Here $r_\mathrm{d}$ is the sound horizon at baryon decoupling, $a_\mathrm{d}$.
The \medi{} is best measured by the difference in distance to photon--baryon decoupling as inferred
from the CMB and to some scale factor $a_m$ later in the matter era from, say, BAO, taking the form
$D_M(a_\mathrm{d}, a_m) / r_\mathrm{d} = 1/\theta_\perp(a_\mathrm{d}) - 1/\theta_\perp(a_m)$.
Since the high-redshift end is supplied by the CMB, of which there is only a single (and exceedingly
precise) realization, the problem reduces to measuring the distance to matter domination, loosely
defined.
We next discuss how the CMB (i.e., the photon distribution) constrains the acoustic feature in the
baryon distribution at decoupling, the measurements of $D_M(a_m) / r_\mathrm{d}$ from BAO data,
and the implications of the drag horizon serving as a distance calibrator.

\subsubsection{Distance to decoupling}
\label{sec:distance-to-decoupling}

We denote the CMB's transverse distance measurement by $\theta_\perp(a_\mathrm{d})$, which is not
the conventional $\theta_s = r_s(a_\star) / D_M(a_\star)$ (with $a_\star$ the scale factor of peak
photon visibility) but rather $r_\mathrm{d} / D_M(a_\mathrm{d})$.
We phrase the CMB's acoustic scale information via the so-called drag horizon, i.e.,
the sound horizon evaluated when baryons (rather than photons) decouple from Thomson scattering,
because it is measured nearly as well as (and is tightly related to~\cite{Lin:2021sfs}) the photon
sound horizon and is directly comparable with BAO measurements from the galaxy
distribution~\cite{Loverde:2024nfi}.
Working with $\theta_\perp(a_\mathrm{d})$ rather than $\theta_s$ is thus significantly more
convenient, though not necessary.
Any model that breaks the strong correlation between the acoustic scale that imprints in the photon
and baryon distributions would be a candidate explanation of the BAO--CMB tension (see
Ref.~\cite{Garny:2025szk} for an example).

The CMB's geometric information is completely insensitive to the postdecoupling expansion history
and is usually quite insensitive to modified physics before decoupling, with a notable exception
being the radiation content (i.e., the fraction of radiation that freely
streams)~\cite{Bashinsky:2003tk, Baumann:2015rya}.
Even in this case, the impact on $\theta_\perp(a_\mathrm{d})$ is marginal, and, moreover, it is
accompanied by a different calibration of the \medi{}; we therefore defer a discussion thereof to
\cref{sec:predecoupling}.
For most of the scenarios we consider, the CMB robustly constrains
$1/\theta_\perp(a_\mathrm{d}) = 94.304 \pm 0.028$, which is also largely uncorrelated with the
\medi{} that it calibrates in \LCDM{}.

\subsubsection{Distance to matter domination}
\label{sec:distance-to-matter-domination}

DESI directly measures distances at redshift $2.330$ via its current Lyman-$\alpha$ forest sample.
DESI DR2's Lyman-$\alpha$ measurement at $z = 2.330$, marginalized over its longitudinal distance,
is $D_M(1/[1 + 2.330]) / r_\mathrm{d} = 38.988 \pm 0.531$~\cite{DESI:2025zpo}, while
combining DR2 BAO with the Alcock--Paczynski (AP) effect from DR1 yields
$D_M(1/[1 + 2.330]) / r_\mathrm{d} = 38.90 \pm 0.38$~\cite{Cuceu:2025nvl}.
Though the resulting $0.7\%$ measurement of the \medi{} is completely independent of cosmological
dynamics (up to the well-established assumptions involved in the derivations of the individual
distances), it is insufficiently precise to be of practical use (having uncertainty larger, for
instance, than the impact of a cosmological constant quoted in \cref{sec:theory-cc}).
Including additional data from lower redshifts enables better joint determinations of
$D_M(a_m) / r_\mathrm{d}$ that can reduce this uncertainty, so long as one specifies a model for the
low-redshift Universe to enable its inference.

\begin{figure}[t!]
\begin{centering}
    \includegraphics[width=\textwidth]{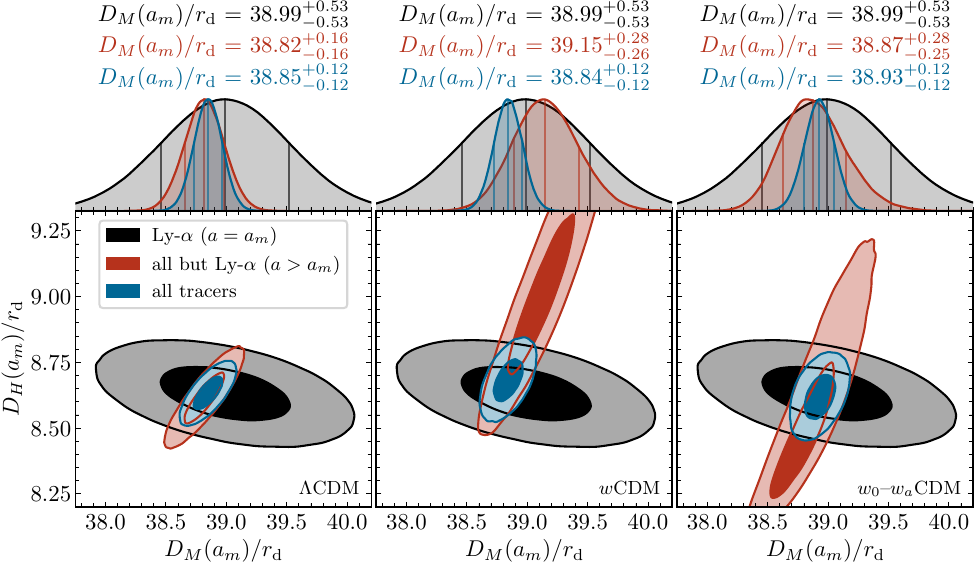}
    \caption{
        Robustness of the jointly inferred distance to DESI DR2's highest-redshift BAO tracer
        (Lyman-$\alpha$ at $z_m = 2.330$~\cite{DESI:2025zpo}) to increasing dynamical freedom of
        dark energy (by panel as labelled).
        The Lyman-$\alpha$ likelihood (whose $1$ and $2 \sigma$ mass levels appear in black)
        combines with the predictions of each model as calibrated to all lower-redshift DESI data
        (red) to yield joint measurements (blue) of the transverse distance
        $D_M(a_m) / r_\mathrm{d}$ with more than four times the precision of the direct one.
        The Lyman-$\alpha$ sample's transverse measurement is in fact almost completely irrelevant:
        its most important role is rather to constrain the expansion rate
        [$D_H(a_m) / r_\mathrm{d}$] at high redshift and therefore how much distance photons
        accumulate from lower redshift (as measured by the observations at lower redshift).
    }
    \label{fig:distance-to-md}
\end{centering}
\end{figure}

Even with relatively arbitrary freedom, lower-redshift data stand to contribute meaningfully to the
inference of $D_M(a_m) / r_\mathrm{d}$.
Transverse distance data constrain the accumulation of distance along the line of sight $\sim
\int_0^a \ud \tilde{a} / \tilde{a} H(\tilde{a})$, while longitudinal data constrain the
instantaneous rate $\sim 1/a H(a)$ of propagated distance, i.e., between DESI's different samples.
As illustrated in \cref{fig:distance-to-md}, the longitudinal measurement limits the allowed range
of densities between DESI's Lyman-$\alpha$ sample and the next highest-redshift sample.
(See \cref{app:numerics} for details on parameter sampling methods.)
Combined, the longitudinal data regularize the amount of distance photons can accumulate between the
Lyman-$\alpha$ sample and the next highest-redshift bin, the distance to which is itself well
measured (the argument continuing by induction).
The longitudinal Lyman-$\alpha$ measurement breaks the degeneracy between $D_M(a_m) / r_\mathrm{d}$
and $D_H(a_m) / r_\mathrm{d}$ from the lower-$z$ data, resulting in a joint constraint on
$D_M(a_m) / r_\mathrm{d}$ more precise than the direct measurement by a factor greater than
four.\footnote{
    Including the aforementioned AP effect measurement from DESI DR1 data~\cite{Cuceu:2025nvl},
    which improves the direct measurement of $D_M(a_m) / r_\mathrm{d}$ to $1\%$ precision,
    has no effect on the jointly inferred measurement---a testament to the irrelevance of
    the direct transverse information.
}

\Cref{fig:distance-to-md} also demonstrates the robustness of this inference of
$D_M(a_m) / r_\mathrm{d}$ to increasing freedom in the dark-energy equation of state (as a proxy for
freedom in the low-redshift expansion history).
The uncertainty is unchanged, while only under the linear-in-$a$ (or $w_0$--$w_a$) equation of state
model~\cite{Chevallier:2000qy, Linder:2002et} does the median shift at all, and only by
$0.66 \sigma$ toward the mean of the direct measurement.
Experiments with fitting an arbitrarily evolving dark energy density [as splines of
$\ln \bar{\rho}_\mathrm{DE}(\ln a)$] yield results quite similar to the $w_0$--$w_a$ result, so long
as no more than one node lies between each DESI measurement (otherwise, the two degrees of freedom
are sufficient to fit two measurements from the next-highest redshift independently of the fit at
lower redshift).

\subsubsection{Drag horizon as units}\label{sec:drag-horizon-as-units}

By and large, acoustic scale measurements themselves are independent of early-Universe
dynamics~\cite{Bernal:2020vbb}, in that the relevant information is robustly compressed by the usage
of the drag horizon as the ``ruler'' for an intrinsically dimensionless observable.
The \LCDM{} prediction for the \medi{} [\cref{eqn:medi-with-cc}] measured from the acoustic scale is
approximately
\begin{align}
    \frac{\chi(a_\mathrm{d}, a_m)}{r_\mathrm{d}}
    &\simeq \frac{
            \left[ 1 - (a_m / \amL)^3 / 14 \right] \sqrt{a_m + a_\mathrm{eq}}
            - \sqrt{a_\mathrm{d} + a_\mathrm{eq}}
        }{
            H_{100} \sqrt{\omega_m} r_\mathrm{d} / 2
        }
    .
    \label{eqn:medi-lcdm-rd}
\end{align}
The dependence of \cref{eqn:medi-lcdm-rd} (and all acoustic scale observables) on absolute densities
and the drag horizon always arises in combinations like
$\omega_m r_\mathrm{d}^2$~\cite{Eisenstein:2004an, Loverde:2024nfi}.
In most of our analyses, we therefore treat the drag horizon as setting units for distances and
densities.
Comparing dimensionless distances that share common units avoids any problems of calibration,
repackaging the dependence of the observable on predecoupling physics in a manner that is maximally
compartmentalized from postdecoupling dynamics.

Measuring dimensionful length scales (from which absolute cosmological densities may be inferred)
with the acoustic scale requires modeling (and calibrating) the drag horizon, which we return to in
\cref{sec:predecoupling}.
Absolute rather than relative densities, however, are only of interest for comparison with the
inference of other, independent observables that do not share the same calibration, such as local
measurements of the Hubble constant~\cite{Riess:2021jrx, Freedman:2024eph, H0DN:2025lyy} or
laboratory limits on the neutrino mass sum~\cite{Esteban:2020cvm, deSalas:2020pgw}.
Notably, the aforementioned strong correlation between the acoustic scale in the photon and galaxy
distributions~\cite{Lin:2021sfs, Loverde:2024nfi} implies that BAO and CMB observations do share a
common ruler; in effectively full generality, their agreement or disagreement may be quantified
without reference to dimensionful values.
A miscalibrated drag horizon therefore cannot (on its own) explain the tension between BAO and CMB
data nor their joint incompatibility with neutrino masses (as has been claimed), a subtlety which we
elaborate on further in \cref{sec:predecoupling}.

\subsection{The BAO--CMB tension as a matter-era distance excess}
\label{sec:medi-tension}

We now show that the matter-era distance interval effectively summarizes the present tension between
DESI's BAO measurements and CMB observations.\footnote{
    Throughout this work, the CMB dataset we employ is always a combination from
    \Planck{}~\cite{Planck:2019nip}, the Atacama Cosmology Telescope~\cite{AtacamaCosmologyTelescope:2025blo}, and
    the South Pole Telescope~\cite{SPT-3G:2025bzu}; see \cref{app:numerics}.
}
This tension may be equivalently understood either by comparing BAO measurements to the low-redshift
distances extrapolated by the CMB or by comparing the matter-era distance interval measured by
the acoustic scale (from both BAO and the CMB) with that predicted by the shape of the CMB
anisotropy spectra.
The former maintains the separation of BAO and CMB likelihoods by studying the CMB's prediction for
$D_M(a_m) / r_\mathrm{d}$ in the form
$1/\theta_\perp(a_\mathrm{d}) - D_M(a_\mathrm{d}, a_m) / r_\mathrm{d}$, which in \LCDM{} is fully
calibrated by physics before decoupling (up to the small impact of dark energy).
From this point of view, our results below show that the tension can be effectively reduced to one
in this single distance as jointly inferred by low-redshift measurements (see
\cref{sec:distance-to-matter-domination}) and as independently predicted by the CMB.

The latter description instead separates the more direct distance measurements from the acoustic
scale by rearranging the comparison to the matter-era distance interval $D_M(a_\mathrm{d}, a_m)$
itself, as motivated by the typical insensitivity of $\theta_\perp(a_\mathrm{d})$ to dynamics after
decoupling and its negligible correlation with the extrapolated \medi{}.
Certain exceptional cases featuring particular effects on the dynamics of the plasma before
decoupling do modify the inference of $\theta_\perp(a_\mathrm{d})$ (see
\cref{sec:high-density-recombination}), for which the previous formulation is more apt, but the
\medi{} isolates the observable that is most sensitive to all postdecoupling (and numerous
predecoupling) extensions of \LCDM{}.
This description thus inherits the advantages of the \medi{}'s analytic tractability and physical
transparency by compartmentalizing model-independent measurements from the acoustic scale and
model-dependent predictions from high-redshift physics.

Moreover, as shown in \cref{sec:theory}, the matter-era distance interval (as predicted by the CMB)
is almost fully determined by the matter density in \LCDM{}, suggesting that the matter density
deficit evident from DESI and CMB data~\cite{Loverde:2024nfi, Lynch:2025ine} implies a matter-era
distance excess.
In \cref{sec:matter-density}, we demonstrate the converse: by applying the analytic results of
\cref{sec:theory} to interpret the joint inference of the late-time matter density from the acoustic
scale, we show that even in \LCDM{} the \medi{} contains most of the information thereof.
A matter-era distance excess is therefore the generalized statement of tension from BAO and CMB
data.
We explain the CMB side of the tension in \cref{sec:deficit-to-excess}, highlighting the role of
massive neutrinos in the predicted \medi{}, and we also discuss the general implications of our
analysis for solutions to the excess.

\subsubsection{Matter density measurements}\label{sec:matter-density}

We now investigate the relative importance of the matter-era distance interval and of individual BAO
distance measurements in determining the matter density at late times.
We derive measurements of the matter density from the \medi{} alone analytically using
\cref{eqn:medi-mr-piecewise}, exploiting the large hierarchy in information to treat a likelihood
for the \medi{} as a posterior for $\omega_m r_\mathrm{d}^2$ that is conditioned on (rather than
jointly distributed with) $a_\mathrm{eq}$, $\amL$, and $\omega_{cb} r_\mathrm{d}^2$, while
still marginalizing over these parameters drawn from independent priors.
In other words, we propagate the uncertainty in these parameters to the matter density
$\omega_m r_\mathrm{d}^2$ inferred from the \medi{} while neglecting the (highly) subdominant
information the likelihood adds for parameters other than $\omega_m r_\mathrm{d}^2$.
We take a brute-force sampling approach, which is less cumbersome than analytically propagating the
uncertainty from the prior distribution but is still analytic in the sense of solving the likelihood
for $\omega_m r_\mathrm{d}^2$ rather than sampling parameters from it (e.g., with Markov chain Monte
Carlo methods).

In more detail, to account for the effects of radiation and cosmological constant, we sample $\amL$
and $a_\mathrm{eq}$ from conservative priors---namely, $\amL \sim \mathcal{N}(0.75, 0.02)$ [double
the uncertainty from DESI DR2 alone and quadruple that when including the CMB's
$\theta_\perp(a_\mathrm{d})$]
and $3401 a_\mathrm{eq} \sim \mathcal{N}(1, 0.032)$ (four times broader than that inferred by CMB
data).
This procedure is additionally conservative in neglecting any correlation between these parameters
and with $\omega_m r_\mathrm{d}^2$.
Combined, these priors propagate just $0.047\%$ uncertainty at any fixed value of
$\omega_m r_\mathrm{d}^2$.
We then sample $\theta_\perp(a_\mathrm{d})$ and $\omega_{cb} r_\mathrm{d}^2$ from a joint CMB prior.
Finally, we draw $\log_{10} a_\times \sim \mathcal{U}(-3, -2.75)$, which contributes negligibly to
the uncertainty, in order to roughly match measurements in \LCDM{} that fix the matter abundance
$a^3 \bar{\rho}_m$ at all redshift.
(The resulting \medi{} is therefore quite insensitive to $\omega_{cb} r_\mathrm{d}^2$.)
We then analytically solve for $\omega_m r_\mathrm{d}^2$ by inverting \cref{eqn:medi-mr-piecewise}
with whatever given sample of $D_M(a_\mathrm{d}, a_m) / r_\mathrm{d}$.

\begin{figure}[t!]
\begin{centering}
    \includegraphics[width=\textwidth]{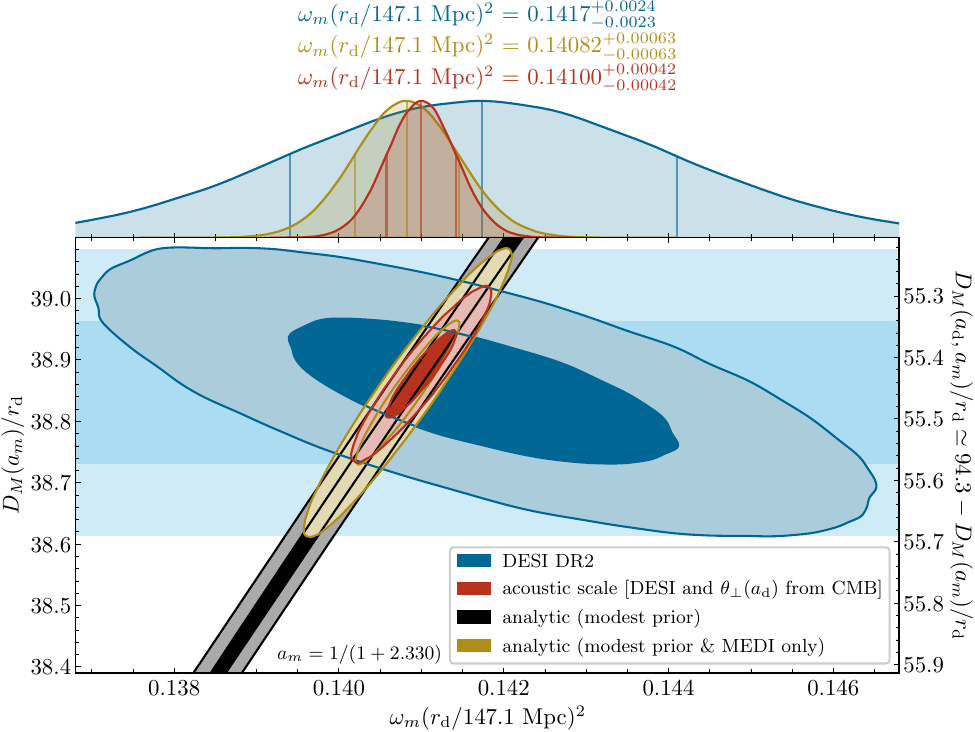}
    \caption{
        Measurements of the late-time matter density $\omega_m r_\mathrm{d}^2$ from the
        acoustic scale and the matter-era distance interval
        $D_M(a_\mathrm{d}, a_m) / r_\mathrm{d} \equiv D_M(a_\mathrm{d}) / r_\mathrm{d} - D_M(a_m) / r_\mathrm{d}$.
        The $1$ and $2 \sigma$ mass levels of the joint distribution over $\omega_m r_\mathrm{d}^2$
        and $D_M(a_m) / r_\mathrm{d}$ deriving from DESI DR2 and from all acoustic scale data (i.e.,
        DESI and the CMB's geometric information) appear in blue and red, respectively.
        Black and gold regions depict results computed analytically with a minimal prior as
        described in the main text, with the \medi{} respectively drawn from a broad, uniform
        distribution and computed as the difference of the CMB's measured
        $1 / \theta_\perp(a_\mathrm{d})$ and DESI's measured $D_M(a_m) / r_\mathrm{d}$ (horizontal
        blue bands); the latter represents an acoustic scale measurement agnostic to dynamics at
        $a > a_m$.
        The right axis depicts the \medi{} corresponding to $D_M(a_m) / r_\mathrm{d}$, taking
        the CMB's mean $1 / \theta_\perp(a_\mathrm{d})$ (neglecting its small uncertainty).
        Comparing the marginal measurements of the matter density demonstrates the critical role of
        the \medi{} and the relative unimportance of DESI's matter density information from late times
        ($a > a_m$).
    }
    \label{fig:matter-density-from-medi}
\end{centering}
\end{figure}

\Cref{fig:matter-density-from-medi} demonstrates how the \medi{} measures the matter density in
\LCDM{}.
We start by analyzing the ``distance to matter domination'' $D_M(a_m) / r_\mathrm{d}$, keeping BAO
and CMB information as separate as possible, and then translate to the acoustic scale measurement of
the \medi{}.
We compare the posterior over $D_M(a_m) / r_\mathrm{d}$ and $\omega_m r_\mathrm{d}^2$ derived
from DESI DR2 data alone and all acoustic scale data [including $\theta_\perp(a_\mathrm{d})$ from
the CMB].
Following the above analytic procedure, we sample the \medi{} from a broad, uniform distribution to
depict the joint distribution of $\omega_m r_\mathrm{d}^2$ and $D_M(a_m) / r_\mathrm{d}$ compatible
with the conservative prior described above.
(One could equivalently draw $\omega_m r_\mathrm{d}^2$ from a uniform distribution and then compute
the \medi{}.)
Finally, we draw values of the \medi{} from the acoustic scale measurement, i.e.,
$1/\theta_\perp(a_\mathrm{d})$ drawn from the CMB prior minus DESI's marginalized measurement of
$D_M(a_m) / r_\mathrm{d}$, and apply the analytic procedure to infer the matter density from the
\medi{} alone.
The right axis of \cref{fig:matter-density-from-medi} illustrates the corresponding values of the
\medi{}, neglecting the subdominant uncertainty in $1/\theta_\perp(a_\mathrm{d})$; the red and blue
posteriors then represent results from all acoustic scale data that respectively model the expansion
history consistently at all redshift and are agnostic to it at $z > z_m = 2.330$.

\Cref{fig:matter-density-from-medi} evidences the critical role of the CMB's geometric information
in the inference of the late-time matter density.
Even in a model with as little freedom as \LCDM{}, with only one additional relevant parameter
($\Lambda$) constrained by $13$ percent-level data points, DESI's measurement of the matter density
is dwarfed by that from the matter-era distance interval alone.
One could interpret this exercise as modeling the expansion history with the matter abundance taking
independent values below and above $z_m$, whose respective measurements are that obtained by DESI
alone and that from the \medi{} alone.
The full measurement from all acoustic scale data (modeled consistently) is only $1.5$ times more
precise than the \medi{}'s, deriving from the mild correlation between DESI's inference of
$\omega_m r_\mathrm{d}^2$ and $D_M(a_m) / r_\mathrm{d}$.
Artificially increasing DESI's preferred matter density would have only a minor effect on the joint
measurement compared with instead increasing its measured distance to $z = 2.330$ (i.e., decreasing
the \medi{}).
Freedom in the equation of state of dark energy, which we discuss further in \cref{sec:dde}, largely
decorrelates the two, such that the \medi{} represents nearly the totality of the information on the
matter density.

\subsubsection{From density deficit to distance excess}\label{sec:deficit-to-excess}

\Cref{sec:matter-density} establishes that analyzing the full set of acoustic scale observations
from BAO and the CMB, in particular through the matter-era distance interval, is key to physically
interpret their combination (or tension).
The remaining information from the CMB---from the shape of the anisotropy spectra---calibrates the
densities in all species relevant around and before photon--baryon decoupling (within any given
model) and thus predicts the \medi{} by extrapolating those densities to late times.
In the \LCDM{} model with massless neutrinos, the CMB measures
$\omega_{cb} (r_\mathrm{d} / 147.1~\mathrm{Mpc})^2 = 0.14224 \pm 0.00061$ and therefore predicts
$D_M(a_\mathrm{d}, a_m) / r_\mathrm{d} = 55.2 \pm 0.11$ as a baseline expectation.
The measurements of the \medi{} in \cref{fig:matter-density-from-medi}, whether marginalized over
DESI's matter density information or not, exceed this prediction by about $1.7\sigma$.
That the tension must be understood as between the acoustic scale and the shape of the CMB rather
than between the CMB and DESI itself is best evidenced by the irrelevance of DESI's independent
inference of the matter density, $\omega_m r_\mathrm{d}^2 = 0.1417 \pm 0.0024$, whose $\pm 1 \sigma$
interval encompasses the CMB's $-5 \sigma$ and $+ 3 \sigma$ limits of $\omega_{cb} r_\mathrm{d}^2$.

\begin{figure}[t!]
\begin{centering}
    \includegraphics[width=\textwidth]{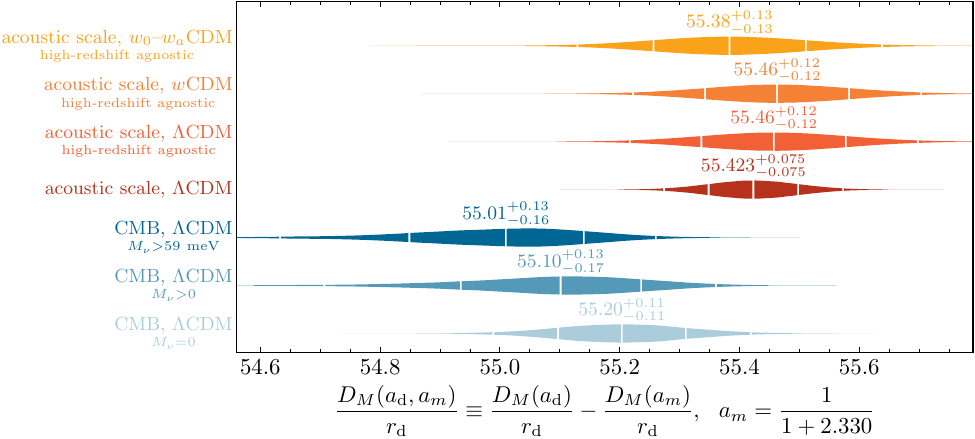}
    \caption{
        Quantification of the matter-era distance excess that underlies the BAO--CMB tension, i.e.,
        between CMB predictions and acoustic scale measurements of the matter-era distance interval.
        Blue posteriors depict \LCDM{} predictions calibrated by CMB temperature and polarization
        data from \Planck{} PR3, ACT DR6, and SPT-3G D1, either fixing the neutrino mass sum to
        zero, marginalizing it over nonnegative values, or marginalizing it over values compatible
        with neutrino oscillations; these results depend only on nongeometric information.
        Red/orange results indicate measurements from the acoustic scale, with the darkest red
        indicating the constraint in standard \LCDM{} from DESI DR2 data and the CMB's geometric
        information [$1 / \theta_\perp(a_\mathrm{d})$] alone, in noticeable tension with the
        prediction from the CMB's nongeometric information.
        Lighter shades display more conservative results that assume nothing about the expansion
        history at $z > z_m$, instead taking $D_M(a_\mathrm{d}) / r_\mathrm{d}$ from the
        CMB's fully marginalized measurement of $1 / \theta_\perp(a_\mathrm{d})$.
        The low-redshift expansion history is then modeled under varying assumptions in order to
        combine all DESI data to measure the ``distance to matter domination''
        $D_M(a_m) / r_\mathrm{d}$, taking a cosmological constant, a dark energy fluid with fixed
        equation of state, or one with equation of state evolving linearly with $a$.
        These results thus represent measurements of the matter-era distance interval that are
        independent of physics between $a_m$ and $a_\mathrm{d}$, which remain in tension with the
        CMB's predictions thereof, especially when massive neutrinos are properly accounted for.
    }
    \label{fig:medi-tension}
\end{centering}
\end{figure}

Comparing the measured \medi{} to CMB predictions in the base \LCDM{} model, however, understates
the severity of the tension, given that neutrinos (with masses compatible with neutrino
oscillations) must increase the late-time matter density by at least a half percent over that from
CDM and baryons alone.
To fully quantify the $\summnu$-marginalized matter-era distance tension,
\cref{fig:medi-tension} compares complete CMB predictions within \LCDM{} (under various assumptions
on neutrino masses) to acoustic scale measurements (under various assumptions on the expansion
history).
In the most minimal and motivated case (a cosmological constant and $\summnu$ compatible with
neutrino oscillations), the tension between the acoustic scale and the prediction extrapolated by
the CMB is $2.6\sigma$, degrading only by about $0.5\sigma$ if taking dynamical dark energy at
$z < z_m$ or allowing sub-$59~\mathrm{meV}$ neutrinos.\footnote{
    The degenerate mass hierarchy is least accurate for mass sums closest to the minimum, which
    dominate the posterior's mass density.
    The effect on the \medi{}, however, is quite small: summing each mass eigenstate's
    contribution using \cref{eqn:medi-mnu-apx}, the modification to the sensitivity coefficient
    multiplying $f_\nu$ is $\sim 10\%$ at most, rapidly diminishing with larger $\summnu$.
    The degenerate hierarchy thus remains suitable, as tested explicitly by
    Ref.~\cite{Herold:2024nvk}.
}
For comparison, the tension between DESI and the CMB in \LCDM{}, quantified as the difference in
posterior mean vectors $(\omega_m r_\mathrm{d}^2, \Omega_m)$ with the summed covariance serving as
the metric, is about the same size.

An analytic approach like that used to measure the late-time matter density in
\cref{sec:matter-density} is less apt to measure neutrino masses, as the \medi{} likelihood peaks at
distances longer rather than shorter than the CMB prediction even for massless neutrinos---that is,
most of its probability mass would be cut off by the prior that $\summnu$ is nonnegative.
We may, however, reliably estimate the increase in tension from any \textit{fixed} neutrino mass sum
by adding \cref{eqn:medi-mnu-apx} to the CMB's aforementioned prediction with massless neutrinos,
and computing the tension with the high-redshift--agnostic acoustic scale measurement of $55.46 \pm
0.12$ in \cref{fig:medi-tension}.\footnote{
    In doing so, we ignore any adjustment to the CMB's inferred $\omega_{cb} r_\mathrm{d}^2$, etc.,
    in response to the suppression of the lensing amplitude by massive neutrinos at late times.
    Marginalizing over $\summnu > 0$ only shifts the CMB's preferred
    $\omega_{cb} r_\mathrm{d}^2$ upward by just $0.2 \sigma$, so this approximation only slightly
    underestimates the increase in tension.
}
Series expanding \cref{eqn:medi-mnu-apx} about $\summnu = 59~\mathrm{meV}$ and taking a fiducial
$r_\mathrm{d} = 147.1~\mathrm{Mpc}$, we find that the matter-era distance tension is about
$\left[ 2.06 + (\summnu - 59~\mathrm{meV}) / 97~\mathrm{meV} \right] \sigma$.
The minimum mass sums for the normal and inverted mass hierarchies thus increase the tension to
$2.1 \sigma$ and $2.5 \sigma$, respectively.

The CMB predictions in \cref{fig:medi-tension} do not include lensing reconstruction data, simply
to avoid adding more late-time information that might be affected by any physics invoked to modify
the predicted \medi{}; we comment on the intrinsic lensing information in temperature and
polarization anisotropies in \cref{sec:lensing-excess}.
The CMB lensing dataset from Ref.~\cite{ACT:2025qjh} only reduces the uncertainty in the CMB
prediction for $\summnu = 0$ from $0.11$ to $0.093$ and has no discernible impact when marginalizing
over nonzero $\summnu$.

The matter-era distance interval enables a generalized description of the geometric sensitivity of
CMB and BAO data to neutrino masses.
In \LCDM{}, the steep increase in the matter fraction (or increase in $\amL$) with $\summnu$
required to fix the distance to last scattering enables photons to travel far enough during dark
energy domination to compensate for the reduced horizon during matter
domination~\cite{Loverde:2024nfi}.
BAO data break this degeneracy largely by directly measuring $\amL$, thereby determining the
relative amount of distance traveled in each era.
The \medi{} generalizes this logic: BAO data at low redshift measure the distance to matter
domination (\cref{sec:distance-to-matter-domination}) and thus also the remaining portion of the
distance to last scattering that accrued deep in the matter era.
The arguments of this section show that the majority of the information (and the present geometric
tension) originates in the latter portion.

Recently, Ref.~\cite{Graham:2025dqn} attempted to quantify the geometric tension between the CMB and
BAO data by introducing an effective, signed neutrino mass parameter.
Their model, however, only modifies low-redshift distances for BAO and not the CMB itself (i.e., not
its own geometric information) and therefore does not capture the main effect (on the \medi{}).
The distance to last scattering in their treatment is a function of $\omega_{cb} r_\mathrm{d}^2$ and
$\Omega_m$ alone as in \LCDM{}; with the former well measured by the shape of the CMB spectra,
$\theta_s$ measures $\Omega_m$ (rather, the dark energy density) independent of this effective
neutrino mass.
Imposing this effective prior on $\Omega_m$ thus artificially sharpens the measurement
of the full late-time matter density (including the effective, neutrinolike contribution) from
low-redshift BAO data.
Such a measurement does not reflect the actual physical origin of the acoustic scale's
sensitivity to neutrino masses (nor any consistent cosmological model\footnote{
    CMB photons and photons emitted by tracers of structure accumulate the same
    distance out to the redshifts observed by BAO surveys.
    In effect, the model of Ref.~\cite{Graham:2025dqn} imposes some modification to the expansion
    history at high redshift that exactly cancels the shift in distance induced by their
    effective neutrino density at low redshift.
}), as underscored by the importance of the matter-era distance interval---namely, the physical
inputs that modify its prediction, rather than the low-redshift BAO distances themselves.

The matter-era distance excess persists even if the acoustic scale measurement is made fully
agnostic to dynamics at $a < a_m$.
When simply taking $D_M(a_\mathrm{d}) / r_\mathrm{d} = 1 / \theta_\perp(a_\mathrm{d})$ as measured
by the CMB independent of all other parameters (rather than jointly modeling and fitting it), the
measured \medi{} only shifts slightly higher (see \cref{fig:medi-tension}), with precision degrading
by a factor of $1.6$ (with size similar to the CMB's uncertainty).
In \LCDM{}, this exercise effectively allows the matter abundance to differ above and below $a_m$,
with the improvement in precision deriving from imposing that the values match as illustrated in
\cref{fig:matter-density-from-medi}.
Furthermore, the robustness of the high-redshift-agnostic \medi{} measurements from the acoustic
scale to dark energy dynamics demonstrates that the excess is far more likely to be explained by
dynamics at high redshift ($z > z_m$) than at low redshift.\footnote{
    The \medi{} measurements in \LCDM{} from DESI DR1~\cite{DESI:2024mwx} and the Sloan Digital Sky
    Survey~\cite{Ross:2014qpa, eBOSS:2020yzd}, using the same CMB prior on
    $\theta_\perp(a_\mathrm{d})$, are $55.46 \pm 0.12$ and $55.24^{+0.14}_{-0.13}$, respectively;
    the high-redshift-agnostic counterparts are $55.49 \pm 0.2$ and $55.11 \pm 0.28$ and are robust
    to dark energy dynamics to an extent similar to that from DESI DR2.
}

\section{High-redshift solutions}
\label{sec:high-redshift-solutions}

Given that DESI itself is compatible with quite a broad range of matter densities (i.e., as inferred
from its low-redshift distances alone), \cref{sec:medi-tension} shows that solving the matter-era
distance excess would leave DESI compatible with rather heavy neutrinos, even if the Universe were
completely described by \LCDM{} at low redshift.
The robustness of the matter-era distance interval measured by the acoustic scale, as evident in
\cref{fig:medi-tension}, leaves little room to resolve or even alleviate the matter-era distance
tension by modifying the distances predicted at $z \leq z_m$.
Conversely, the sensitivity of the CMB's prediction to assumptions on neutrino masses reflects the
\medi{}'s importance in constraining the expansion history after photon--baryon decoupling more
broadly.
The baseline expectation for the \medi{} from the CMB is also sensitive to dynamics before
decoupling insofar as they determine the densities in baryons and CDM (relative to the sound
horizon).

We now consider what nonstandard physics before (\cref{sec:predecoupling}) and after
(\cref{sec:postdecoupling}) decoupling may modify the CMB prediction and resolve the matter-era
distance tension.
We summarize their implications for neutrino masses in \cref{sec:mnu-vs-highz}.
Finally, in \cref{sec:dde-alternatives} we show that dynamical dark energy addresses the
matter-era distance excess not through its dynamics at late times (as already evident in
\cref{fig:medi-tension}) but at high redshift.
We demonstrate that the alternative solutions to the excess from
\cref{sec:predecoupling,sec:postdecoupling} reduce the evidence for dark energy dynamics and
identify the features in DESI's data that underlie the residual preference.

\subsection{Predecoupling physics}\label{sec:predecoupling}

In general, parametrizing the overall, dimensionless amplitude of acoustic scale observables in
terms of $\omega_m r_\mathrm{d}^2$ (rather than the more conventional $h r_\mathrm{d}$) more
transparently identifies its dependence on early-Universe physics (rather than dark
energy)~\cite{Eisenstein:2004an, Loverde:2024nfi}.
This advantage is especially clear for the matter-era distance interval, as the distance photons
travel during matter domination is (by definition) independent of dark energy and the present-day
expansion rate.
In \LCDM{}, the CMB measures $\omega_{cb} r_\mathrm{d}^2$ even better than it measures
$\omega_{cb}$, setting a baseline expectation for the \medi{} (the $\summnu = 0$ result in
\cref{fig:medi-tension}).
In this section, we generalize this statement to models that modify the CMB's calibration of the
\medi{}
and explain why the actual value of the drag horizon is not uniquely relevant to the present tension
between DESI's measurements and the CMB, as has been suggested previously~\cite{Pogosian:2024ykm,
Jhaveri:2025neg, Mirpoorian:2025rfp}.
We explain what features of predecoupling physics genuinely modify acoustic scale predictions,
following arguments from Refs.~\cite{Baryakhtar:2024rky, Loverde:2024nfi}.

We begin by reviewing how the CMB calibrates the density in matter components.
The CMB anisotropy spectra are dimensionless observables and therefore only measure density
ratios~\cite{Eisenstein:2004an, Hu:2004kn, Weinberg:2008zzc, Baryakhtar:2024rky}; furthermore, the
anisotropies generated at last scattering depend on density ratios evaluated at that epoch, not the
present~\cite{Baryakhtar:2024rky, Costa:2025kwt}.
The shape of the temperature and polarization spectra measures the matter-to-radiation ratio
$\bar{\rho}_{cb}(a) / \bar{\rho}_r(a)$ around last scattering, primarily via the radiation-driving
and early integrated Sachs--Wolfe effects~\cite{Hu:1996vq, Hu:1996mn, Hu:2004kn, Weinberg:2008zzc,
Ilic:2020onu, Baryakhtar:2024rky, Costa:2025kwt}.
This information is effectively summarized by the radiation-matter ratio evaluated at peak
visibility,
\begin{align}
    x_\mathrm{eq}
    &\equiv \frac{\bar{\rho}_r(a_\star)}{\bar{\rho}_m(a_\star)}
    = \frac{\omega_r}{\omega_{cb} a_\star}
    = \frac{a_\mathrm{eq}}{a_\star}
    \label{eqn:x-eq}
    ,
\end{align}
which, for standard matter and radiation, reduces to the amount of expansion between
matter--radiation equality and last scattering.
The acoustic peak structure and damping tail also measure the baryon-to-photon
ratio~\cite{Hu:1995en, Hu:1996vq, Hu:1996mn}
\begin{align}
    R_\star
    &\equiv \frac{3 \bar{\rho}_b(a_\star)}{4 \bar{\rho}_{\gamma}(a_\star)}
    \propto \frac{3 \omega_b a_\star}{4 \omega_\gamma}
    \label{eqn:R-star}
    .
\end{align}
The CMB precisely measures $R_\star$ and $x_\mathrm{eq}$ (respectively reaching
$0.44\%$ and $0.77\%$ precision from the latest combination of \Planck{}, ACT, and SPT temperature
and polarization data~\cite{Planck:2018vyg, AtacamaCosmologyTelescope:2025blo, SPT-3G:2025bzu}) and does so robustly in a
range of extended cosmologies~\cite{Baryakhtar:2024rky, Loverde:2024nfi}.

The sound horizon depends on the baryon and CDM densities only via $R_\star$ and
$x_\mathrm{eq}$~\cite{Eisenstein:1997ik, Baryakhtar:2024rky, Loverde:2024nfi}:
\begin{align}
    r_s(a)
    &= \frac{a_\star}{\sqrt{\omega_r}}
        \frac{2 \sqrt{ x_\mathrm{eq} / 3 R_\star} }{H_{100} / c}
        \ln \left(
            \frac{
                \sqrt{R_\star} \sqrt{x + x_\mathrm{eq}}
                + \sqrt{1 + R_\star x}
            }{
                1 + \sqrt{R_\star x_\mathrm{eq}}
            }
        \right)
    ,
    \label{eqn:rs-in-matter--radiation-ito-wr-R_star}
\end{align}
where $x \equiv a / a_\star$.\footnote{
    The analogous equation in Ref.~\cite{Loverde:2024nfi}, Eq.~(2.7), omitted the subscript
    $\star$ in the leading factor of $a$.
}
Its dimensionful value is thus determined by the calibration of the radiation density
$\omega_r \equiv a_\star^4 \bar{\rho}_r(a_\star) / 3 H_{100}^2 \Mpl^2$, i.e., the density in
relativistic species at last scattering translated to the present,\footnote{
    The present-day radiation density proves most convenient simply because the present-day CMB
    temperature is measured and the total relativistic density at last scattering is quantified
    relative to the photon density.
    Written in terms of $T_\star$ and $T_0$,
    $r_s(a) \propto H_{100} \Mpl / T_0 T_\star \cdot \sqrt{\omega_\gamma / \omega_r}$.
} but this dependence is
absorbed into a dimensionless ratio in acoustic scale observables [i.e.,
$D_M(a) / r_\mathrm{d}$ and $D_H(a) / r_\mathrm{d}$].
Evaluating the drag horizon $r_\mathrm{d} \equiv r_s(a_\mathrm{d})$ from
\cref{eqn:rs-in-matter--radiation-ito-wr-R_star} also requires
the ratio of the temperatures of photon and baryon decoupling,
$x_\mathrm{d} = a_\mathrm{d} / a_\star = T_\star / T_\mathrm{d}$, which in \LCDM{} is strongly
constrained ($1.0270 \pm 0.0003$) and also varies little in extensions~\cite{Lin:2021sfs,
Loverde:2024nfi}.
For this reason, the drag scale $D_M(a_\mathrm{d}) / r_\mathrm{d}$ provides nearly as precise a
summary statistic for the CMB's geometric information as the photon sound horizon does via
$D_M(a_\star) / r_s(a_\star)$.

Only with knowledge of the amount of expansion since last scattering (and the dilution rate of
densities with redshift) can density ratios at last scattering be translated to late
times.\footnote{
    Further converting ratios to absolute densities relies on the present energy density of CMB
    photons, i.e., the measured CMB temperature, and also the amount of additional density in
    relativistic species around last scattering (whether the prediction for Standard Model neutrinos,
    i.e., $N_\mathrm{eff}$, or otherwise)~\cite{Baryakhtar:2024rky, Loverde:2024nfi}; as mentioned
    previously, calibrating these densities is not necessary to test CMB predictions against
    acoustic scale observations.
}
For standard matter content, these density-ratio measurements set a baseline expectation for the
\medi{} [\cref{eqn:medi-mr}] with $\omega_m r_\mathrm{d}^2$ determined by
$a_\star^3 \bar{\rho}_{cb}(a_\star) r_\mathrm{d}^2
= a_\star^3 \bar{\rho}_r(a_\star) r_\mathrm{d}^2 / x_\mathrm{eq}$.
[We consider modifications to the standard $(a / a_\star)^{-3}$ extrapolation in
\cref{sec:postdecoupling}.]
Allowing for an instantaneous change in the pre- and postdecoupling matter abundances ($\omega_{cb}$
and $\omega_m$), the prediction for the \medi{} in a Universe with arbitrary amounts of matter and
radiation reduces to
\begin{align}
    \frac{\chi(a_\mathrm{d}, a_m)}{r_\mathrm{d}}
    &= \sqrt{\frac{\omega_{cb}}{\omega_m}}
        \left(
            \sqrt{x_m + x_\mathrm{eq}}
            - \sqrt{x_\mathrm{d} + x_\mathrm{eq}}
        \right)
        \sqrt{3 R_\star}
        \ln \left[
            \frac{
                \sqrt{R_\star} \sqrt{x_\mathrm{d} + x_\mathrm{eq}}
                + \sqrt{1 + R_\star x_\mathrm{d}}
            }{
                1 + \sqrt{R_\star x_\mathrm{eq}}
            }
        \right]^{-1}
    \label{eqn:medi-lcdm}
    ,
\end{align}
which depends on the temperature and densities at recombination only in the form of dimensionless
parameters that are well constrained by independent information in the shape of the CMB spectra
($R_\star$, $x_\mathrm{eq}$, and $x_\mathrm{d}$).
Therefore, adjusting the density in \textit{any} single component of the standard cosmological model
cannot alter the prediction \cref{eqn:medi-lcdm} (or any acoustic scale observable) without
modifying the fit to the CMB spectra.
(All acoustic scale observables depend on predecoupling physics in this manner, not just the
\medi{}.)
Otherwise, \cref{eqn:medi-lcdm} only depends on $x_m = a_m / a_\star$, where $a_m$ is determined,
e.g., by the measured redshift of a galaxy sample (or whatever redshift a collection of BAO
measurements best constrain), which is unrelated to recombination physics.
We next discuss the implications of these insights for modified recombination and high-density
recombination models.

\subsubsection{High-temperature recombination, or low-temperature present}\label{sec:high-redshift-recombination}

The most trivial means to alter the relationship between density ratios at recombination and at the
present is to change how much expansion elapsed in between, $a_0 / a_\star = T_\star / T_0$.
Such a possibility either requires that the CMB temperature is mismeasured~\cite{Ivanov:2020mfr} or
that nonstandard recombination physics modifies the temperature at last scattering
$T_\star$~\cite{Hart:2019dxi, Sekiguchi:2020teg, Jedamzik:2020krr, Baryakhtar:2024rky,
Lynch:2024gmp, Lynch:2024hzh, Mirpoorian:2024fka}.
Modifying the redshift of recombination is most effectively realized microphysically via a
hyperlight scalar field that modulates the electron mass in time~\cite{Baryakhtar:2024rky,
Baryakhtar:2025uxs}, which leaves few to no signatures in the CMB anisotropies as generated at
last scattering; introducing arbitrary additional freedom in the recombination
history~\cite{Lynch:2024gmp, Lynch:2024hzh, Mirpoorian:2024fka} does not appear to introduce any
additional qualitative features.
Varying the CMB temperature from its measured value also modifies $a_\star$ without disrupting
density ratios at decoupling and thus modifies acoustic scale observables in an identical manner.

Since $\omega_{cb} r_\mathrm{d}^2 \propto a_\star$ at fixed $x_\mathrm{eq}$ and $R_\star$ (with no
other parameter dependence), modified recombination changes the predicted \medi{} relative to the
drag horizon, as encoded by $x_m = a_m / a_\star$ in \cref{eqn:medi-lcdm}.
In \LCDM{}, the reduction in $\omega_m r_\mathrm{d}^2$, combined with the strongly correlated
reduction in $\Omega_m \propto (\omega_m r_\mathrm{d}^2)^5$ at fixed $\theta_s$, fully mediates the
tension between DESI and the CMB~\cite{Baryakhtar:2024rky, Loverde:2024nfi}.
More generally, early recombination (or ``late present,'' i.e., smaller $T_0$) mediates the
matter-era distance excess by reducing the sound horizon squared more so than it increases the
matter density at any fixed redshift.
From \cref{eqn:medi-lcdm}, $\chi(a_\mathrm{d}, a_m) / r_\mathrm{d} \propto 1 / \sqrt{a_\star}$;
a reduction in $a_\star$ by about a percent accommodates the half-percent increase in the \medi{}
preferred by DESI, with nonzero neutrino masses allowed via a further reduction by
about $1 / \sqrt{1 + f_\nu}$~\cite{Baryakhtar:2024rky, Loverde:2024nfi}.

More quantitatively, we may estimate the measurement of $a_\star / a_0$ from the \medi{}
semianalytically by computing
$a_\star / a_{\star, \mathrm{fid}} = \omega_{cb} r_\mathrm{d}^2 / \omega_m r_\mathrm{d}^2$, with the
denominator measured by the acoustic scale as in \cref{fig:matter-density-from-medi}, the
numerator taken from a CMB posterior, and $a_{\star, \mathrm{fid}}$ the scale factor of last
scattering in standard cosmology.
This quantity corresponds to the ratio of the electron mass at early and late times, if it evolves,
or the inverse of the CMB temperature divided by its measured value, if supposed to be different.
The result, $1 + \left( 0.88 \pm 0.52 \right) \times 10^{-2}$, closely matches that from a full fit to
the CMB and DESI DR2 in the varying electron mass scenario,
$1 + \left( 0.76 \pm 0.48 \right) \times 10^{-2}$.
When additionally varying the neutrino mass sum (allowing any nonnegative value),
the result shifts to $1 + \left( 1.1^{+0.62}_{-0.54} \right) \times 10^{-2}$, while the 95th
percentile of the marginal posterior over $\summnu$ is $160~\mathrm{meV}$
(see \cref{fig:mnu-high-z-sols}), slightly tighter than the limit of $200~\mathrm{meV}$ derived in
Ref.~\cite{Loverde:2024nfi} for \Planck{} CMB data alone with DESI DR1.

\subsubsection{High-density recombination}\label{sec:high-density-recombination}

When the amount of expansion between last scattering and today is unchanged, and to the extent that
the CMB measures the density in baryons and CDM relative to the total radiation density alone, a
change in the calibrated, absolute density has no effect on acoustic scale observables.
Namely, fixing $x_\mathrm{eq}$ requires that $\omega_{cb}$ increases by exactly the amount
$r_\mathrm{d}^2 \propto 1/\omega_r$ decreases.
This feature is manifest in \cref{eqn:medi-lcdm} in its lack of dependence on the total radiation
density, reflecting that all acoustic scale observables (and CMB anisotropies) are dimensionless.
The insensitivity of acoustic scale observables to the radiation density explains why BAO data
contribute comparatively little to CMB constraints on the effective number of relativistic neutrinos
$N_\mathrm{eff}$.

Strictly speaking, however, the invariance of the primary CMB at fixed $x_\mathrm{eq}$ and
$R_\star$ only holds when additionally fixing the diffusion rate per $e$-fold~\cite{Hou:2011ec,
Baryakhtar:2024rky, Saravanan:2025cyi} and the fraction of the total radiation density that is
collisionless~\cite{Ge:2022qws, Saravanan:2025cyi} (which is respected by the high-redshift
recombination models discussed in \cref{sec:high-redshift-recombination}).
[\Cref{eqn:rs-in-matter--radiation-ito-wr-R_star,eqn:medi-lcdm} are correct nonetheless because they
depend only on the cosmological background.]
In practice, CMB preferences for $\omega_{cb}$ are driven by a tradeoff between fixing
$x_\mathrm{eq}$ and the fraction of pressure-supported matter
$f_b = \bar{\rho}_b(a_\star) / \bar{\rho}_{cb}(a_\star)$~\cite{Saravanan:2025cyi},
while its measurements of $\omega_b$ compromise between fixing the acoustic peak structure ($R_\star$)
and the damping tail (unless introducing additional parameter freedom).
Combined, the baryon density increases marginally with the radiation density, while the density in
baryons and CDM combined increases sublinearly with the radiation density.
Reference~\cite{Saravanan:2025cyi} shows that, when strictly adding radiation on top of the Standard
Model prediction, the changes to inferred density ratios lead the CMB to predict smaller values of
$\omega_{cb} r_\mathrm{d}^2$ that are more consistent with DESI's BAO data (i.e., that alleviate the
matter-era distance excess).

\begin{figure}[t!]
\begin{centering}
    \includegraphics[width=\textwidth]{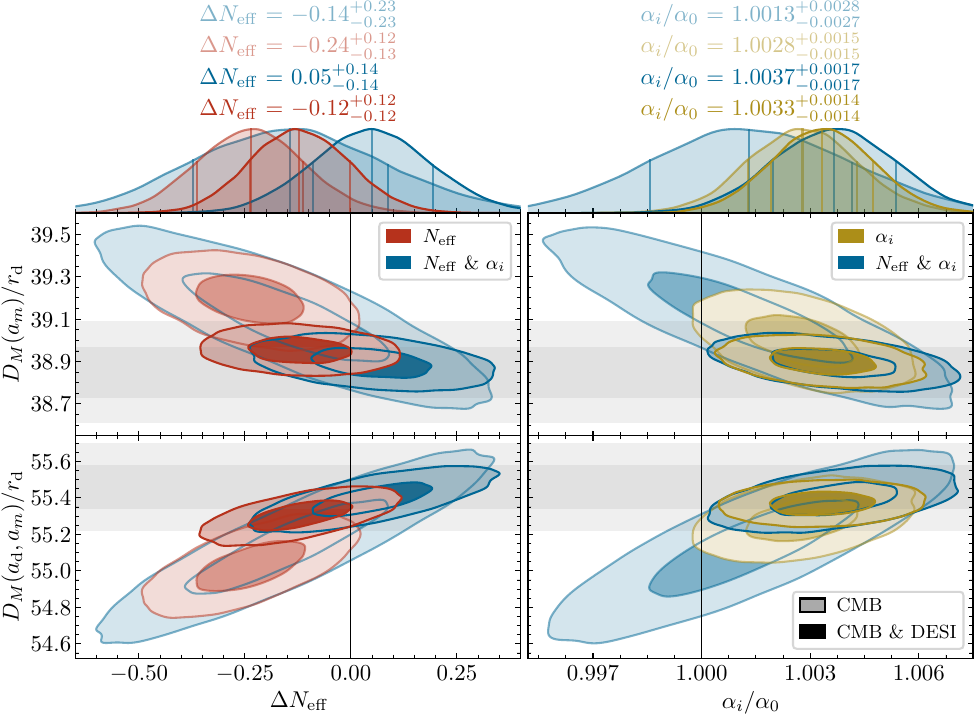}
    \caption{
        Concordance of CMB predictions and acoustic scale measurements in a selection of scenarios
        modifying physics before decoupling: varying the effective number of neutrino species
        $N_\mathrm{eff}$ (red), the early-time value of the fine-structure constant (gold), or both
        (blue).
        Middle and bottom panels display the joint posterior over each parameter and the ``distance
        to matter domination'' $D_M(a_m) / r_\mathrm{d}$ and the matter-era distance interval
        $D_M(a_\mathrm{d}, a_m) / r_\mathrm{d}$, respectively; up to the small effect of
        $\Delta N_\mathrm{eff}$ on the inferred $D_M(a_\mathrm{d}) / r_\mathrm{d}$, both rows
        contain the same information.
        Horizontal shaded regions depict the $1$ and $2 \sigma$ limits of acoustic scale
        measurements from \cref{fig:distance-to-md,fig:medi-tension}.
        Current CMB data alone prefer $N_\mathrm{eff}$ below the Standard Model prediction, which
        exacerbates the matter-era distance tension, but prefer $\alpha_i$ larger
        than the present-day value by $\gtrsim 2 \sigma$, which alleviates the tension.
        Results fix the neutrino mass sum to zero to isolate early-time effects on extrapolated
        distances; since CMB predictions at best match acoustic scale measurements in each scenario,
        none can fully reconcile the matter-era distance excess for arbitrarily large $\summnu$.
        Varying the helium yield to absorb the impact of $\Delta N_\mathrm{eff}$ on diffusion has
        no substantive impact on acoustic scale predictions.
    }
    \label{fig:neff-alpha}
\end{centering}
\end{figure}

Moreover, the fraction of radiation that freely streams affects the propagation of acoustic waves,
modulating their amplitude and phase~\cite{Bashinsky:2003tk, Baumann:2015rya}.
\Cref{fig:neff-alpha} shows that current CMB data prefer values lower than the prediction of
Standard Model neutrinos~\cite{AtacamaCosmologyTelescope:2025nti} (i.e., low- rather than
high-density recombination), whose net impact is to exacerbate the matter-era distance excess.
On the one hand, the reduced free-streaming fraction shifts the acoustic peaks to higher multipoles,
which is compensated for by decreasing the distance to last scattering [and therefore also
$D_M(a_\mathrm{d}) / r_\mathrm{d}$].
[Since the phase shift affects the CMB's inference of $\theta_\perp(a_\mathrm{d})$, we present both
the \medi{} and $D_M(a_m) / r_\mathrm{d}$ in \cref{fig:neff-alpha}, with the latter being the more
relevant prediction from the CMB to compare with DESI data.]
On its own, a reduced phase shift would then alleviate the matter-era distance tension; however, its
effect is outweighed by the increase in $\omega_{cb} r_\mathrm{d}^2$.

With the photon density (i.e., the density in collisional radiation) fixed, however, varying the
free-streaming fraction via the number of effective degrees of freedom in radiation is quite limited
by the correlated impact on diffusion damping.
Extending the parameter freedom to include the helium yield from nucleosynthesis is a common choice
to absorb some of the impact on diffusion.
We find that the helium yield has little impact on acoustic scale predictions, while instead
varying the fine-structure constant~\cite{Hannestad:1998xp, Kaplinghat:1998ry, Hart:2017ndk,
Baryakhtar:2024rky, Baryakhtar:2025uxs} (which also modulates the diffusion rate) has a more
substantial effect. In fact, a time-varying fine-structure constant alone resolves the matter-era
distance excess; \cref{fig:neff-alpha} shows that current CMB data prefer a larger early-time value
at the $2 \sigma$ level~\cite{AtacamaCosmologyTelescope:2025nti}, which can be realized with a
hyperlight scalar field~\cite{Baryakhtar:2024rky, Baryakhtar:2025uxs}.\footnote{
    We find that the fine-structure constant and helium yield are strongly degenerate in CMB
    predictions, simultaneously in their impact on diffusion and
    polarization~\cite{Baryakhtar:2024rky}, which in fact allows for a range of recombination
    redshifts nearly as wide as that from electron-mass variations.
    The degenerate combination, however, is not well aligned with the predicted variation of the
    helium yield with $\alpha$ from nucleosynthesis calculations~\cite{Baryakhtar:2025uxs}.
    For this reason, and because it offers no relevant phenomenology not captured by a varying
    electron mass, we do not explore this scenario further, though it would be interesting to
    understand the physical mechanism for the degeneracy.
}
That said, the results in \cref{fig:neff-alpha} take massless neutrinos and only just resolve the
excess, so at best they modestly alleviate geometric limits on neutrino masses.

The extent to which the CMB permits different density ratios depends in general upon the detailed
impact of new components on the evolution of the photon--baryon plasma.
For instance, the impact of strongly interacting (as opposed to collisionless) dark radiation on CMB
anisotropies is more effectively absorbed by shifts to \LCDM{} parameters, permitting better
compatibility with DESI BAO data~\cite{Saravanan:2025cyi}.
Unlike free-streaming radiation (studied in \cref{fig:neff-alpha}), the decrease in
$\omega_{cb} r_\mathrm{d}^2$ with increasing radiation density is accompanied by a decrease rather
than an increase in $D_M(a_\mathrm{d}) / r_\mathrm{d}$ (due to the phase shift as discussed above).
For the same increase in radiation density, new fluidlike radiation therefore more greatly
reduces the BAO--CMB tension than new free-streaming radiation.

While the analytic results above [\cref{eqn:rs-in-matter--radiation-ito-wr-R_star,eqn:medi-lcdm}]
are only exact in standard cosmology (with arbitrary radiation density), other classes of
high-density recombination are subject to similar considerations.
For instance, CMB fits with early dark energy require $\omega_c$ to increase with the EDE
fraction~\cite{Poulin:2018cxd, Vagnozzi:2021gjh, Poulin:2023lkg, Vagnozzi:2023nrq}, which has been
argued to preserve the early integrated Sachs--Wolfe feature on moderate angular
scales~\cite{Vagnozzi:2021gjh, Poulin:2023lkg}; this suggests that the matter--radiation ratio at
recombination (as encoded by $x_\mathrm{eq}$) generalizes to a matter--nonmatter ratio.
References~\cite{Chaussidon:2025npr, SPT-3G:2025vyw} indeed showed that EDE has little effect on the
BAO--CMB tension, as the inference of $\omega_{cb} r_\mathrm{d}^2$ from the CMB is only marginally
altered (see in particular Fig.~2 of Ref~\cite{SPT-3G:2025vyw}, which quoted only marginal
improvements to $2.5\sigma$ from $2.8 \sigma$ in \LCDM{}).\footnote{
    Though marginal summaries may understate the impact of early dark energy due to prior volume
    effects~\cite{Smith:2020rxx, Herold:2021ksg}, Fig.~2 of Ref.~\cite{SPT-3G:2025vyw} visually shows
    that the gradient of the CMB posterior in the $\Omega_m$--$\omega_m r_\mathrm{d}^2$ plane
    with respect to the EDE fraction is small in magnitude and, more importantly, not well aligned
    with the direction of tension.
    Moreover, an improvement in joint fit to CMB and BAO data (as cited in, e.g.,
    Ref.~\cite{Chaussidon:2025npr}) does not uniquely identify a reduction in tension rather than an
    unrelated improvement in fit to either individual dataset, especially given that early dark
    energy improves the fit to CMB data alone by nearly as
    much~\cite{AtacamaCosmologyTelescope:2025nti, SPT-3G:2025vyw}.
    Furthermore, dynamical dark energy provides a substantially larger improvement in joint fit
    despite having little direct effect on the CMB~\cite{Chaussidon:2025npr}.
}
Given the typically large sound speed of its fluid perturbations~\cite{Poulin:2023lkg}, early dark
energy is likely more akin to free-streaming than fluidlike radiation.

We conclude by commenting on analyses of acoustic scale data that attempt to be agnostic to
recombination-era physics by treating $r_\mathrm{d}$ as a free parameter~\cite{Pogosian:2024ykm,
GarciaEscudero:2025lef}.
These analyses impose a prior on the matter density inferred from the CMB in
\LCDM{}~\cite{Pogosian:2024ykm} or constrain it using other datasets modeled within
\LCDM{}~\cite{GarciaEscudero:2025lef, Sharma:2025iux} and are therefore not truly agnostic to
dynamics at early times, but merely to the calibration of $r_\mathrm{d}$~\cite{Sharma:2025iux}.
This choice does not capture nonstandard cosmologies: the CMB measures density ratios, and a change
to the absolute density at decoupling, whether because it occurs early or late or in the presence of
additional dark components, is required to change $r_\mathrm{d}$.
A larger inferred CDM density, as generally encoded by the CMB's requirement to fix $x_\mathrm{eq}$,
is well established in early recombination via varying constants~\cite{Sekiguchi:2020teg,
Baryakhtar:2024rky} and phenomenological modifications to the ionization
history~\cite{Lynch:2024gmp}, as well as high-density recombination via extra
radiation~\cite{Hou:2011ec} or other species like early dark energy~\cite{Poulin:2018cxd,
Vagnozzi:2021gjh, Poulin:2023lkg}.
These analyses are thus not meaningfully agnostic to recombination-era physics, but rather they
simply exclude (almost all) CMB data.
Constraining $\omega_{cb}$ in this manner is especially problematic for the inference of neutrino
masses, given that the neutrino density is geometrically determined via the difference between the
matter abundance long after decoupling (measured by the acoustic scale) and before decoupling
($\omega_{cb}$).
Likewise, identifying the suppression of structure by massive neutrinos requires knowledge of the
transfer function, i.e., $k_\mathrm{eq} \propto \omega_{cb} / \sqrt{\omega_r}$ (as well as the
primordial initial conditions).
The CMB contains by far the most information on $\omega_{cb}$; its variation with $r_\mathrm{d}^2$
is thus always key to neutrino mass measurements.
The approach taken in \cref{sec:matter-density} and Ref.~\cite{Loverde:2024nfi}, of simply
parametrizing densities in units $\propto 1 / r_\mathrm{d}^2$, incurs no such inconsistencies and
maintains maximal generality, beyond which one must specify (not ignore) the relationship between
the drag scale and early-time densities and ratios thereof.

\subsubsection{Optical depth, or the CMB lensing excess}\label{sec:lensing-excess}

We now discuss how the optical depth to reionization $\tau_\mathrm{reio}$---that is, supposing its
value is underestimated by large-scale CMB polarization data---alleviates the BAO--CMB tension, as
noted in Refs.~\cite{Sailer:2025lxj, Jhaveri:2025neg}.
We first explain the effective equivalence of the optical depth and a modulation of CMB lensing
on small scales.
Beyond dynamical effects around last scattering, the joint density in baryons and CDM determines the
turnover scale in the transfer function,
$k_\mathrm{eq} = a_\mathrm{eq} H(a_\mathrm{eq}) = \omega_{cb} / \sqrt{\omega_r / 2} \cdot H_{100}$.
A higher CDM density (i.e., $x_\mathrm{eq}$) can explain the preference for excess lensing of the
CMB's acoustic peaks~\cite{Planck:2018vyg, SPT-3G:2024atg, AtacamaCosmologyTelescope:2025blo, SPT-3G:2025bzu}, though not
without degrading the fit at large angular scales that are relatively unaffected by lensing.
Namely, the more density in matter compared to radiation, the smaller the turnover scale and the
larger amplitude of late-time structure at fixed wave number.
This tension between low and high multipoles~\cite{Planck:2016tof} can be mediated by enhanced
structure growth after decoupling, as frequently tested via a phenomenological rescaling of the
lensing amplitude $A_L$~\cite{Calabrese:2008rt}.

Given that the overall amplitude of the CMB anisotropy spectra at $\ell \gg 30$ is suppressed by
late-time reionization by a factor $e^{- 2 \tau_\mathrm{reio}}$, the CMB lensing excess is
equivalent to the preference for larger $\tau_\mathrm{reio}$ inferred indirectly from CMB
lensing~\cite{Planck:2018vyg, Loverde:2024nfi} than that measured from features in polarization on
large angular scales.
That is, increasing the optical depth and the amplitude of the primordial power spectrum in tandem
preserves the overall amplitude of the CMB while enhancing the amplitude of large-scale structure
itself.
A higher lensing amplitude permits lower CDM densities that better fit the CMB on unlensed scales,
i.e., $\ell \lesssim 1000$ or so.
In \LCDM{}, a decreased CDM density is compensated for by an increased dark energy density to fix
$\theta_s$, resulting in the exacerbated decrease in the matter fraction $\Omega_m$
correlated with $\tau_\mathrm{reio}$ that was noted in Ref.~\cite{Sailer:2025lxj}.

To explain the relevance to the BAO--CMB tension, Ref.~\cite{Jhaveri:2025neg} claimed that lensing
contributes significantly to the CMB's calibration of the sound horizon and that freeing the lensing
amplitude (via the optical depth, for instance) therefore mediates tension by weakening constraints
on $r_\mathrm{d}$.
However, neither the premise nor the conclusion is the case.
First, removing lensing information has no appreciable effect on the uncertainty in $r_\mathrm{d}$,
which increases only from $0.17\%$ to $0.19\%$ by excluding CMB lensing reconstruction data and to
$0.21\%$ by additionally marginalizing over the amplitude of two-point lensing (via a
phenomenological rescaling $A_\mathrm{smear}$).
Second, the decrease in $\omega_c$ with increasing $A_L$ or $\tau_\mathrm{reio}$ increases the sound
horizon $\propto x_\mathrm{eq}^{0.25} \propto \omega_c^{-0.21}$ at fixed $R_\star$ (i.e., the
Universe is larger at decoupling), which on its own decreases the CMB-calibrated \medi{} in
\cref{fig:medi-tension} (though only by a tenth of a percent) and exacerbates rather than alleviates
the tension.
As emphasized in \cref{sec:drag-horizon-as-units,sec:high-density-recombination} and
Ref.~\cite{Loverde:2024nfi}, the sound horizon only determines the units with which densities are
measured by the acoustic scale; a small rescaling of units cannot explain the sign of the matter-era
distance tension (i.e., the matter density deficit).
The salient effect of the lensing amplitude is the decrease in $\omega_{cb}$ that overcompensates
for the resulting increase in $r_\mathrm{d}^2$, yielding a smaller $\omega_{cb} r_\mathrm{d}^2$ and
larger \medi{}.

\begin{figure}[t!]
\begin{centering}
    \includegraphics[width=\textwidth]{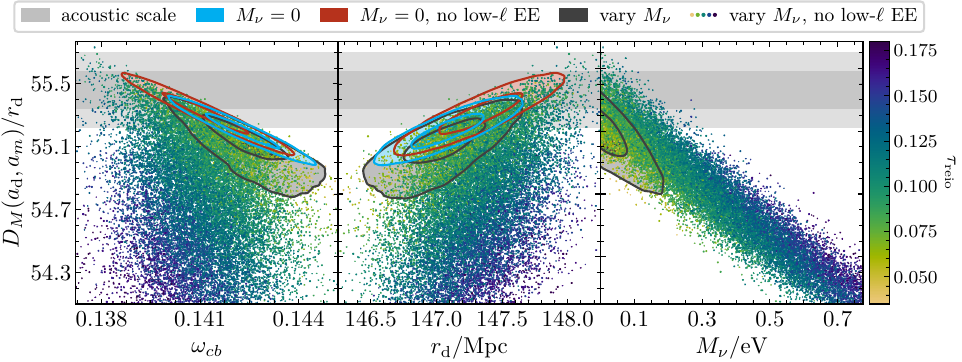}
    \caption{
        Posterior distribution over the matter-era distance interval and the density in
        baryons and CDM, the drag horizon, and the neutrino mass sum.
        Results that vary $\summnu$ and exclude low-$\ell$ polarization data are
        displayed as scatter colored by the optical depth, with the $1$ and $2 \sigma$
        mass levels thereof that include all CMB data appearing in black.
        The same mass levels of posteriors that fix $\summnu = 0$ and either exclude and
        include low-$\ell$ polarization data appear in red and blue, respectively.
        Horizontal shaded regions depict the $1$ and $2 \sigma$ limits of the \medi{}
        as measured by the acoustic scale (see \cref{fig:medi-tension}).
    }
    \label{fig:medi-tau-mnu}
\end{centering}
\end{figure}

\Cref{fig:medi-tau-mnu} demonstrates the positive correlation between the \medi{} and the drag
horizon calibrated by the CMB, despite the former's inverse dependence on the latter, and a stronger
anticorrelation with $\omega_{cb}$.
The larger values of $\tau_\mathrm{reio}$ permitted by excluding \Planck{}'s large-scale
polarization data allow for lower $\omega_{cb}$ and thus predict larger
$D_M(a_\mathrm{d}, a_m) / r_\mathrm{d} \propto 1 / \sqrt{\omega_{cb}} r_\mathrm{d}$, despite the
slight, correlated increase in $r_\mathrm{d}$.
This trend leaves room for slightly heavier neutrinos (whose increased suppression of structure is
also accommodated by a yet-larger optical depth), but not enough to fully resolve the tension, as
the CMB still underpredicts the \medi{} even with massless neutrinos; see
\cref{fig:mnu-high-z-sols}.
Note that including low-$\ell$ polarization data but marginalizing over $A_\mathrm{smear}$ indeed
yields nearly identical results, with $\ln A_\mathrm{smear} / 2$ taking the place of
$\tau_\mathrm{reio}$.

\subsection{Postdecoupling physics}\label{sec:postdecoupling}

We now take the CMB's calibration of $\omega_{cb} r_\mathrm{d}^2$ for granted and consider physics
after photon--baryon decoupling that modify the extrapolated matter-era distance interval.
With spatial curvature an exception treated separately in \cref{sec:theory-curvature}, such
scenarios may be studied via (presumably perturbatively small) modifications to the expansion
history during the matter era.
The variation of the comoving distance $\delta \chi(a)$ with small relative variations in the
energy density $\delta \ln \bar{\rho}(\ln \tilde{a})$ about the density $\bar{\rho}_{mr}$ in matter
and radiation is
\begin{align}
	\frac{\delta \chi(a)}{\delta \ln \bar{\rho}(\ln \tilde{a})}
    = - \int_0^a \frac{\ud \ln a'}{a' H(a')}
        \frac{\delta \ln H(a')}{\delta \ln \bar{\rho}(\ln \tilde{a})}
    = - \frac{1}{2 \tilde{a} H(\tilde{a})}.
\end{align}
We take $\ln \tilde{a}$ as the coordinate to obtain the sensitivity coefficient per $e$-fold.
Writing $H_{mr}(a) = \sqrt{\bar{\rho}_{mr}(a) / 3 \Mpl^2}$ and recalling \cref{eqn:medi-mr},
the sensitivity of the \medi{} evaluates to
\begin{align}
	\frac{\delta \ln \chi(a_\mathrm{d}, a_m)}{\delta \ln \bar{\rho}(\ln a)}
    \simeq - \frac{1}{2 a H_{mr}(a) \chi_{mr}(a_\mathrm{d}, a_m)}
    &= - \frac{1}{4}
        \frac{a}{\sqrt{a + a_\mathrm{eq}}}
        \frac{
            1
        }{
            \sqrt{a_m + a_\mathrm{eq}} - \sqrt{a_\mathrm{d} + a_\mathrm{eq}}
        }
    .
    \label{eqn:medi-sensitivity-general-rho}
\end{align}
In general, the \medi{}'s sensitivity (per $e$-fold) to small modifications to
the energy budget scales with $1 / a H_{mr}(a) \propto \sqrt{a}$ (for $a \gg a_\mathrm{eq}$), i.e.,
earlier modifications are modestly suppressed relative to later ones (when the comoving horizon is
larger).
Integrating yields the leading-order correction to the matter--radiation result,
\begin{align}
    \chi(a_\mathrm{d}, a_m)
    \simeq
        \chi_{mr}(a_\mathrm{d}, a_m)
        - \int_{a_\mathrm{d}}^{a_m} \frac{\ud \ln a}{2 a H_{mr}(a)}
        \frac{\bar{\rho}(a) - \bar{\rho}_{mr}(a)}{\bar{\rho}_{mr}(a)}
    .
    \label{eqn:medi-correction-general-rho}
\end{align}
This correction is naturally just the perturbation to the energy density integrated over
conformal time $\tau = \int \ud \ln a / a H$; which integration variable is most convenient (scale
factor, time, or conformal time) depends on the particular application, though distances are
observed at known redshift.

\begin{figure}[t!]
\begin{centering}
    \includegraphics[width=6in]{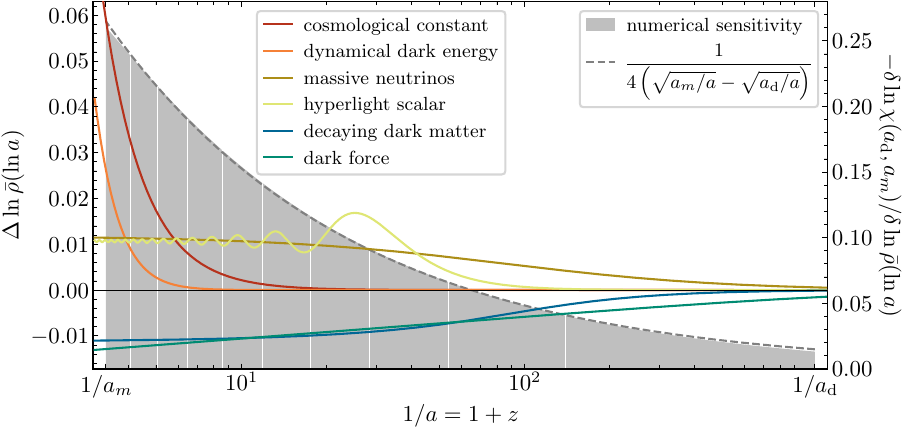}
    \caption{
        Sensitivity of the matter-era distance interval $\chi(a_\mathrm{d}, a_m)$
        [with $a_m = 1/(1 + 2.330)$] to modifications to the expansion history.
        The grey filled region depicts the sensitivity coefficient per $e$-fold to variations
        $\delta \ln \bar{\rho}(\ln a)$ (right axis), divided into ten equal-weight intervals.
        The dashed grey line depicts an analytic approximation thereof
        [\cref{eqn:medi-sensitivity-general-rho}].
        Colored lines (left axis) display the relative change to the energy density
        $\Delta \ln \bar{\rho}(\ln a)$ incurred by various extensions of or additions to a
        matter--radiation Universe, each discussed in detail in the main text.
        Parameters for all scenarios are chosen to match the fiducial $0.44\%$ change in the
        \medi{} incurred by a cosmological constant ($\summnu = 157~\mathrm{meV}$, $1.06\%$ of dark
        matter in a hyperlight scalar, $1.33\%$ of dark matter decaying, and a dark force strength
        $2.94 \times 10^{-3}$ that of gravity), except for dynamical dark energy; in that case,
        parameters are chosen to be representative of preferences from contemporary data (see
        \cref{sec:dde-alternatives} for discussion).
    }
    \label{fig:medi-sensitivity}
\end{centering}
\end{figure}

In \cref{fig:medi-sensitivity} we compare the sensitivity of the \medi{}
[\cref{eqn:medi-sensitivity-general-rho}] to the variations in energy density of a variety of
postdecoupling scenarios, similar to an analysis of the sound horizon and diffusion scale in
Ref.~\cite{Knox:2019rjx}.
\Cref{fig:medi-sensitivity} depicts the relatively greater sensitivity of the \medi{} to later eras
when the Universe is largest; nearly $90\%$ of the weight lies below redshift $100$ and $50\%$ below
$10$.
On top of that, the various scenarios we consider exhibit greater deviations at lower redshifts
(though in some cases controlled by a second model parameter), making their effects well
approximated by their impact on an otherwise matter-dominated Universe.

In the remainder of this section, we study the impact of these models on the matter-era distance
interval (with massive neutrinos discussed in \cref{sec:mnu} and hyperlight scalars in
\cref{app:scalars}) analytically and numerically.
We calculate their effect on the \medi{} analytically from \cref{eqn:medi-correction-general-rho}
(with details relegated to \cref{app:analytic} and spatial curvature treated separately), derive
constraints from measurements of the \medi{} alone using the semianalytic procedure presented in
\cref{sec:matter-density}, and compare those results to fully numerical posterior distributions.
For the latter, we use DESI DR2 data and CMB priors on $x_\mathrm{eq}$ and $R_\star$
for simplicity of illustration (though neglecting modifications to the growth of structure has a
small or negligible effect for all cases except decaying dark matter).
In constructing the CMB priors, we continue to exclude lensing reconstruction data (which, in the
case of dark forces in \cref{sec:dark-forces}, is for technical reasons discussed in
Ref.~\cite{Costa:2025kwt}).
In all cases, for analytic estimates we conservatively quantify the matter-era distance tension with
minimum-mass neutrinos, i.e., we add a $\summnu = 60~\mathrm{meV}$ contribution
[\cref{eqn:medi-mnu-apx}] to the $\summnu = 0$ result in \cref{fig:medi-tension}, yielding
$\Delta D_M(a_\mathrm{d}, a_m) / r_\mathrm{d} = 0.36 \pm 0.16$.
We also comment on geometric degeneracies with varying (rather than fixed) neutrino masses and what
influence each extension has on the growth of structure and CMB lensing excess, with complete
analyses of CMB and BAO data presented in \cref{sec:mnu-vs-highz}.

\subsubsection{Decaying dark matter}\label{sec:ddm}

A straightforward means to address the matter density deficit, and therefore the matter-era distance
tension, is to deplete a subcomponent of dark matter via decays into, say, dark radiation on
timescales long compared to the age of the Universe at decoupling but short compared to its present
age~\cite{Turner:1984ff, BOSS:2014hhw, Poulin:2016nat, Bringmann:2018jpr, McCarthy:2022gok,
Lynch:2025ine, Montandon:2026vuc}.
Reference~\cite{Lynch:2025ine} explored how a compensatory decrease in the dark matter abundance,
matching the late-time mass density in neutrinos and timed around their nonrelativistic transition,
alleviates the incompatibility of BAO and CMB data with nonzero neutrino masses.
Though the CMB's preference for excess lensing curtails the model's potential to alleviate the
matter-era distance excess~\cite{Lynch:2025ine}, we review this proposal as an application of the
\medi{} and our analytic methods.

Using the perturbative approach of \cref{eqn:medi-correction-general-rho}, \cref{app:ddm}
analytically derives the impact of the decay of a dark matter subcomponent into dark radiation
(accounting for the decay products and the noninstantaneous transition at leading order), which is
well approximated by
\begin{align}
    \frac{\Delta \chi(a_\mathrm{d}, a_m)}{1 / H_{100} \sqrt{\omega_m}}
    &\simeq
        f_\mathrm{ddm} \sqrt{a_m}
        \left(
            1
            + \sqrt{\frac{a_\mathrm{dd}}{a_m}}
            \left[
                \frac{1}{10} \left( \frac{a_\times}{a_\mathrm{dd}} \right)^2
                - \sqrt{\frac{a_\times}{a_\mathrm{dd}}}
                - \frac{\Gamma(5/3)}{\sqrt{a_\times / a_\mathrm{dd}}}
            \right]
            + \Gamma(5/3) \frac{a_\mathrm{dd}}{a_m}
        \right)
    \label{eqn:medi-ddm-apx}
    .
\end{align}
Here
$f_\mathrm{ddm} = \lim_{a \to 0} \bar{\rho}_\mathrm{ddm}(a) / \bar{\rho}_m(a)$ is the fraction of
all matter (rather than dark matter, for simplicity) that decays and $a_\mathrm{dd}$ the scale
factor of decay, defined as when the age of the Universe
$t(a_\mathrm{dd}) \simeq 2 a_\mathrm{dd}^{3/2} / 3 \sqrt{\omega_m} H_{100}$ (in matter domination)
matches the lifetime $1 / \Gamma_\mathrm{dd}$.
Choosing $a_\times = 1.12 a_\mathrm{dd}$ minimizes errors (see \cref{app:ddm} for details).
Finally, the fiducial value of the matter density $\omega_m$ in \cref{eqn:medi-ddm-apx} is set to
the density in CDM and baryons inferred by the CMB, since the shape of the acoustic peaks is
sensitive to the matter density before decoupling and therefore, for the parameters we consider,
before decay.

\begin{figure}[t!]
\begin{centering}
    \includegraphics[width=\textwidth]{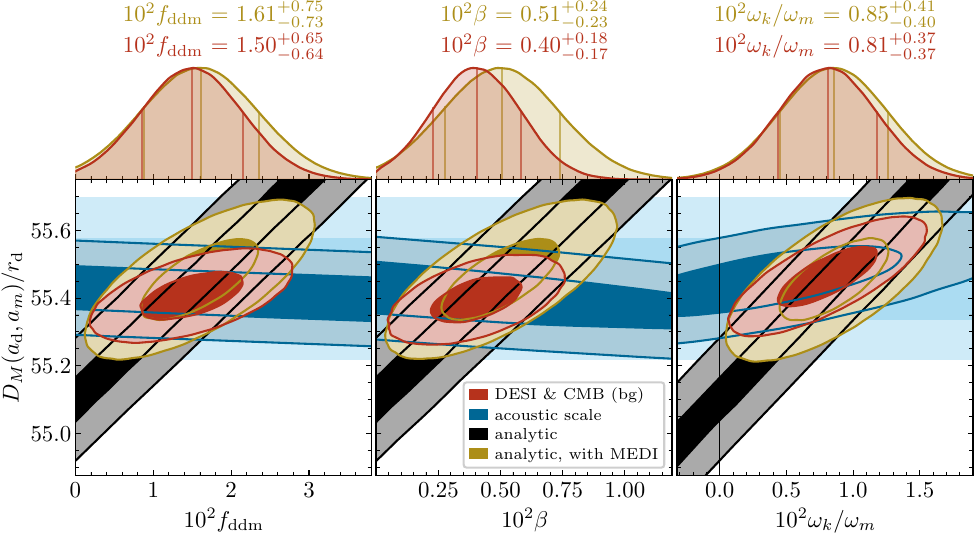}
    \caption{
        Acoustic scale constraints on postdecoupling modifications to the background cosmology
        from a fraction $f_\mathrm{ddm}$ of dark matter decaying at redshift $\sim 150$ (left),
        a scalar-mediated long-range force acting on dark matter with strength $\beta$ relative
        to gravity (middle), or nonzero curvature with density relative to that in matter
        $\omega_k / \omega_m$ (right).
        Each panel depicts the distribution of the matter-era distance interval with each model's
        new parameter, deriving from acoustic scale data from DESI DR2 and the CMB (blue) and
        combined with a CMB prior on the CDM and baryon densities at recombination (red), from this
        latter CMB prior combined with a weak one on $a_\mathrm{eq}$ and $\amL$ (black, same as in
        \cref{fig:matter-density-from-medi}), and from this prior combined with a fully marginalized
        measurement of the \medi{} (gold); the latter two are derived with the analytic
        approximations from this section and the semianalytic procedure described in
        \cref{sec:matter-density}.
        The \medi{} demonstrably contains most of the information for each extension parameter, with
        the discarded information being just its correlation with the \medi{} measured by the
        acoustic scale (blue).
        Moreover, the $> 2 \sigma$ preference in each case derives from mediating the matter-era
        distance tension.
    }
    \label{fig:high-z-sols}
\end{centering}
\end{figure}

For our fiducial $a_m = 1 / (1 + 2.330)$, the sensitivity coefficient is
$\partial \ln \chi(a_\mathrm{d}, a_m) / \partial \ln (1 + f_\mathrm{ddm}) \simeq 0.4$ for decay
occurring at redshifts around $150$, from which we estimate that
the matter-era distance tension would be resolved by a fraction $10^2 f_\mathrm{ddm} = 1.6 \pm 0.7$
of dark matter decaying.
This estimate agrees closely with the result derived from the semianalytic procedure outlined in
\cref{sec:matter-density}, as depicted in \cref{fig:high-z-sols}.
Moreover, \cref{fig:high-z-sols} shows that the complete result (from all acoustic scale data and
without analytic approximations) is only about $15\%$ more precise and shifted by a bit more than a
tenth of a standard uncertainty.
The improvement derives from the correlation of the \medi{} measured by the acoustic scale with
$f_\mathrm{ddm}$, which, since we impose a CMB prior on $\omega_{cb} r_\mathrm{d}^2$, merely
reflects its correlation with the late-time matter density $\omega_m r_\mathrm{d}^2$ evident in
\cref{fig:matter-density-from-medi}.
Our simple approach of using the fully marginalized \medi{} measurement in semianalytic results is
conservative (e.g., with respect to possible dynamics of dark energy), though only intended for
illustration.

The results in \cref{fig:high-z-sols} neglect decaying dark matter's effects on the growth of
structure, which ultimately hinder its potential to alleviate neutrino mass
limits~\cite{Lynch:2025ine, Montandon:2026vuc}.
That said, our analytic results make the geometric degeneracy between the two transparent:
the functional dependence of \cref{eqn:medi-ddm-apx} on $a_\mathrm{dd} / a_m$ is identical to
that of \cref{eqn:medi-mnu-apx} on $a_\nu / a_m$ for massive neutrinos, with numerical coefficients
differing by $\sim 10\%$.
Prior work discussed how, when taking $f_\mathrm{ddm} = f_\nu$, choosing $\Gamma_\mathrm{dd}$ to
time decays at~\cite{Lynch:2025ine} or before~\cite{Montandon:2026vuc} $a_\nu$ largely cancels their
effects on the expansion history; however, the cancellation cannot be made exact in $\bar{\rho}(a)$
at all times, nor in distances out to epochs before their transitions (namely, the distance to last
scattering).
The analytic results \cref{eqn:medi-mnu-apx,eqn:medi-ddm-apx} enable a slightly different choice of
$a_\mathrm{dd}$ in terms of $a_\nu$ that cancels their most important observable effect (i.e., on
the \medi{}); however, without any direct probes of the expansion history so deep in the matter era,
nor any subpercent measurement of the matter density at late times (independent of the \medi{}),
there is no particular need to fix $f_\mathrm{ddm} = f_\nu$ and match the \medi{} by adjusting
$a_\mathrm{dd}$ in this manner.
Indeed, other observables like CMB lensing cannot be matched, even with a fixed expansion history:
by reducing the density in clustering matter, the decay of dark matter suppresses perturbations in
the density field $\delta \rho$ by $1 - f_\mathrm{ddm}$.
Subsequently, the energy density in the decay products transiently suppresses the growth rate.
The CMB lensing excess is therefore exacerbated in proportion to the degree to which decaying dark
matter alleviates the matter-era distance excess.

\subsubsection{Dark forces}\label{sec:dark-forces}

The salient feature of decaying dark matter is not that the present-day matter density differs from
that inferred from the primary CMB, but rather that the Universe dilutes faster on average during
matter domination.
The \medi{} increases if dark matter redshifts faster than $a^{-3}$ by whatever means---for
instance, via the mass evolution characteristic of dark matter subject to a scalar-mediated
long-range force~\cite{Archidiacono:2022iuu, Bottaro:2023wkd, Bottaro:2024pcb}.
This nontrivial background evolution is a main signature of such dark forces in linear cosmology,
with the largest effect being the accumulated reduction in dark matter's energy density as its mass
decays with time~\cite{Archidiacono:2022iuu, Costa:2025kwt}.
Given that the acoustic scale mostly measures the matter density via its integrated effect on the
\medi{} rather than its instantaneous value at late times, DESI's sensitivity to dark matter's
modified rate of redshift in the last $e$-fold of expansion ($z < z_m$) is far subdominant to the
\medi{}'s sensitivity to the preceding six $e$-folds.
References~\cite{Bottaro:2024pcb, Costa:2025kwt} indeed demonstrated a marginal preference for
nonzero force strengths from CMB and DESI data; we now show that these results are reproduced by the
\medi{} alone.

Building on analytic results from Ref.~\cite{Costa:2025kwt} (leaving details to
\cref{app:dark-forces}), the sensitivity of the \medi{} to a scalar-mediated force (whose range is
not much smaller than the size of the Universe at $a_m$) acting on a fraction $f_\chi$ of all matter
with strength $\beta$ relative to gravity is approximately
\begin{align}
    \frac{\chi(a_\mathrm{d}, a_m)}{\chi_{mr}(a_\mathrm{d}, a_m)} - 1
    &\simeq \frac{\beta f_\chi^2}{2}
        \frac{
            \sqrt{a_m} \left[ \ln (a_m / 4 a_\mathrm{eq}) - 4 / 3 \right]
        }{\sqrt{a_m + a_\mathrm{eq}} - \sqrt{a_\mathrm{d} + a_\mathrm{eq}}}
    \label{eqn:medi-lrf-apx}
    .
\end{align}
The sensitivity coefficient for $a_m = 1 / (1 + 2.330)$ is
$\partial \ln \chi(a_\mathrm{d}, a_m) / \partial \ln (1 + \beta f_\chi^2) \simeq 2.3$.
Since the dark matter mass begins evolving before decoupling (around equality), the appropriate
choice of matter density in the fiducial \medi{} $\chi_{mr}(a_\mathrm{d}, a_m)$ is less trivial than
for decaying dark matter; Reference~\cite{Costa:2025kwt} showed that the CMB best measures the dark
matter density around decoupling, corresponding roughly to
$\lim_{a \to 0} a^3 \bar{\rho}_{\chi b}(a) \left( 1 - \beta f_\chi^2 \right)$.
To respect this distinction, when drawing values of $\omega_{cb} r_\mathrm{d}^2$ from a CMB prior
for \LCDM{} we set $\omega_m r_\mathrm{d}^2$ in $\chi_{mr}(a_\mathrm{d}, a_m) / r_\mathrm{d}$ to
$\omega_{cb} r_\mathrm{d}^2 / \left( 1 - \beta f_\chi^2 \right)$, reducing the sensitivity
coefficient of the \medi{} to $\beta f_\chi^2$ by one half.

With the force acting on all of dark matter ($f_\chi \simeq 0.84$), we may estimate that the
matter-era distance tension is resolved by a force strength $10^2 \beta = 0.5 \pm 0.23$, which again
reproduces the result in \cref{fig:high-z-sols} employing the semianalytic procedure from
\cref{sec:matter-density}.
Accounting for the correlation of the \medi{} measured by the acoustic scale with $\beta$ has a
marginally more significant effect than for decaying dark matter (see \cref{fig:high-z-sols}), since
the mass evolution impacts low-redshift distances less trivially than merely shifting the late-time
matter density.
These results also reproduce full parameter inference from Ref.~\cite{Costa:2025kwt}, which
obtained $10^2 \beta = 0.3^{+0.16}_{-0.15}$ with $\summnu$ fixed to zero, for which the tension is
$\Delta D_M(a_\mathrm{d}, a_m) / r_\mathrm{d} = 0.26 \pm 0.16$; our semianalytic result in this case
is $10^2 \beta = 0.37 \pm 0.23$, while the numerical result from the acoustic scale and the CMB prior
is $0.3^{+0.17}_{-0.16}$.

The precise consistency of our results with those of Ref.~\cite{Costa:2025kwt}, despite ignoring
effects on perturbations, is no coincidence: the enhanced growth rate of dark matter overdensities
due to the long-range force is precisely compensated by the faster decay of the background density,
leaving all the gravitational effects of dark matter perturbations effectively
unchanged~\cite{Costa:2025kwt}.
Dark forces mediated by light scalars therefore have no impact on the CMB lensing excess,
contradicting claims from Ref.~\cite{Graham:2025dqn}, but rather only alleviate neutrino mass limits
via background effects~\cite{Costa:2025kwt}.
That said, dark forces are not subject to the same penalty as decaying dark matter from the CMB's
preference for excess lensing.
Reference~\cite{Costa:2025kwt} and \cref{fig:mnu-high-z-sols} show that dark forces remove most of
DESI's impact on neutrino mass limits, allowing mass sums nearly as large as allowed by the CMB
alone.

\subsubsection{Spatial curvature}\label{sec:theory-curvature}

The \medi{} is the most sensitive geometric probe of spatial curvature.
In a nonflat Universe where the transverse and comoving distances do not coincide, one may still
compute the difference $D_M(a_\mathrm{d}, a_m) \equiv D_M(a_\mathrm{d}) - D_M(a_m)$---it just
does not reduce to a single integral of $1 / a H$.
Curvature most affects the furthest distances, as
$D_M / \chi \approx 1 - \left( \chi / R_k \right)^2 / 6$
where again $R_k^2 = - 1 / \omega_k H_{100}^2$~\cite{Hogg:1999ad}.
In particular, the distance to last scattering [and therefore the \medi{} $D_M(a_\mathrm{d}, a_m)$]
is parametrically more sensitive to curvature than low-redshift (nearby) distances, e.g., six times
more so than DESI's furthest measurement (at $z = 2.330$).
The value of high-redshift BAO in mitigating the sensitivity of curvature measurements to dark
energy was noted long ago~\cite{Knox:2005hx}; the arguments of \cref{sec:theory},
particularly the improvement in $D_M(a_m) / r_\mathrm{d}$ from an ensemble of BAO measurements that
is robust to dark energy dynamics, indicate that the \medi{} represents the majority of the
information even from all data.
Reference~\cite{Chen:2025mlf} explained how curvature's impact on the distance to last scattering is
preferred in combination with DESI and slightly alleviates the incompatibility with massive
neutrinos; here we show that the matter-era distance excess underlies these results, and we augment
their discussion of parameter degeneracies with an analytic derivation from the \medi{}.

Curvature affects distances both in the comoving--transverse distance relation and in the Friedmann
equation.
To show that the former far dominates over the latter, we first note that, since
$\bar{\rho}_k \propto a^{-2}$, the shift in the comoving \medi{} is
\begin{align}
    \frac{\Delta \chi(a_\mathrm{d}, a_m)}{1 / H_{100} \sqrt{\omega_m}}
    \simeq - \frac{a_m^{3/2} \omega_k}{3 \omega_m}
    .
\end{align}
Rewriting the total change, including the leading-order correction of $D_M[\chi]$, gives
\begin{align}
    \frac{D_M(a_\mathrm{d}, a_m)}{\chi_{mr}(a_\mathrm{d}, a_m)} - 1
    &\simeq \frac{a_m \omega_k}{6 \omega_m}
        \left(
            4 \left[ \frac{H_{100} \chi(a_\mathrm{d}, a_m)}{2 \sqrt{a_m} / \sqrt{\omega_m}} \right]^2
            \frac{
                \chi(a_\mathrm{d})^3 - \chi(a_m)^3
            }{\left[ \chi(a_\mathrm{d}) - \chi(a_m) \right]^3}
            - \left[ \frac{H_{100} \chi(a_\mathrm{d}, a_m)}{2 \sqrt{a_m} / \sqrt{\omega_m}} \right]^{-1}
        \right)
    \label{eqn:medi-curvature}
    .
\end{align}
Since the quantity in parentheses multiplies the small parameter $\omega_k / \omega_m$, it may be
approximated with fiducial values.
Taking measured values from the acoustic scale of
$1 / \theta_\perp(a_\mathrm{d}) \approx 94.3$ and $1 / \theta_\perp(a_m) \approx 38.9$ [for our usual
$a_m = 1 / (1 + 2.330)$], the difference of distances cubed divided by the cubed difference is
about $4.6$, while the combination $\sqrt{\omega_m} H_{100} \chi(a_\mathrm{d}, a_m) / 2 \sqrt{a_m}$
is about $0.93$.
Curvature's impact through the Friedmann equation [the second term in parentheses in
\cref{eqn:medi-curvature}] is thus about $15$ times less important than its
effect on the transverse distance at fixed $\chi$ (the first term)---not surprising, given that a
subpercent contribution to the total density today corresponds to a yet-smaller fractional
contribution during matter domination that diminishes at higher redshift with $1/(1 + z)$.

The coefficient of $\omega_k / \omega_m$ in \cref{eqn:medi-curvature} is about $0.75$, so mediating
the matter-era distance excess requires negative curvature (i.e., with positive density parameter).
We estimate $\omega_k / \omega_m = 0.87 \pm 0.39$, which matches the semianalytic result in
\cref{fig:high-z-sols} precisely; the full, numerical result is hardly altered by including rather
than marginalizing over the correlation between the \medi{} and $\omega_k$.
Taking $\Omega_m \approx 0.295$ to convert the latter from $\omega_k / \omega_m$ to $\Omega_k$
yields $10^3 \Omega_k = 2.4 \pm 1.1$, quite consistent with the full result from DESI and CMB data
of $2.3 \pm 1.1$~\cite{DESI:2025zgx}.

Our analytic results identify a degeneracy with neutrino masses that agrees with
Ref.~\cite{Chen:2025mlf}, who report its slope as $\Delta f_\nu / \Delta \Omega_k = 3.5$:
from \cref{eqn:medi-mnu-apx,eqn:medi-curvature}, the sensitivity coefficients of the \medi{} to
$f_\nu$ and $\Omega_k$ are about $- 0.4$ (for a single massive eigenstate) and $2.5$, respectively,
so we obtain a slope of $2.5 \Omega_k / 0.4 f_\nu \approx 3.3$
for $\summnu = 60~\mathrm{meV}$ and $\Omega_k = 2.4 \times 10^{-3}$ as above.
Separately, Ref.~\cite{Graham:2025dqn} claimed that curvature addresses the BAO--CMB tension by
increasing the total density at decoupling, mimicking matter.
As \cref{eqn:medi-curvature} shows, curvature's contribution to the energy density is at all times
entirely negligible relative to its impact on the conversion from comoving to transverse distance,
and also has the opposite sign; its density at decoupling would not be relevant anyway compared to
late times (see \cref{fig:medi-sensitivity}), and even if it were, the relative contribution from
curvature is reduced by a factor of $1 + z_\mathrm{d}$ at decoupling compared to today, not enhanced
as they write.

The preferred curvature fraction above reduces the amplitude of CMB lensing by just over a percent
(about double its impact on the \medi{}), which increases the CMB lensing excess by a small amount
relative to the $\gtrsim 3\%$ precision with which current data constrain phenomenological
rescalings of the lensing amplitude.
Curvature thus alleviates neutrino mass limits solely by mediating the matter-era distance excess,
with a penalty from the lensing excess that increases as greater curvature is invoked to accommodate
neutrinos heavier than $60~\mathrm{meV}$.

\subsection{Implications for neutrino masses}\label{sec:mnu-vs-highz}

To further illustrate the importance of the matter-era distance excess for current cosmological
inference, we quantify the degree to which some of the high-redshift solutions discussed in
\cref{sec:predecoupling,sec:postdecoupling} relax neutrino mass limits.
\Cref{sec:mnu} establishes the primary role of the \medi{} in geometric measurements of neutrino
masses; with a new-physics explanation of the excess, neutrino masses are then constrained almost
entirely by the amplitude of late-time structure, including whatever impact the invoked new physics
has on structure growth.
In this section, we derive neutrino mass constraints in these various extensions of the cosmological
model and summarize discussion throughout \cref{sec:predecoupling,sec:postdecoupling} on their
interplay.

\begin{figure}[t!]
\begin{centering}
    \includegraphics[width=\textwidth]{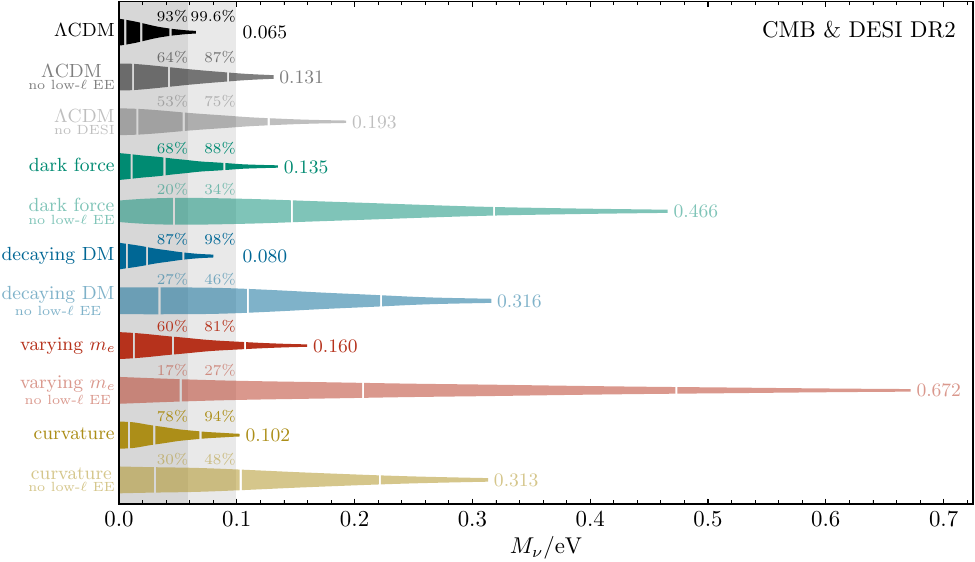}
    \caption{
        Impact of high-redshift solutions to the matter-era distance excess on neutrino mass limits
        from DESI DR2 and CMB temperature and polarization data, comparing dark forces
        (\cref{sec:dark-forces}), decaying dark matter (\cref{sec:ddm}), high-redshift recombination
        via a time-varying electron mass (\cref{sec:high-redshift-recombination}), and spatial
        curvature (\cref{sec:theory-curvature}) to \LCDM{} results.
        Each posterior is truncated at the 95th percentile (which is labeled), white lines mark
        the median and $\pm 1 \sigma$ quantiles, and the percent of the posterior that is
        incompatible with each mass hierarchy (vertical grey regions) is also labeled.
        Results either include (opaque) or omit (transparent) large-scale polarization data; the
        latter effectively removes the information on neutrino masses from the amplitude of
        structure.
        The \LCDM{} result without large-scale polarization data illustrates the limited impact of
        optical depth measurements on the matter-era distance excess (\cref{sec:lensing-excess}) as,
        despite the simultaneous resolution of the CMB lensing excess, the 95th percentile relaxes
        only halfway to the result that excludes DESI data.
        Spatial curvature and (especially) decaying dark matter further suppress the lensing
        amplitude and therefore only have a marginal effect; a dark force and a time-varying
        electron mass, which have no direct impact on CMB lensing, are most successful in mediating
        the matter-era distance excess and alleviating neutrino mass limits.
        In these scenarios, upper limits are substantially increased by additionally omitting
        large-scale polarization data (which mitigates the impact of the CMB lensing excess).
    }
    \label{fig:mnu-high-z-sols}
\end{centering}
\end{figure}

\Cref{fig:mnu-high-z-sols} displays posterior distributions over the neutrino mass sum derived from
DESI DR2 BAO data combined with CMB temperature and polarization data and marginalized over the
parameters of various high-redshift solutions to the matter-era distance excess.
As anticipated from \cref{sec:high-redshift-recombination} and shown previously in
Refs.~\cite{Baryakhtar:2024rky,Loverde:2024nfi}, high-redshift recombination via a time-varying
electron mass is the most effective means to remove geometric information on neutrino masses.
Dark forces are nearly as effective since, as shown in Ref.~\cite{Costa:2025kwt} and reviewed in
\cref{sec:dark-forces}, they have minimal direct impact on CMB lensing.
These two scenarios remove nearly all of the penalty on neutrino masses from adding DESI data to CMB
data.
Decaying dark matter, on the other hand, exacerbates the lensing excess by reducing the density in
clustering matter and suppressing the growth rate via its relativistic decay products (see
\cref{sec:ddm} and Ref.~\cite{Lynch:2025ine}); spatial curvature suppresses the lensing amplitude to
a slightly lesser extent (\cref{sec:theory-curvature} and Ref.~\cite{Chen:2025mlf}).
Decaying dark matter and curvature thus only marginally relax neutrino mass limits.

In all scenarios, but especially for high-redshift recombination and a dark force, dropping
large-scale CMB polarization data (which measure the optical depth) further relaxes limits by
allowing the primordial power spectrum to have a larger amplitude, enabling greater suppression of
structure by free-streaming neutrinos (and decaying dark matter or negative spatial curvature).
The CMB lensing amplitude remains a point of tension in all models, underscoring the importance of
robust measurements of the optical depth.
Removing the optical depth constraint alone, despite its potential to not just allow for greater
neutrino free-streaming but also partly mediate the matter-era distance excess (see
\cref{sec:lensing-excess}), is less impactful than a varying electron mass and comparably so to a
dark force.\footnote{
    Our result excluding low-$\ell$ polarization data, which uses high-$\ell$ data from \Planck{},
    ACT, and SPT, is consistent with that using just \Planck{} PR4~\cite{Sailer:2025lxj}
    or PR3~\cite{Jhaveri:2025neg} and is slightly tighter than that from PR3 and DESI
    DR1~\cite{Loverde:2024nfi}.
}
This comparison suggests that solving the geometric (matter-era distance) tension alone more
effectively addresses the neutrino mass tension than solving the lensing excess alone.
Regardless, more complex reionization histories or large-scale primordial power spectra only achieve
a small fraction of the requisite increase in the optical depth~\cite{Jhaveri:2025neg}; another
possibility is birefringence induced by a hyperlight axion~\cite{Namikawa:2025doa}, though it has
not yet been tested in a full parameter analysis.

\subsection{High-redshift alternatives to dynamical dark energy}\label{sec:dde-alternatives}

Current low-redshift distance measurements, which directly trace the late-time expansion history, do
not provide compelling evidence for dynamical dark energy on their own~\cite{DESI:2025zgx,
DES:2025sig}.
Conversely, the CMB is only weakly sensitive to the low-redshift expansion history and, insofar as
dark energy is concerned, effectively provides nothing more than an integral constraint via the
distance to last scattering.
Yet the strongest evidence for dynamical dark energy emerges when combining distances with the CMB.
However, \cref{fig:medi-tension} demonstrates that this improvement in joint fit cannot derive from
dynamical dark energy's impact on the \textit{measured} matter-era distance interval---i.e., on
$D_M(a_m) / r_\mathrm{d}$ inferred from DESI data alone.

In this section, we show that the joint evidence for dark energy dynamics in fact derives from its
impact on distances at \textit{high} redshift.
That is, dynamical dark energy mediates the BAO--CMB tension specifically through the extrapolation
of linear-in-$a$ approximations of the equation of state to high redshift, where a series expansion
in small $1 - a$ is least valid and the inferred ``phantomlike'' behavior most severe.\footnote{
    Our analysis thus identifies the physical mechanism underlying the observation in
    Ref.~\cite{Yang:2026yaq} that the phantom transition is necessary to alleviate neutrino mass
    limits.
}
In turn, we show that any high-redshift solution to the matter-era distance excess, like those
discussed in \cref{sec:high-redshift-solutions}, eliminates the CMB's contribution to the preference
for dynamical dark energy.
We then identify the residual and more direct evidence for dark energy dynamics in DESI's
low-redshift distances---that is, the features a microphysical model would have to reproduce, on top
of addressing the matter-era distance excess, in order to match the evidence attributed to the
phenomenological model.

\subsubsection{Dynamical dark energy and the \medi{}}\label{sec:dde}

We apply \cref{eqn:medi-correction-general-rho} to generalize the impact of a cosmological constant
derived in \cref{sec:theory-cc} to dark energy with a varying equation of state.
For tractability we assume radiation and dark energy are never simultaneously relevant.
The correction is integrable for a constant equation of state $w_\mathrm{DE}$ at $z > z_m$, i.e.,
$\bar{\rho}_\mathrm{DE}(a) / \bar{\rho}_m(a) = \bar{\rho}_\mathrm{DE}(a_m) / \bar{\rho}_m(a_m) \cdot (a / a_m)^{- 3 w_\mathrm{DE}}$.
In this limit,\footnote{
    In \cref{eqn:medi-mr-de-correction} we dropped the lower limit of the integral in
    \cref{eqn:medi-correction-general-rho}, but the correction would be early-time dominated were
    $w_\mathrm{DE} > -1/6$ (and is logarithmically sensitive when the inequality is saturated).
    Excluding this possibility is a more restrictive condition than requiring that dark energy
    redshift more slowly than matter, but extrapolations of a linear-in-$a$ equation of state from
    current data strongly preclude such a regime.
}
\begin{align}
    \frac{\chi(a_\mathrm{d}, a_m)}{\chi_{mr}(a_\mathrm{d}, a_m)}
    - 1
    \simeq
        - \frac{\bar{\rho}_\mathrm{DE}(a_m) / \bar{\rho}_m(a_m)}{14 - 12 [1 + w_\mathrm{DE}(a_m)]}
    &\simeq
        - \frac{(a_m / \amde)^{- 3 w_\mathrm{DE}}}{2 - 12 w_\mathrm{DE}}
    .
    \label{eqn:medi-mr-de-correction}
\end{align}
To make contact with the result for a cosmological constant [\cref{eqn:medi-with-cc}], the second
expression in \cref{eqn:medi-mr-de-correction} writes
$\bar{\rho}_\mathrm{DE}(a) / \bar{\rho}_m(a) = (a / \amde)^{- 3 w_\mathrm{DE}}$ where
the matter and dark energy densities cross at $a = \amde$.\footnote{
    The equality scale factor $\amde$, which is uniquely specified by
    $\Omega_m$ in \LCDM{} and measured as $0.751 \pm 0.010$ by DESI, is a more robust
    summary statistic than $\Omega_m$ with dynamical dark energy, with
    $\amde = 0.748_{-0.017}^{+0.016}$.
}
Approximating the equation of state as constant with value $w_\mathrm{DE}(a_m)$ for $z > z_m$ is
more than sufficient to qualitatively understand the effect in the
$w_\mathrm{DE}(a) = w_0 + (1 - a) w_a$ model, since the linearized equation of state has undergone
$70\%$ of its evolution toward its asymptotic value $w_\infty = w_0 + w_a$ by $z_m = 2.330$.
While \cref{eqn:medi-mr-de-correction} only captures the instantaneous rate of change in its density
[$w_\mathrm{DE}(a_m)$] at the endpoint of the integral, its effect is generally largest at the
latest times, particularly for fits to current data that extrapolate
$w_\mathrm{DE}(a \ll 1) \ll -1$.

For parameters preferred by current data, with $w_\infty \ll -1$ but $\amde$ largely unchanged in
fit compared to \LCDM{}, \cref{eqn:medi-mr-de-correction} shows that the dominant impact of
dynamical dark energy is to rapidly decrease its instantaneous contribution to the \medi{} at
$a > a_m$ compared to a cosmological constant, as illustrated in \cref{fig:medi-sensitivity}.
Namely, the extrapolation of the preferred phantomlike behavior both suppresses the numerator
in \cref{eqn:medi-mr-de-correction} (i.e., the dark energy density decreases with redshift)
and enhances the denominator (i.e., its density decreases more rapidly with redshift).
For instance, with $w_\mathrm{DE}(a_m) \approx -2$ inferred by DESI alone, the fiducial half-percent
suppression of the \medi{} by a cosmological constant is reduced by a factor of $30$.
Since the matter-era distance excess (for $\summnu = 60~\mathrm{meV}$) is just over half a percent,
dynamical dark energy thus resolves the BAO--CMB tension by eliminating its impact on distances
during matter domination ($z > z_m$).\footnote{
    In fact, the suppression of the dark energy density is relevant at lower redshift, too:
    in the fits to current data, the equation of state crosses $-1$ slightly before
    matter--dark-energy equality, increasing not only the ``predicted'' \medi{} but also the
    ``measured'' distance to matter domination $D_M(a_m) / r_\mathrm{d}$ as evident in
    \cref{fig:distance-to-md}.
    Both effects alleviate the matter-era distance excess, though the latter is subdominant.
}

\subsubsection{Reducible and irreducible evidence}\label{sec:dde-evidence}

To retain a Bayesian approach in quantifying preferences for new physics, we focus on posterior
distances, i.e., what for sufficiently normal distributions would be the norm of parameter (vector)
differences with the inverse covariance matrix serving as the metric.
Since some posteriors are not well approximated as normal, we estimate the distance from
a cosmological constant by computing the smallest mass level that encloses the point
$[w(z=0.5), w_a] = [-1, 0]$ and converting it to the corresponding number of $\sigma$s for a
two-dimensional normal distribution.
DESI reports a number of frequentist statistics~\cite{DESI:2025zgx}, which should be equivalent for
sufficiently well-constrained posteriors~\cite{Herold:2025hkb}, but are less straightforward to use
for interpreting and disentangling preferences for dark energy dynamics and the high-redshift
extensions we consider.

Second, we continue to fix a neutrino mass sum of $60~\mathrm{meV}$ to be minimally compatible with
neutrino oscillation constraints, and also to avoid the complication of CMB lensing's independent
disfavor of the suppression of structure from heavier neutrinos (i.e., marginalized over geometric
effects~\cite{Loverde:2024nfi}).
There is nuance in generalizing fixed-$\summnu$ results, in particular because the matter-era
distance tension must ultimately be addressed in the context of an \textit{a priori} unknown
neutrino mass sum.
For instance, $\summnu$ is conventionally restricted only to be nonnegative, so when most of the
posterior mass is below $60~\mathrm{meV}$ (as inferred by DESI and the CMB within \LCDM{}),
marginalizing over $\summnu$ reduces the apparent tension.
Conversely, marginalizing over neutrino mass sums larger than $60~\mathrm{meV}$ strengthens the
matter-era distance tension.

Finally, to minimize computational cost and enable counterfactual tests (that artificially shift
parameters preferred by the CMB), we employ CMB priors in place of likelihoods; following arguments
and evidence from \cref{sec:postdecoupling} and Refs.~\cite{Baryakhtar:2024rky, Loverde:2024nfi,
Lynch:2025ine, Costa:2025kwt}, a prior parametrized in terms of $\theta_s$, $R_\star$,
$x_\mathrm{eq}$ (see \cref{sec:predecoupling}) generalizes primary CMB information extremely well
from \LCDM{} to all the scenarios we consider.
Likewise, Ref.~\cite{DESI:2025zgx} established their efficacy under dynamical dark energy, which
follows from its negligible ($\lesssim 1\%$) impact on CMB lensing for parameters preferred by DESI
and the CMB.
We exclude lensing reconstruction, which would increase the joint preference to about $3 \sigma$,
for reasons discussed around \cref{fig:medi-tension}.
The considered extensions (including dynamical dark energy) again have a minimal impact on the
integrated Sachs-Wolfe effect and CMB lensing, with the exception of decaying dark matter (see
Ref.~\cite{Montandon:2026vuc} for a dedicated analysis).
Differences at the level of tenths of a standard deviation, whether from the choice of statistic,
neutrino mass sum, or dataset, are unimportant for our purpose, which is to assess the differential
impact of the matter-era distance excess and alternative explanations thereof on preferences for dark
energy dynamics.

\begin{figure}[t!]
\begin{centering}
    \includegraphics[width=\textwidth]{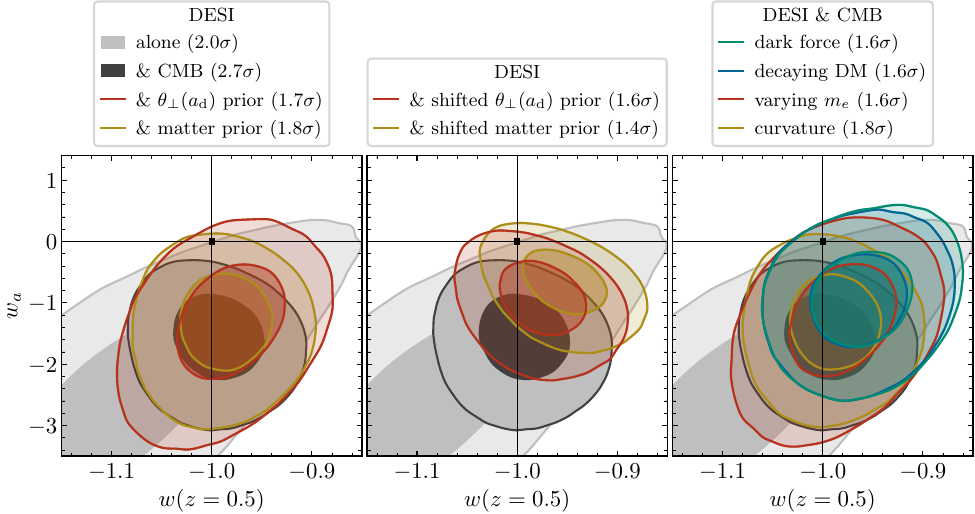}
    \caption{
        Impact of the matter-era distance excess (and alternative explanations thereof) on dynamical
        dark energy preferences from DESI DR2 and CMB data, presented as $w_a$ and the equation of
        state at redshift $0.5$ (close to the pivot redshift in all cases and much less correlated
        with $w_a$ than $w_0$ is) for visual clarity.
        Contours indicate 1 and $2 \sigma$ levels of the 2D distributions (i.e., the $39.3\%$ and
        $86.5\%$ mass levels thereof) to facilitate reading off posterior distances from
        $w(a) = -1$ (which are also noted in the legends).
        Each panel shows standard results from DESI alone (light grey) and with a CMB prior (dark
        grey).
        Including a CMB prior with only geometric
        [$\theta_\perp(a_\mathrm{d})$, gold] or only nongeometric (baryon and CDM densities, red)
        information (left panel) is insufficient to realize the CMB's added evidence; both are
        required to manifest the matter-era distance excess.
        Taking the full CMB prior but artificially shifting the mean of either
        $\theta_\perp(a_\mathrm{d})$ (gold) or the CDM density (red) to remove the matter-era
        distance excess (middle panel) eliminates the added preference from the CMB.
        Any of the high-redshift solutions presented in \cref{sec:high-redshift-solutions} (right
        panel) achieve a similar effect.
    }
    \label{fig:dde-vs-high-z}
\end{centering}
\end{figure}
To test that the CMB's added evidence for dark energy dynamics derives uniquely from its
high-redshift resolution of the matter-era distance excess, \cref{fig:dde-vs-high-z} shows that
combining DESI data with a CMB prior on either the CDM and baryon densities (which calibrates the
CMB's prediction for the \medi{}) or $1 / \theta_\perp(a_\mathrm{d})$ (which provides the
high-redshift anchor for the acoustic scale's measurement of the \medi{}) is insufficient to
reproduce the preference.
(See Ref.~\cite{Wang:2025mqz} for a similar analysis.)
In fact, adding either the CMB's geometric or nongeometric information only \textit{reduces}
the preference compared to DESI's alone.
Moreover, the added evidence also vanishes when artificially shifting the central value of
$\theta_\perp(a_\mathrm{d})$ or the CDM density by the precise amount to remove the matter-era
distance excess.

These tests establish the importance of dynamical dark energy's high-redshift behavior and,
moreover, that any other solution to the matter-era distance excess is a viable
alternative explanation of the BAO--CMB tension.
\Cref{fig:dde-vs-high-z} shows that a number of the models discussed in
\cref{sec:predecoupling,sec:postdecoupling} indeed remove the CMB's added evidence for dynamical
dark energy by independently resolving the matter-era distance excess.
Observe, for instance, that marginalizing over curvature has the same effect as removing the CMB's
geometric information, while modified recombination via a varying electron mass
(which allows nearly arbitrary values of $\omega_m r_\mathrm{d}^2$) is tantamount to removing the
CMB's nongeometric information.\footnote{
    Reference~\cite{Mirpoorian:2025rfp} previously proposed modified recombination as an alternative
    to dynamical dark energy, though they misattributed its effect to the change in $r_\mathrm{d}$
    itself.
}
Decaying dark matter and dark forces achieve a similar effect, but only in the direction of
decreasing $\omega_m r_\mathrm{d}^2$ and therefore increasing $w_a$ closer to zero.

\begin{figure}[t!]
\begin{centering}
    \includegraphics[width=\textwidth]{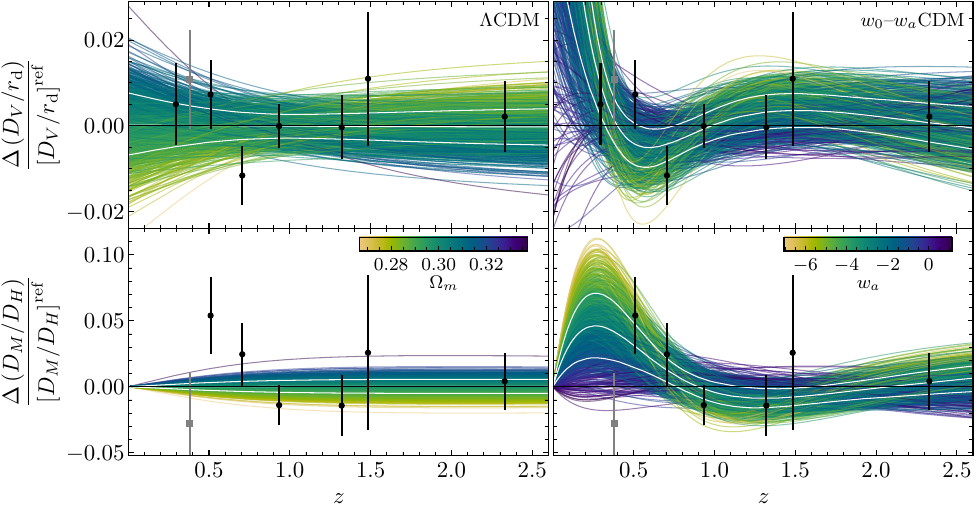}
    \caption{
        Features in DESI DR2 data that drive its preferences for dynamical dark energy over
        cosmological constant.
        Each panel depicts $1000$ posterior samples from DESI alone for each model as residuals
        compared to DESI's best-fit \LCDM{} model, with DESI DR2 data superimposed in black;
        each curve is colored by its value of the present matter fraction $\Omega_m$
        for \LCDM{} (left panels) or $w_a$ for $w_0$--$w_a$CDM (right panels).
        White lines depict the corresponding median and $\pm 1 \sigma$ quantiles.
        While \LCDM{} provides a perfectly adequate fit to the isotropic BAO scale
        $D_V / r_\mathrm{d} \equiv 1 / \theta_\mathrm{BAO}$ [\cref{eqn:theta-bao-DV}, top panels],
        it has extremely little freedom to accommodate the variations in the Alcock--Paczynski
        distortion $D_M / D_H$ (bottom panels), which varies with $\Omega_m$ alone in \LCDM{} and
        approaches $z$ for small $z$ in all cosmologies.
        In contrast, dynamical dark energy fits its low-redshift trend nearly perfectly with
        decreasing $w_a$.
        Since DESI DR2 does not supply a measurement of $D_M / D_H$ from its lowest redshift sample,
        we display the lowest-redshift measurement from the Sloan Digital Sky
        Survey~\cite{eBOSS:2020yzd} in grey, as it opposes this trend.
    }
    \label{fig:desi-lcdm-w0wa-spaghetti}
\end{centering}
\end{figure}

\begin{figure}[t!]
\begin{centering}
    \includegraphics[width=\textwidth]{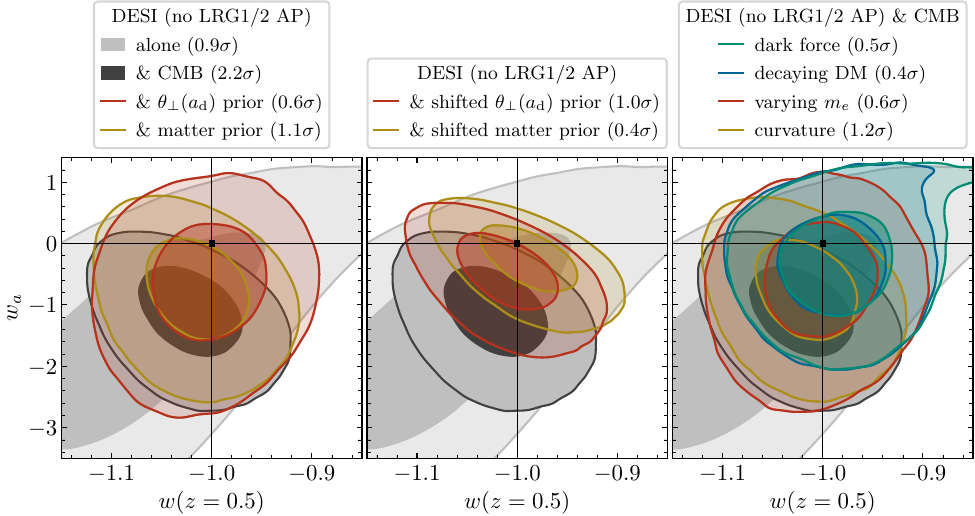}
    \caption{
        Same as \cref{fig:dde-vs-high-z}, but taking DESI's measurements of $D_V$ alone for its
        $z = 0.51$ and $0.706$ bins (i.e., marginalizing the respective likelihoods over
        $D_M / D_H$).
        With this subset of DESI data, solutions to the matter-era distance excess reduce the
        joint preference for dynamical dark energy to the $0.5 \sigma$ level.
    }
    \label{fig:dde-vs-high-z-drop-lrg12-ap}
\end{centering}
\end{figure}

Though the evidence for dynamical dark energy deriving from the matter-era distance excess has quite
a number of microphysically grounded alternatives (those discussed in
\cref{sec:predecoupling,sec:postdecoupling}) that do not rely on the extrapolation of
phenomenological models, the residual preference thereof is never much lower than that from DESI
data alone.
This irreducible evidence derives from DESI's more direct constraints on the low-redshift expansion
history as traced by its full set of observations at $z < z_m$.
\Cref{fig:desi-lcdm-w0wa-spaghetti} demonstrates that this evidence is most uniquely represented in
the Alcock--Paczynski distortion $D_M / D_H$, a ratio which asymptotes to $z$ for small redshift in
all cosmologies.
For further illustration, \cref{fig:dde-vs-high-z-drop-lrg12-ap} displays the same analyses as
\cref{fig:dde-vs-high-z} with the $D_M / D_H$ information removed from the DESI likelihoods at
redshifts $0.51$ and $0.706$.
DESI's independent preference for dynamics drops from $2 \sigma$ to $1 \sigma$, while that of the
DESI subset combined with CMB data (driven by the matter-era distance excess) persists at
$2.2 \sigma$.
Just as in \cref{fig:dde-vs-high-z}, the preference emerges only if sufficient CMB information is
included to realize the matter-era distance excess; alternative explanations thereof reduce the joint
evidence to as low as $0.4 \sigma$.

Reproducing the sharp feature at low redshift in DESI's data requires a rather extreme change to the
expansion history at very late times---one that \LCDM{} cannot accommodate.\footnote{
    The lack of low-redshift variability in $D_M / D_H$ in \LCDM{}, and its role in dynamical dark
    energy preferences, is also reflected in the direction in the $D_M(z)$--$D_H(z)$ plane best constrained by
    $\theta_\perp(a_\mathrm{d})$ in \LCDM{}~\cite{Efstathiou:2025tie}; indeed, in \LCDM{} this
    direction and $D_M / D_H$ are nearly aligned.
}
That this feature is fit so well by a phenomenological model like $w_0$--$w_a$ (with no theoretical
prior to regulate its predictions) might suggest that it is relatively susceptible to low-redshift
(high-leverage) statistical fluctuations.
The tests shown in \cref{fig:dde-vs-high-z-drop-lrg12-ap} do not call these data into question, but
merely identify the observations most responsible for the dynamical dark energy preference.
Indeed, were the physical behavior of the dynamical dark energy fit realized in a microphysical
model at both low and high redshift, the feature in $D_M / D_H$ and the matter-era distance excess
could be taken as consilient evidence.
While DESI DR2 does not derive a measurement of $D_M / D_H$ from its lowest redshift sample, the
lowest-redshift measurement from the Sloan Digital Sky Survey~\cite{eBOSS:2020yzd} opposes the trend
in the $w_0$--$w_a$ fit to DESI; future anisotropic measurements at low redshift, from both DESI and
other surveys with different sky coverage, may therefore be decisive in differentiating dynamical
dark energy from solutions to the matter-era distance excess alone.

\section{Summary and conclusions}
\label{sec:conclusions}

Recent observations position the acoustic scale as a precise and robust observable for constraining
the expansion history, with future surveys promising multifold improvements, especially at high
redshift.
In this work, we developed an analytic description of high-redshift distances that cleanly
compartmentalizes the acoustic scale's sensitivity to the dark-energy- and matter-dominated eras.
We conclude by summarizing the matter-era distance interval and its implications for the BAO--CMB
tension, neutrino masses, dynamical dark energy, and tests of the cosmological model going forward.

\subsection{The matter-era distance interval}

The CMB's acoustic scale information in the angular extent of the sound horizon is the second
best-measured number in cosmology (after the present-day CMB temperature), providing the most
precise and robust constraint on the integrated evolution of the dark energy density.
Any cosmological model that expands the freedom of this quantity must therefore contend with
parameter degeneracies with dark energy.
With low-redshift acoustic scale measurements reaching subpercent precision, however, current data
stand to simultaneously constrain both dark energy and a broad range of new physics at higher
redshift.

In \cref{sec:theory}, we showed how combining acoustic scale measurements from low-redshift surveys
and from the CMB cancels off a nearly optimal amount of their sensitivity to dark energy (even if
just a cosmological constant).
The difference in distances (measured relative to the sound horizon) to photon--baryon decoupling as
constrained by the CMB and to the highest redshift accessible to large-scale structure surveys---the
``matter-era distance interval''---represents the vast majority of the information on the interim
expansion history.
While DESI's Lyman-$\alpha$ forest sample, with effective redshift $2.330$, enables a direct and
model-independent measurement of the distance to matter domination (per se), \cref{sec:inference}
showed that all of DESI's BAO data combine to measure the distance to $z = 2.330$ with more than
quadruple the precision; this result, moreover, is largely robust to dark energy dynamics.
The \medi{} derived in this manner therefore represents not just a critical component of the
information in acoustic scale data but also a constraint on high-redshift dynamics that is agnostic
to low-redshift dynamics.

As a first application, \cref{sec:matter-density} demonstrated that the \medi{} represents the key
contribution of the CMB's geometric information to measurements of the late-time matter density from
the acoustic scale.
Without it, DESI presently constrains $\omega_m r_\mathrm{d}^2$ (i.e., the matter density in
acoustic scale units) with $1.7\%$ precision.
Even marginalizing over DESI's constraint on the matter density at $z < 2.330$, the \medi{} alone
measures the matter density (say, between $z = 2.330$ and decoupling) with nearly four times the
precision.
The combined measurement (i.e., taking the same matter abundance at all epochs) is just $1.5$ times
more precise than the latter, deriving from the modest correlation of the jointly inferred distance
to $z = 2.330$ with the late-time matter density.

As a corollary, the \medi{} generalizes arguments of Ref.~\cite{Loverde:2024nfi} on how cosmological
distances measure (or constrain) neutrino masses in \LCDM{}; moreover, it represents the
\textit{direct} measurement of neutrino masses from the expansion history.
That is, attributing to massive neutrinos any reduction in the measured \medi{} is no less
``direct'' than is attributing any reduction in the measured amplitude of late-time structure.
In both cases, the measurement is made relative to the CMB's massless-neutrino prediction
extrapolated from the conditions (the density and spatial distribution of matter) it infers at last
scattering, when neutrinos are relativistic.
And in both cases observations exhibit an excess where there should be a deficit.

The results in \cref{sec:matter-era-distance-interval} are general to physics before decoupling and
show that the parameter combination that encodes the \medi{}'s early-time dependence is
$\omega_{cb} r_\mathrm{d}^2$, the density of baryons and CDM in acoustic scale units, not the
density or drag horizon themselves.
\Cref{sec:predecoupling} explains how the \medi{} predicted by the CMB depends on density ratios at
last scattering, with \cref{eqn:medi-lcdm} applicable in any cosmology whose early-time expansion
history is well described by radiation and matter.
We then reviewed how high-redshift recombination modulates the extrapolation of density ratios from
last scattering to late times (and thus modifies the \medi{}), with little impact on primary CMB
anisotropies; high-density recombination, on the other hand, largely affects the \medi{} only via
small shifts in inferred early-time density ratios and is therefore more limited by CMB constraints.

By sequestering the dominant effects of dark energy, the \medi{} depends on high-redshift physics in
a manner that is not only physically transparent but also analytically tractable.
Indeed, we derived accurate analytic approximations for a variety of high-redshift effects beyond
\LCDM{}, including the effects of massive neutrinos (and their noninstantaneous nonrelativistic
transition), modified dynamics of dark matter (nontrivial redshifting or subcomponents that decay or
emerge after decoupling), spatial curvature, and the small impact of (possibly dynamical) dark
energy at high redshift.
Using these analytic results, in \cref{sec:postdecoupling} we showed that \medi{} measurements
robustly approximate constraints deriving from the full set of acoustic scale observations.

\subsection{Acoustic scale tensions}

The \medi{} provides a unified description of the growing tension between CMB and low-redshift BAO
observations, as explained in \cref{sec:medi-tension}.
In models whose postdecoupling expansion history is well described by \LCDM{}, the tension reduces
to a late-time matter density deficit, which has been identified as the origin of the geometric
incompatibility of CMB and BAO data with massive neutrinos~\cite{Loverde:2024nfi, Lynch:2025ine}.
Namely, the aforementioned measurement of the matter density from the acoustic scale is more than
$2 \sigma$ below the CMB's preferred matter density (including neutrinos with the minimum mass sum
allowed by neutrino oscillations).
However, there is no evidence of a matter density deficit strictly at the redshifts DESI itself
observes ($z \leq 2.330$)---its data alone are compatible with substantially larger matter
densities.
By showing that the \medi{} contains the vast majority of the acoustic scale's information on the
matter density, \cref{sec:matter-density} demonstrated that the matter-era distance excess underlies
the matter density deficit (rather than vice versa).
This analysis thus identifies the true tension as one at higher redshift: the acoustic scale (from
DESI and the CMB) measures (i.e., with reasonable robustness to the cosmological model) a matter-era
distance interval more than $2 \sigma$ larger than that predicted by the CMB---that is, based on its
geometry-independent, early-time information on the density in CDM and baryons.
When marginalizing over neutrino masses compatible with neutrino oscillation
data~\cite{deSalas:2020pgw, Esteban:2020cvm}, the matter-era distance tension is $2.6 \sigma$.

While \LCDM{} fits both DESI and CMB data well individually, the matter-era distance excess
represents the coarse-graining of the physics between decoupling and late times that prevents their
reconciliation.
The physical insight afforded by the \medi{} facilitates a broad and analytically grounded
assessment of potential solutions to the BAO--CMB tension (and identifies deficiencies in prior
interpretations of the tension and its implications).
\Cref{sec:high-redshift-solutions} explained how the \medi{} underpins the mechanism by which
previous specific proposals address the tension, including early
recombination~\cite{Baryakhtar:2024rky}, dark forces~\cite{Bottaro:2024pcb, Costa:2025kwt}, decaying
dark matter~\cite{Lynch:2025ine}, and spatial curvature~\cite{Chen:2025mlf}.
In \cref{sec:predecoupling} we also explored the (rather limited, except for high-redshift
recombination) extent to which predecoupling physics mediates the matter-era distance tension.

\Cref{sec:lensing-excess} also identified the physical mechanism by which freeing the optical depth
to reionization (i.e., from its measurements with large-scale polarization) alleviates the BAO--CMB
tension~\cite{Loverde:2024nfi, Sailer:2025lxj, Jhaveri:2025neg}, though only by about $0.6 \sigma$
when the neutrino mass sum is fixed to zero; the \medi{} thus remains poorly fit.
Excluding low-$\ell$ polarization data is far more effective at resolving the CMB lensing excess,
thereby removing the lensing-amplitude constraint on neutrino masses.
Though doing so reduces the matter-era distance excess to below $2 \sigma$ when fixing
$\summnu = 60~\mathrm{meV}$, it also makes the CMB compatible with substantially heavier neutrinos,
for which the \medi{} is underpredicted by an even greater absolute amount (but with a marginally
decreased significance of $2.3 \sigma$).
\Cref{fig:mnu-high-z-sols} in \cref{sec:mnu-vs-highz} shows that neutrino masses are more strongly
penalized by the matter-era distance excess when the lensing excess is resolved (by freeing the
optical depth) than vice versa.
Moreover, without low-$\ell$ polarization data, solutions to the matter-era distance excess
dramatically alleviate neutrino mass limits, reinforcing that the optical depth is of subdominant
importance to the geometric tension.
Nevertheless, a robust measurement of the optical depth, insofar as it informs the neutrino mass
sum allowed by CMB lensing, is vital to quantifying the matter-era distance excess.

Beyond new physics at high redshift, an analytic understanding of the tension motivates
counterfactual tests to identify the causal relationship between observations and parameter
constraints.
For instance, merely shifting the CMB's acoustic scale measurement removes the tension (though we
are unaware of any mechanism to realize the requisite $\sim 10 \sigma$ shift).
Likewise, increasing just the longitudinal distance measurement from DESI's Lyman-$\alpha$ sample by
$\sim 2 \sigma$ could possibly achieve the requisite shift in $D_M(a_m) / r_\mathrm{d}$; however, in
\LCDM{} the measurement of $D_M(a_m) / r_\mathrm{d}$ is not shifted even by excluding Lyman-$\alpha$
entirely.\footnote{
    We also checked that \medi{} measurements are robust to leaving out any one of DESI's other
    tracers, a finding congruent with the insensitivity of $D_M(a_m) / r_\mathrm{d}$ to dark energy
    dynamics.
}
Were the longitudinal Lyman-$\alpha$ measurement not so consistent with the \LCDM{} result
calibrated by all other tracers, it would supply more evidence for dynamical dark energy (as evident
in \cref{fig:distance-to-md}) and significantly alter the joint constraint on
$D_M(a_m) / r_\mathrm{d}$.
This measurement likely has the most leverage among DESI's data, since it informs the density
between redshifts $\sim 1.5$ and $2.330$ that determines how much comoving distance accumulates
beyond that measured by lower-redshift tracers.
The BAO shift due to nonlinear corrections~\cite{deBelsunce:2024rvv, Sinigaglia:2024kub,
Hadzhiyska:2025cvk}, which is not currently included in DESI's analysis of the Lyman-$\alpha$
forest~\cite{DESI:2025zpo}, could be a relevant systematic, though it is not expected to be large
enough to appreciably affect the matter-era distance excess.

\subsection{Implications for dark energy}

The matter-era distance excess underlies the implications of the BAO--CMB tension for not just
neutrino masses (and other high-redshift physics) but also dynamical dark energy, putting the two on
common physical footing and clarifying their relationship.
Contrary to what one might expect, dark energy dynamics do not address the excess by shifting DESI's
jointly inferred distance to matter domination $D_M(a_m) / r_\mathrm{d}$ (see
\cref{fig:distance-to-md}) but rather by evacuating dark energy at high redshift.
In the phenomenological $w_0$--$w_a$ model, current data prefer the dark energy density to increase
as the Universe expands at early times: by reaching a significantly lower density at $a < a_m$,
these dynamics eliminate the half-percent suppression of the \medi{} from a cosmological constant
(see \cref{fig:medi-sensitivity}), nearly enough to resolve the matter-era distance excess.
Our analytic description in \cref{sec:dde} pinpoints this solution's reliance on the extrapolated
and most physically implausible features of the canonical models of the dark-energy equation of
state, i.e., a leading-order Taylor expansion in $1 - a$.
This resolution to the excess is therefore not strictly guaranteed in any dark energy model that
resembles the phenomenological fit at low redshift.

As a corollary, the alleviation of the neutrino mass tension in dynamical dark energy
scenarios~\cite{DESI:2024mwx, DESI:2025zgx} should also not be taken for granted.
Our findings---namely, the negligible impact on the inferred distance to matter domination---refute
the presumption that dark energy dynamics relax neutrino mass limits strictly through low-redshift
effects.
The key role of dark energy's high-redshift behavior in modulating neutrino mass inference calls for
more careful scrutiny of their interplay in microphysical models of dark energy; it also suggests
that cosmological measurements of neutrino masses may be more robust to dark energy than
phenomenological analyses imply.

The quantitative and counterfactual tests in \cref{sec:dde-evidence} confirm that the CMB's role in
the preference for dynamical dark energy is to introduce the matter-era distance excess, contrary to
the claims in Ref.~\cite{DESI:2025zgx} that its role is to measure the present-day matter density or
matter fraction (neither of which the CMB directly measures with relevant precision).
Rather, the CMB calibrates the \textit{early}-time densities in CDM and baryons and therefore a
baseline prediction of the distance photons traverse in matter domination (see
\cref{sec:medi-tension}).
The CMB itself allows for the late-time matter abundance to vary much more broadly, should it be
increased by massive neutrinos (\cref{sec:mnu}) or hyperlight scalars (\cref{app:scalars}) or
decreased by a subcomponent decaying (\cref{sec:ddm}) or by a scalar-mediated dark force
(\cref{sec:dark-forces}); the present-day matter fraction varies even more widely if dark energy
evolves.
The importance of the matter-era distance interval also underscores the danger in conflating the
matter fraction $\Omega_m$ and the matter density $\omega_m \equiv \Omega_m h^2$: since the
densities of each matter component are conventionally independent parameters, the matter fraction in
reality parametrizes \textit{dark energy}.
These distinctions are key to identifying alternative explanations of the data, and
\cref{fig:dde-vs-high-z} shows that indeed any alternative explanation of the matter-era distance
excess removes the CMB's added evidence for dark energy dynamics: both in general and in specific
models, removing the matter-era distance excess reduces the preference to below the $1.7 \sigma$
level.

The irreducible preference for dark energy dynamics from DESI is driven by features at low redshift
that could plausibly be interpreted as sample variance that is overfit by the relatively arbitrary
low-redshift dynamics allowed by the $w_0$--$w_a$ model (see \cref{fig:desi-lcdm-w0wa-spaghetti}).
(We do not explore the role of supernova distances, partly to limit scope and partly due to the
incoherent implications of current datasets and their sensitivity to revised
calibration~\cite{DES:2025sig, Hoyt:2026fve}, which largely removed their contribution to the
evidence.)
The physical effects that yield evidence for phenomenological models must also be contextualized in
their underlying microphysical implementations.
Given that this model only otherwise improves the joint BAO--CMB fit through its extrapolated and
nominally unphysical dynamics at high redshift, this irreducible evidence must be weighed against
microphysically grounded, alternative explanations of the matter-era distance excess.

That said, a microphysical model that simultaneously explains both the low-redshift Alcock--Paczynski
distortion feature and the matter-era distance tension would be favored over the high-redshift
scenarios discussed in \cref{sec:high-redshift-solutions} that explain only the latter.
While coupled quintessence---namely, the dark-force mediator from \cref{sec:dark-forces} augmented
with a nonflat potential~\cite[e.g.,][]{Khoury:2025txd, Bedroya:2025fwh}---can explain both
features, its explanations of each rely on separate components of the model (the potential shape and
the dark matter coupling, respectively) and thus lack the (presumably accidental) consilience that
strengthens the evidence for the $w_0$--$w_a$ explanation.
The most important theoretical input needed from fundamental models of dynamical dark energy is
therefore not how their low-redshift behavior maps into the $w_0$--$w_a$ parameter space but rather
whether they can simultaneously explain the low-redshift feature in $D_M / D_H$ and the matter-era
distance excess.

\subsection{Outlook}

The matter-era distance interval highlights the key role played by high-redshift distance
measurements in constraining cosmology.
Moreover, the current matter-era distance tension underscores the need to study extensions to the
cosmological model in tandem with neutrino masses, in particular models for which low-redshift
distances are informative.
Fortuitously, the \medi{} is analytically tractable and supports a physically transparent
interpretation of current data and their potential implications for the cosmological model.
Robust explanatory power derives from an analytic (or at least semianalytic) understanding of the
full physics of the problem to inform the interpretation of high-dimensional likelihoods over
parameters, the combination thereof from multiple observations, and their posterior marginalization.

In the near term, the matter-era distance excess represents a focal point for future updates from
DESI.
Stratifying DESI's high-redshift tracers into multiple bins, or incorporating new tracers at
$z \sim 2$~\cite{Bault:2026ffr}, could also be valuable in testing the jointly inferred distance to
matter domination.
Alternatively calibrated distances, like those from supernovae, could also be incorporated to at
least further constrain the relative evolution of distances with redshift.
Naturally, acoustic scale measurements from the rest of the sky would provide independent
measurements of the distance to matter domination, though they would be perhaps more valuable in
testing whether the low-redshift evidence for dynamical dark energy
(\cref{fig:desi-lcdm-w0wa-spaghetti}) is simply sample variance.
The \medi{} also identifies unique physics opportunities in the high-redshift BAO observations that
would be possible with extensions to DESI and stage-5 spectroscopic surveys~\cite{Sailer:2021yzm,
DESI:2022lza, Schlegel:2022vrv, Ebina:2024ojt, Spec-S5:2025uom, Feder:2025ffm}.
BAO distances at yet-higher redshift could provide multiple measurements of the matter-era distance
interval that better disentangle the dwindling influence of dark energy on distances, enabling
robust measurements of neutrino masses or other early-Universe physics with nontrivial redshift
dependence.
It would also be valuable to pursue geometric neutrino mass measurements that are completely
independent of dark energy.

Despite its critical importance in constraining cosmological parameters, the matter-era distance
interval lacks discriminatory power: a single number is insufficient to differentiate multiple
effects on the high-redshift expansion history.
While future surveys may measure multiple intervals, the growth of structure remains (as always) an
independent probe.
Of course, many models in the class we considered in this work have largely scale-independent
effects on observable scales.
Decisive evidence for physics beyond \LCDM{}---in tandem with a neutrino mass measurement---may
ultimately only be possible for those most predictive models whose signatures in large-scale
structure are strongly correlated with their effects on the expansion history.

\begin{acknowledgments}
We thank Simone Ferraro, Lloyd Knox, Marilena Loverde, Gabe Lynch, Noah Sailer, Murali Saravanan,
Kendrick Smith, Martin White, and Matias Zaldarriaga for discussions and feedback.
Research at Perimeter Institute is supported in part by the Government of Canada through the
Department of Innovation, Science and Economic Development and by the Province of Ontario through
the Ministry of Colleges and Universities.
This work made use of a number of open-source software packages~\cite{Foreman-Mackey:2012any,
Hogg:2017akh, Foreman-Mackey:2019, corner, Harris:2020xlr, Virtanen:2019joe, Hunter:2007ouj,
hoyer2017xarray, arviz_2019, Meurer:2017yhf, cmasher, jax2018github, kidger2021on}.
\end{acknowledgments}

\appendix

\section{Calculational details}\label{app:calculational-details}

\subsection{Analytic results}\label{app:analytic}

In this appendix we outline the approximations made in deriving the analytic results quoted in
\cref{sec:medi-tension,sec:postdecoupling}.
In all cases, small perturbations to the matter-era distance interval are calculated from
\cref{eqn:medi-correction-general-rho} as
\begin{align}
    \Delta \chi(a_\mathrm{d}, a_m)
    \equiv \chi(a_\mathrm{d}, a_m) - \chi_{mr}(a_\mathrm{d}, a_m)
    &\simeq
        - \frac{1}{2} \int_{\tau(a_\mathrm{d})}^{\tau(a_m)} \ud \tau
        \frac{\bar{\rho} - \bar{\rho}_{mr}}{\bar{\rho}_{mr}}
    \label{eqn:medi-correction-general-rho-tau}
    \\
    &\simeq
        - \frac{1}{2 H_{100} \sqrt{\omega_{cb}}}
        \int_{a_\mathrm{d}}^{a_m} \frac{\ud a}{\sqrt{a + a_\mathrm{eq}}}
        \frac{\bar{\rho} - \bar{\rho}_{mr}}{\bar{\rho}_{mr}}
    \label{eqn:medi-correction-general-rho-simple}
    .
\end{align}
For cases that introduce a new physical timescale, results that account for the residual effect of
radiation after equality (which precedes decoupling) are not as tractable, in which case we replace
$a_\mathrm{eq}$ with zero and $\bar{\rho}_{mr}$ with $\bar{\rho}_{cb}$.
The various models we study also impact the drag horizon, but we only consider regimes in which
deviations in the expansion history become most important well after decoupling, in which case
the effect on $r_\mathrm{d}$ is smaller than that on $\chi(a_\mathrm{d}, a_m)$ by at least an order
of magnitude (as could be anticipated from \cref{fig:medi-sensitivity}).
That said, computing the change to $r_\mathrm{d}$ would only require a straightforward modification
of the calculations presented before (by including the plasma sound speed in the integrand).

\subsubsection{Massive neutrinos}\label{app:mnu}

\Cref{sec:mnu} presents a simplified, piecewise approximation to the \medi{} that treats neutrinos'
nonrelativistic transition as instantaneous, which we now improve upon by accounting for their
rather gradual transition.
We first write the perturbation to the density in a Universe with fully relativistic neutrinos in
the form
\begin{align}
    \frac{\bar{\rho} - \bar{\rho}_{mr}}{\bar{\rho}_{mr}}
    &= \frac{\bar{\rho}_{\nu, \mathrm{rel}}}{\bar{\rho}_{mr}}
        \frac{\bar{\rho}_\nu - \bar{\rho}_{\nu, \mathrm{rel}}}{\bar{\rho}_{\nu, \mathrm{rel}}}
    = \frac{f_\nu a_\nu}{a + a_\mathrm{eq}}
        \frac{\bar{\rho}_\nu - \bar{\rho}_{\nu, \mathrm{rel}}}{\bar{\rho}_{\nu, \mathrm{rel}}}
    ,
    \label{eqn:mnu-pert-ito-neutrino-pert}
\end{align}
where $a_\nu \equiv \left\langle a p / m_\nu \right\rangle = \left\langle p / T_\nu \right\rangle T_{\nu, 0} / m_\nu$ and $f_\nu = m_\nu \bar{n}_\nu / \bar{\rho}_m$ as in \cref{sec:mnu}.
Analytic expressions for \cref{eqn:mnu-pert-ito-neutrino-pert} are not possible for general
$T_\nu / m_\nu$ but are in the nonrelativistic and relativistic limits.
Since the energy density in a single neutrino eigenstate is
\begin{align}
    \bar{\rho}_\nu
    &= 2 \int \frac{\ud^3 p}{(2 \pi)^3} \frac{\sqrt{p^2 + m_\nu^2}}{e^{p / T_\nu} + 1}
    = 2 \int \frac{\ud^3 p}{(2 \pi)^3} \frac{1}{e^{p / T_\nu} + 1}
        \cdot
        \begin{dcases}
            p + m_\nu^2 / 2 p
            , & m_\nu \ll T_\nu,
            \\
            m_\nu
            , & m_\nu \gg T_\nu,
        \end{dcases}
\end{align}
keeping only the term leading-order in $m_\nu / T_\nu$ in each limit,
the perturbation to the energy density takes the form
\begin{align}
    \frac{\bar{\rho}_\nu - \bar{\rho}_{\nu, \mathrm{rel}}}{\bar{\rho}_{\nu, \mathrm{rel}}}
    &\simeq
        \begin{dcases}
            \frac{1}{2}
            \frac{\left\langle T_\nu / p \right\rangle}{\left\langle p / T_\nu \right\rangle}
            \left( \frac{m_\nu}{T_\nu} \right)^2
            ,
            & m_\nu \ll T_\nu,
            \\
            \frac{1}{\left\langle p / T_\nu \right\rangle}
            \frac{m_\nu}{T_\nu}
            - 1.
            \\
        \end{dcases}
\end{align}
Here angled brackets denote phase-space averages, which are written such that they each evaluate to
a single number: $\left\langle T_\nu / p \right\rangle \approx 0.456144$
and $\left\langle p / T_\nu \right\rangle \approx 3.15137$.

To proceed, we separate the integral \cref{eqn:medi-correction-general-rho-simple} into these two
regimes at some scale factor $a_\times$ near $T_{\nu, 0} / m_\nu$ (specified below), yielding
\begin{align}
    \frac{\Delta \chi(a_\mathrm{d}, a_m)}{1 / H_{100} \sqrt{\omega_{cb}}}
    &\simeq
        - f_\nu
        \left(
            \frac{\left\langle T_\nu / p \right\rangle \left\langle p / T_\nu \right\rangle}{2}
            \left[
                \frac{a^2 - 4 a a_\mathrm{eq} - 8 a_\mathrm{eq}^2}{3 a_\nu \sqrt{a + a_\mathrm{eq}}}
            \right]_{a_\mathrm{d}}^{a_\times}
            +
            \left[
                \frac{a + 2 a_\mathrm{eq} + a_\nu}{\sqrt{a + a_\mathrm{eq}}}
            \right]_{a_\times}^{a_m}
        \right)
        ,
    \label{eqn:medi-mnu}
\end{align}
where $a_\nu \equiv \left\langle a p / m_\nu \right\rangle = \left\langle p / T_\nu \right\rangle T_{\nu, 0} / m_\nu$ as in \cref{sec:mnu}.
We do not distinguish between neutrino eigenstates above since in practice we take a degenerate
hierarchy, but one could do so by summing over the contribution of each.
We are free to choose $a_\times$ different from $a_\nu$ to minimize errors over some range of
masses; we find $a_\times \approx 1.32 a_\nu$ to be roughly optimal for $z_m = 2.330$, yielding
subpercent errors in the sensitivity coefficient for
$0.02 \lesssim \summnu / \mathrm{eV} \lesssim 0.3$.
The only other approximation errors in \cref{eqn:medi-mnu} are $\mathcal{O}(f_\nu)$
[i.e., in the small correction itself, which is the $\mathcal{O}(f_\nu^2)$ error made in
\cref{eqn:medi-correction-general-rho-simple} to begin with], which is only $\gtrsim 10\%$ for
$\summnu \gtrsim 1.3~\mathrm{eV}$, and those when $a_\nu$ is not much greater than $a_m$.
The further simplifications made in \cref{eqn:medi-mnu-apx} by taking $a_\mathrm{eq}$ and
$a_\mathrm{d}$ negligible compared to $a_m$ and $a_\nu$ only marginally increase this error.

\subsubsection{Hyperlight scalar fields}\label{app:scalars}

The gravitational effects of sufficiently light scalar fields closely resemble those of massive
neutrinos~\cite{Amendola:2005ad, Marsh:2011bf, Hlozek:2014lca, Hlozek:2016lzm,
Baryakhtar:2024rky}---in particular, misaligned scalars that begin oscillating after decoupling and
are therefore hyperlight, i.e., have mass
$H_0 \ll m_\phi < H(a_\star) \approx 3 \times 10^{-29}~\mathrm{eV} \ll H(a_\mathrm{eq})$ and
can only be a subcomponent of dark matter.
At early times while frozen, they contribute to the stress-energy tensor as (a negligible amount of)
dark energy and subsequently redshift like matter once oscillating; in the hyperlight mass range,
they increase the matter abundance after decoupling like massive neutrinos.
They also do not cluster on observable scales~\cite{Hu:2000ke, Amendola:2005ad}.
Such light scalars cannot be both nonrelativistic and a thermal relic, for which reason their
late-time abundance is determined by not just their mass (as for neutrinos) but also their
(nonthermal) initial conditions.
As such, hyperlight scalars make for a similar but distinct application of the approximation
\cref{eqn:medi-mr-piecewise} of the \medi{}, in which $\omega_m \geq \omega_{cb}$ and $a_\times$
are separate free parameters.
Like for massive neutrinos, the simple piecewise treatment of \cref{eqn:medi-mr-piecewise} provides
a good estimate; however, the result may in fact be computed analytically with the perturbative
method described in \cref{sec:postdecoupling}, using analytic solutions for the scalar dynamics in a
matter-dominated Universe.

Unlike neutrino masses (which identify a special temperature and therefore scale factor),
the scalar's mass introduces a physical timescale to the problem, so we take a pure-matter Universe.
Hyperlight scalars whose mass satisfies $H(a_m) \ll m_\phi \ll H(a_\mathrm{eq})$ begin oscillating
when $m_\phi = 3 H(a_\mathrm{osc}) / 2$.
The homogeneous value of a hyperlight scalar field initially misaligned at $\bar{\phi}_i$ evolves
according to
$\bar{\phi}(t) / \bar{\phi}_i = \sin (m_\phi t) / m_\phi t = \sin (m_\phi t) / [a(t) / a_\mathrm{osc}]^{3/2}$~\cite[see, e.g.,][]{Baryakhtar:2024rky}.
Here $a_\mathrm{osc} \simeq \sqrt[3]{9/4} \left( m_\phi / \sqrt{\omega_{cb}} H_{100} \right)^{-2/3}$
is defined such that, at $a \gg a_\mathrm{osc}$, the scalar's energy density
$\bar{\rho}_\phi = \left( \ud \bar{\phi} / \ud t \right)^2 / 2 + m_\phi^2 \bar{\phi}^2 / 2$
oscillates about $m_\phi^2 \bar{\phi}_i^2 / 2 (a / a_\mathrm{osc})^{3}$, where
$m_\phi^2 \bar{\phi}_i^2 / 2$ is the early-time energy density in the frozen condensate.
As such, at late times the scalar comprises a fraction
$f_\phi = 3 \left( \bar{\phi}_i / \sqrt{2} \Mpl \right)^2 / 4$ of the total matter density.

The correction to the \medi{} from a hyperlight scalar,
\begin{align}
    \frac{\Delta \chi(a_\mathrm{d}, a_m)}{1 / H_{100} \sqrt{\omega_{cb}}}
    &\simeq
        - \frac{1}{2} \int_{a_\mathrm{d}}^{a_m}
        \frac{\ud a}{\sqrt{a}}
        \frac{\bar{\rho}_\phi}{\bar{\rho}_{cb}}
    \simeq
        - \frac{f_\phi}{2} \int_{a_\mathrm{d}}^{a_m}
        \frac{\ud a}{\sqrt{a}}
        \frac{(a / a_\mathrm{osc})^3 \bar{\rho}_\phi}{m_\phi^2 \bar{\phi}_i^2 / 2}
    ,
    \label{eqn:medi-scalar-general}
\end{align}
has an unilluminating form in terms of special functions, but for $H(a_m) \ll m_\phi \ll
H(a_\mathrm{d})$ we need only its asymptotic forms.
The final result,
\begin{align}
\begin{split}
    \frac{\Delta \chi(a_\mathrm{d}, a_m)}{1 / H_{100} \sqrt{\omega_{cb}}}
    &\simeq - f_\phi \bigg[
            \sqrt{a_m} \left(
                1
                + \frac{\cos(2 m_\phi t_m) / 15 - \sin^2(m_\phi t_m) / 5}{(a_m / a_\mathrm{osc})^{3}}
            \right)
    \\ &\hphantom{ {}={} - f_\phi \bigg[ }
            - \frac{\sqrt{3 a_\mathrm{osc}} \Gamma(1/3)}{5 \sqrt[3]{2}}
            - \sqrt{a_\mathrm{d}} \frac{(a_\mathrm{d} / a_\mathrm{osc})^{3}}{7}
        \bigg]
    ,
    \label{eqn:medi-scalar}
\end{split}
\end{align}
with $m_\phi t_m = (a_m / a_\mathrm{osc})^{3/2}$, is accurate to $\mathcal{O}(f_\phi)$ [i.e., the
subleading corrections to \cref{eqn:medi-correction-general-rho-simple}],
but incurs increasingly large errors as $m_\phi$ approaches
$H(a_\mathrm{eq}) \simeq 2.2 \times 10^{-28}~\mathrm{eV}$ from neglecting radiation's impact on
\cref{eqn:medi-scalar-general} both explicitly and on the solution $\bar{\phi}(t)$.
The dependence of \cref{eqn:medi-scalar} on $a_\mathrm{d}$ is negligible (and subdominant to the
errors from neglecting radiation).
\Cref{eqn:medi-scalar} also breaks down for $m_\phi < 3 H(a_m)$, since $a_m$ is no longer much
greater than $a_\mathrm{osc}$; the sinusoidal terms in \cref{eqn:medi-scalar} improve the accuracy
for masses just above $3 H(a_m)$ by accounting for the initial oscillations of $\bar{\rho}_\phi$ but
do not qualitatively alter the result.
Dropping the aforementioned terms yields
\begin{align}
    \frac{\Delta \chi(a_\mathrm{d}, a_m)}{1 / H_{100} \sqrt{\omega_{cb}}}
    &\simeq - f_\phi \sqrt{a_m} \left[
            1
            - \sqrt{a_\mathrm{osc} / a_m} \Gamma(1/3) \sqrt{3} / 5 \sqrt[3]{2}
        \right].
    \label{eqn:medi-scalar-apx}
\end{align}
The numerical coefficient multiplying $\sqrt{a_\mathrm{osc} / a_m}$ is about $- 0.74$, whose
difference from unity (as would be derived in a piecewise approximation to the \medi{}) accounts for
the scalar's impact at early times before it rapidly oscillates.

Because their phenomenology is so similar to massive neutrinos', with their suppression of structure
growth even sharing a close quantitative relationship to their impact on the \medi{}, we do not
present a dedicated analysis of hyperlight scalars.
The analytic results \cref{eqn:medi-mnu-apx,eqn:medi-scalar-apx} allow for a straightforward
translation of constraints on $f_\nu$ to $f_\phi$ as a function of the scalar mass.

\subsubsection{Decaying dark matter}\label{app:ddm}

\Cref{sec:ddm} discusses decaying dark matter as a nearly instantaneous reduction in the matter
density; we now account for both the relativistic decay products and the early phase of decay.
The decay of nonrelativistic dark matter into dark radiation (whose energy densities are labeled
$\bar{\rho}_\mathrm{ddm}$ and $\bar{\rho}_\mathrm{ddr}$, respectively) with general rate
$\Gamma_\mathrm{dd}(t)$ is described by the Boltzmann equations
\begin{subequations}
\begin{align}
	\dd{\left( a^3 \bar{\rho}_\mathrm{ddm} \right)}{t}
	&= - \Gamma_\mathrm{dd} a^3 \bar{\rho}_\mathrm{ddm}
	\\
	\dd{\left( a^4 \bar{\rho}_\mathrm{ddr} \right)}{t}
	&= a \Gamma_\mathrm{dd} a^3 \bar{\rho}_\mathrm{ddm}
    ,
\end{align}
\end{subequations}
which have formal solutions
\begin{subequations}\label{eqn:rho-ddm-ddr-formal}
\begin{align}
	a(t)^3 \bar{\rho}_\mathrm{ddm}(t)
	&= a(t_i)^3 \bar{\rho}_\mathrm{ddm}(t_i)
        \exp \left[ - \int_{t_i}^t \ud t' \, \Gamma_\mathrm{dd}(t') \right]
    \label{eqn:rho-ddm-formal}
	\\
	a(t)^4 \bar{\rho}_\mathrm{ddr}(t)
	&= a(t_i)^4 \bar{\rho}_\mathrm{ddr}(t_i)
        + a(t_i)^3 \bar{\rho}_\mathrm{ddm}(t_i)
        \int_{t_i}^t \ud t' \,
        a(t') \Gamma_\mathrm{dd}(t')
        \exp \left[ - \int_{t_i}^{t'} \ud t'' \, \Gamma_\mathrm{dd}(t'') \right]
    .
    \label{eqn:rho-ddr-formal}
\end{align}
\end{subequations}
With a time-independent decay rate, the integral in \cref{eqn:rho-ddm-formal} is trivial, but the
outer integral in \cref{eqn:rho-ddr-formal} is not, as it depends explicitly on both time and scale
factor.
That is, we took $\Gamma_\mathrm{dd}$ to be a physical time scale, so as in \cref{app:scalars} we
restrict results to matter domination in which $a \propto t^{2/3}$.
The remaining integral has the asymptotic expansions of the form
\begin{subequations}\label{eqn:ddr-integral-asymptotics}
\begin{align}
    \int \ud x \, x^n e^{- x}
    &= - \Gamma(1 + n)
        + x^{n}
        \left[
            \frac{x}{1 + n}
            - \frac{x^{2}}{2 + n}
            + \mathcal{O}(x^{3})
        \right]
    \\
    &= e^{- x}
        \left[
            -1
            - \frac{n}{x}
            + \mathcal{O}(x^{-2})
        \right]
    ,
\end{align}
\end{subequations}
where $n = 2/3$ and $x = \Gamma_\mathrm{dd} t$ in our case.

Like for massive neutrinos in \cref{app:mnu}, we resort to splitting the integral at $a_\times$ into
the two regimes of \cref{eqn:ddr-integral-asymptotics} and choose an optimal value for $a_\times$
after the fact.
Taking $\bar{\rho}_\mathrm{ddr}(t_i) = 0$, $t_i \to 0$, and
$a(t) / a_\mathrm{dd} = (\Gamma_\mathrm{dd} t)^{2/3}$ (which defines $a_\mathrm{dd}$),
\cref{eqn:rho-ddm-ddr-formal} becomes
\begin{subequations}
\begin{align}
    \bar{\rho}_\mathrm{ddm}(a)
    &= \frac{a(t_i)^3 \bar{\rho}_\mathrm{ddm}(t_i)}{a^3}
        e^{- (a / a_\mathrm{dd})^{3/2}}
    \\
    \bar{\rho}_\mathrm{ddr}(a)
    &= \frac{a_\mathrm{dd} a(t_i)^3 \bar{\rho}_\mathrm{ddm}(t_i)}{a^4}
        \begin{dcases}
            3 (a / a_\mathrm{dd})^{5/2} / 5
            , & a \ll a_\mathrm{dd},
            \\
            \Gamma(5/3) - e^{- (a / a_\mathrm{dd})^{3/2}}
            , & a \gg a_\mathrm{dd}.
        \end{dcases}
\end{align}
\end{subequations}
The perturbation to the matter density (where $\bar{\rho}_m$ is that if the decaying species were
stable) thus has early- and late-time behavior given by
\begin{align}
    \frac{\bar{\rho} - \bar{\rho}_m}{\bar{\rho}_m}
    = f_\mathrm{ddm}
        \begin{dcases}
            - 2 (a / a_\mathrm{dd})^{3/2} / 5
            , & a \ll a_\mathrm{dd},
            \\
            - 1
            + \frac{\Gamma(5/3)}{a / a_\mathrm{dd}}
            , & a \gg a_\mathrm{dd},
        \end{dcases}
\end{align}
where $f_\mathrm{ddm} = \lim_{a \to 0} \bar{\rho}_\mathrm{ddm}(a) / \bar{\rho}_m(a)$
is the fraction of matter (by energy density) that decays.
\Cref{eqn:medi-correction-general-rho-simple} evaluates to
\begin{align}
    \frac{\Delta \chi(a_\mathrm{d}, a_m)}{1 / H_{100} \sqrt{\omega_m}}
    &\simeq
        f_\mathrm{ddm}
        \left(
            \frac{\sqrt{a_\mathrm{dd}}}{10}
            \left[
                \left( \frac{a}{a_\mathrm{dd}} \right)^2
            \right]_{a_\mathrm{d}}^{a_\times}
            +
            \left[
                \sqrt{a}
                \left(
                    1
                    + \frac{\Gamma(5/3)}{a / a_\mathrm{dd}}
                \right)
            \right]_{a_\times}^{a_m}
        \right)
    ,
    \label{eqn:medi-ddm}
\end{align}
which, when taking $a_\times = 1.12 a_\mathrm{dd}$, is accurate at the percent level for
$a_\mathrm{dd} \lesssim a_m / 2$ (up to corrections next-to-leading in $f_\mathrm{ddm}$), with a
marginal increase in error as $a_\mathrm{dd}$ approaches $a_\mathrm{eq}$.
Here $\omega_m = \lim_{a \to 0} a^3 \bar{\rho}_m(a) / 3 H_{100}^2 \Mpl^2$.
Taking $a_\mathrm{d} / a_\mathrm{dd}$ to zero yields \cref{eqn:medi-ddm-apx}.

Because the decay is exponentially fast, soon after $\Gamma_\mathrm{dd} t = 1$ the Universe is well
described by decoupled matter and radiation (in different amounts than before decay).
Therefore, a treatment of the postdecay regime that does not rely on a perturbative expansion in
small $f_\mathrm{ddm}$ simply takes the standard matter--radiation result \cref{eqn:medi-mr} with
$\omega_{cb}$ multiplied by $1 - f_\mathrm{ddm}$ and $a_\mathrm{eq}$ augmented to account for the
relativistic decay products via
$a_\mathrm{eq} = [\left. a_{\mathrm{eq}} \right\vert_{f_\mathrm{ddm} = 0} + \Gamma(5/3) f_\mathrm{ddm} a_\mathrm{dd}] / ( 1 - f_\mathrm{ddm} )$.

\subsubsection{Dark forces}\label{app:dark-forces}

\Cref{sec:dark-forces} studies dark matter subject to a long-range force mediated by a linear
coupling to a light scalar, in which the homogeneous dynamics of the scalar induce the dark matter
particle mass to evolve with time.
Employing analytic results from Ref.~\cite{Costa:2025kwt}, we derive the analytic result for
\cref{eqn:medi-correction-general-rho-tau} quoted in \cref{sec:dark-forces}.
The homogeneous mode of the mediator $\bar{\varphi}$ evolves as
$\bar{\varphi}(a) = \bar{\varphi}_i - f_\chi d_{m_\chi}^{(1)} (1 - 1 / y + 2 \ln y)$ where
$2 y = 1 + \sqrt{1 + a / a_\mathrm{eq}}$,
while the dark matter mass $m_\chi$ evolves in proportion to
$e^{d_{m_\chi}^{(1)} (\bar{\varphi} - \bar{\varphi}_i)} \approx 1 + d_{m_\chi}^{(1)} (\bar{\varphi} - \bar{\varphi}_i)$.
Here $f_\chi$ is the fraction of all matter coupled to $\varphi$ and $d_{m_\chi}^{(1)}$ the linear
coupling coefficient thereof; the force strength relative to gravity is
$\beta = ( d_{m_\chi}^{(1)} )^2$.
Note that $y = 1 + \tau / 4 \sqrt{2} [a_\mathrm{eq} H_\mathrm{eq}]^{-1}$, i.e., the dynamics are
expressed uniquely in terms of conformal time; likewise, in a matter--radiation Universe,
$\tau / [a_\mathrm{eq} H_\mathrm{eq}]^{-1} = 2 \sqrt{2} \left( \sqrt{1 + a / a_\mathrm{eq}} - 1 \right)$.
As such, \cref{eqn:medi-correction-general-rho-tau} may be integrated analytically without
approximation [beyond working at $\mathcal{O}(\beta f_\chi^2)$].

The perturbation to the energy density is
\begin{align}
    \frac{\bar{\rho}(a) - \bar{\rho}_{mr}(a)}{\bar{\rho}_{mr}(a)}
    = \frac{
            f_\chi \left[ m_\chi(a) / m_\chi(a_i \to 0) - 1 \right]
        }{
            1 + a_\mathrm{eq} / a
        }
        + \frac{ \left( \ud \bar{\varphi} / \ud \ln a \right)^2}{3},
\end{align}
the two terms encoding the evolution of $m_\chi$ and the contribution of the mediator's kinetic
energy.
Taking a cue from the analytic solution for $\bar{\varphi}$, it proves expedient to change variables
to $y$ itself, such that
\begin{align}
\begin{split}\label{eqn:medi-lrf}
    \frac{\Delta \chi(a_\mathrm{d}, a_m)}{1 / H_{100} \sqrt{\omega_{cb}}}
    &\simeq \sqrt{a_\mathrm{eq}} \beta f_\chi^2
        \Bigg[
            \frac{2 y (2 \ln y - 1)}{1 - 1 / 2 y}
            - \frac{2 y}{3}
            - 4 \ln y
            + 1
            + \frac{4 y - 1}{3 y^2 (1 - 1 / 2 y)}
        \Bigg]_{y(a_\mathrm{d})}^{y(a_m)}
    .
\end{split}
\end{align}
The sensitivity coefficient
$\partial \ln \chi(a_\mathrm{d}, a_m) / \partial \left( 1 + \beta f_\chi^2 \right)$
evaluates to $2.2867$ for $a_m = 1 / (1 + 2.330)$, which matches the numerical result up to
corrections at next-to-leading order in $\beta f_\chi^2$.
Since our interest is $a_m \gg a_\mathrm{d} > a_\mathrm{eq}$ [and thus $y(a_m) \gg 1$],
\cref{eqn:medi-lrf-apx} presents the leading-order behavior of \cref{eqn:medi-lrf},
neglecting the lower boundary of the integral as well.
This result could be further generalized to mediators with masses larger than $H_0$ using analytic
solutions from Ref.~\cite{Costa:2025kwt}, accounting for the mediator's late-time abundance with a
calculation analogous to that for uncoupled scalars in \cref{app:scalars}.

\subsection{Numerical implementation}\label{app:numerics}

The combination of CMB data we use includes foreground-marginalized ACT DR6 temperature and
polarization data (with likelihood implemented in \textsf{candl}~\cite{Balkenhol:2024sbv}) and a
subset of \Planck{} PR3 observations~\cite{Planck:2019nip} prescribed by Ref.~\cite{AtacamaCosmologyTelescope:2025blo} for
combination with ACT, restricted to multipoles $\ell \leq 1000$ in temperature and $\leq 600$ in
polarization and temperature-polarization cross correlation and including the $\ell < 30$
temperature and polarization likelihoods from PR3.
Specifically, we use the foreground-marginalized \texttt{Plik\_lite} variants with
\texttt{Commander} and \texttt{SimAll} for $\ell < 30$~\cite{Planck:2018vyg, Planck:2019nip, clipy}.
We truncate the ACT likelihood at $\ell = 4000$ to mitigate needless computational cost (see
Ref.~\cite{Costa:2025kwt}).
Following Ref.~\cite{SPT-3G:2025bzu}, we also include SPT-3G D1~\cite{SPT-3G:2025bzu} data without
further amendments.
We include no lensing reconstruction data.

For models that significantly modify dynamics before decoupling (those considered in
\cref{sec:predecoupling}), we use the Boltzmann code \textsf{CLASS}~\cite{Blas:2011rf,
Lesgourgues:2011re}; we use \textsf{CAMB}~\cite{Lewis:1999bs} for the varying neutrino mass results
presented in \cref{fig:medi-tension}.
We employ precision settings as enumerated in Appendix D of Ref.~\cite{Costa:2025kwt}.
We perform parameter sampling with \textsf{emcee}~\cite{Foreman-Mackey:2012any, Hogg:2017akh,
Foreman-Mackey:2019}, running long enough to obtain at least $10^4$ independent samples.
We take standard priors for \LCDM{} parameters:
$\omega_b \sim \mathcal{U}(0.005, 0.035)$,
$\omega_c \sim \mathcal{U}(0.01, 0.25)$,
$100 \theta_s \sim \mathcal{U}(0.9, 1.1)$,
$\tau_\mathrm{reio} \sim \mathcal{U}(0.02, 0.2)$,
scalar spectral amplitude $A_s$ via $\ln(10^{10} A_s) \sim \mathcal{U}(1.61, 3.91)$,
scalar spectral tilt $n_s \sim \mathcal{U}(0.8, 1.2)$,
and $\summnu / \mathrm{eV} \sim \mathcal{U}(0, 1.5)$.
Here $\mathcal{U}(a, b)$ specifies a uniform distribution between $a$ and $b$
and $\mathcal{N}(\mu, \sigma)$ a normal distribution with mean $\mu$ and standard deviation $\sigma$.
For the extensions in \cref{sec:predecoupling}, we sample
$\Delta N_\mathrm{eff} \sim \mathcal{U}(-2, 3)$ and the early-time values of the electron mass
and fine structure constant relative to their present values,
$m_{e, i} / m_{e, 0} \sim \mathcal{U}(0.7, 1.3)$
and
$\alpha_i / \alpha_{0} \sim \mathcal{U}(0.7, 1.3)$.
When varying either of the latter, we replace $\omega_b$ with
$\tilde{\omega}_b = \omega_b / \left( m_{e, i} \alpha_i^2 / m_{e, 0} \alpha_0^2 \right)$ and
likewise for $\omega_c$, each sampled from the same prior as above.
The posterior over these parameters is more efficiently sampled because the rescaling removes the
degeneracy with $a_\star$ [per \cref{eqn:R-star,eqn:x-eq}].

For all other results that do not rely on a Boltzmann code, we employ a prior on the angular extent
of the sound horizon and the baryon and CDM densities (via $R_\star$ and $x_\mathrm{eq}$, as defined
in \cref{sec:predecoupling}) derived from \LCDM{} fit to the aforementioned primary CMB data.
The background evolution in such cases is implemented with \textsf{JAX}~\cite{jax2018github} with
ordinary differential equation solvers supplied by \textsf{Diffrax}~\cite{kidger2021on}.
In these cases, we sample
$\omega_b \sim \mathcal{U}(0.02, 0.025)$,
$\omega_c \sim \mathcal{U}(0.01, 0.5)$,
and $\omega_\mathrm{DE} \sim \mathcal{U}(0.0, 1.0)$.
The narrower prior on the baryon density, which is nearly centered on the values preferred by the
CMB (and BBN) with width of $30$--$40$ $\sigma$, minimizes changes to the recombination history
(which we do not model) while still allowing for substantial variation in $r_\mathrm{d}$.

The extensions in \cref{sec:predecoupling} sample the curvature density
$\omega_k \sim \mathcal{U}(-0.5, 0.5)$,
the would-be present density in decaying dark matter
$\tilde{\omega}_\mathrm{ddm} \equiv \lim_{a \to 0} a^3 \bar{\rho}_\mathrm{ddm} / 3 H_{100}^2 \Mpl^2 \sim \mathcal{U}(0.0, 0.1)$,
or the analogously defined density in coupled dark matter
$\tilde{\omega}_\chi \sim \mathcal{U}(0.01, 0.5)$ and the coupling strength relative to gravity
$\beta \sim \mathcal{U}(10^{-6}, 0.1)$.
For dynamical dark energy (\cref{sec:distance-to-matter-domination,sec:dde-evidence}), we sample the
present equation of state
$w_0 \sim \mathcal{U}(-3, 1)$ and, for the linear-in-$a$ model,
$w_\infty = \lim_{a \to 0} w(a) = w_0 + w_a \sim \mathcal{U}(-20, 0)$.
Unlike those in Ref.~\cite{DESI:2025zgx}, these priors do not truncate a nonnegligible fraction of
the posterior in the $w_a$ direction.

\bibliography{references,manual}

\end{document}